\documentclass[12pt]{article}
\usepackage{amssymb,amsmath,amsthm,mathtools,mathrsfs,latexsym,amsxtra,graphicx,appendix,setspace,subfig,fix-cm,color,cite,adjustbox,tabularx,comment,afterpage,float,caption,hyperref,scalerel}

\numberwithin{equation}{section}

\usepackage{titlesec}
\titleformat{\paragraph}
{\normalfont\normalsize\bfseries}{\theparagraph}{1em}{}
\titlespacing*{\paragraph}{0pt}{3.25ex plus 1ex minus .2ex}{1.5ex plus .2ex}

\newcommand{\beq}{\begin{equation}}
\newcommand{\be}{\begin{equation}}
\newcommand{\ee}{\end{equation}}
\newcommand{\bea}{\begin{eqnarray}}
\newcommand{\eea}{\end{eqnarray}}

\newcommand{\pa}{\partial}

\newcommand{\nn}{\nonumber}

\newcommand{\ti}[1]{\tilde{#1}}

\newcommand{\zb}{\bar{z}}

\newcommand{\vs}[1]{\vspace{#1pt}}
\newcommand{\hs}[1]{\hspace{#1pt}}
\newcommand{\nmax}{n_{\rm max}}
\newcommand{\Nmax}{N_{\rm max}}
\newcommand{\Nmin}{N_{\rm min}}
\newcommand{\bz}{\mathbb{Z}}

\begin{document} 

\begin{titlepage}
\thispagestyle{empty}
\begin{center}

\hfill \\
\hfill \\
\vskip 0.75in
{\Large 
{\bf Covering space maps for $n$-point functions \vspace*{0.5cm} \\ with three long twists}
}\\

\vskip 0.4in

{\large Benjamin A. Burrington${}^{a}$ and Ida G.~Zadeh${}^{b}$
}\\
\vskip 4mm

${}^{a}$
{\it Department of Physics and Astronomy, Hofstra University, Hempstead, NY 11549, USA} \vskip 1mm
${}^{b}$
{\it Mathematical Sciences and STAG Research Centre, University of Southampton, Highfield, Southampton SO17 1BJ, U.K.} \vskip 1mm
\end{center}

\vskip 0.35in

\begin{center} {\bf Abstract } \end{center}
We consider correlation functions in symmetric product orbifold CFTs on the sphere, focusing on the case where all operators are single-cycle twists, and the covering surface is also a sphere.  We directly construct the general class of covering space maps where there are three twists of arbitrary lengths, along with any number of twist-2 insertions.  These are written as a ratio of sums of Jacobi polynomials with $\Delta N+1$ coefficients $b_N$.  These coefficients have a scaling symmetry $b_N\rightarrow \lambda b_N$, making them naturally valued in $\mathbb{CP}^{\Delta N}$.  We explore limits where various ramified points on the cover approach each other, which are understood as crossing channel specific OPE limits, and find that these limits are defined by algebraic varieties of $\mathbb{CP}^{\Delta N}$.  We compute the expressions needed to calculate the group element representative correlation functions for bare twists.  Specializing to the cases $\Delta N=1,2$, we find closed form for these expressions which define four- and five-point functions of bare twists. 
\vfill

\end{titlepage}

\setcounter{page}{1}
\setcounter{tocdepth}{2}

\tableofcontents
\thispagestyle{empty}

\vspace{2cm}
\newpage

\section{Introduction and summary}\label{section_intro}
\setcounter{page}{1}
\hypersetup{pageanchor=false}

Symmetric product orbifold conformal field theory (CFT) is a prominent ingredient in the study of string theory, holographic correspondence, as well as studies of black hole microstates \cite{Dixon:1985jw,Dixon:1986jc,Dijkgraaf:1989hb,Vafa:1995zh,Vafa:1995bm,Strominger:1996sh,Dijkgraaf:1996xw,Maldacena:1997re,Giveon:1998ns,Elitzur:1998mm,deBoer:1998us,Maldacena:1999bp,Seiberg:1999xz,deBoer:1999gea,Larsen:1999uk,David:1999ec,Dijkgraaf:2000fq,Argurio:2000tb,Gukov:2004ym,Avery:2010qw} -- see \cite{Aharony:1999ti,David:2002wn} for reviews. In particular, various formulations of the AdS$_3$/CFT$_2$ correspondence incorporate families of the symmetric product orbifold CFTs, distinct by the seed theories on which the symmetric group acts, as either the dual CFT or a point on its moduli space \cite{Gaberdiel:2014cha,Gaberdiel:2015mra,Eberhardt:2017pty,Datta:2017ert,Giribet:2018ada,Eberhardt:2018sce,Eberhardt:2018ouy,Eberhardt:2019qcl,Eberhardt:2019niq,Belin:2019rba,Eberhardt:2019ywk,Belin:2020nmp,Eberhardt:2020akk,Eberhardt:2020bgq,Gaberdiel:2020ycd,Balthazar:2021xeh,Eberhardt:2021vsx,Gaberdiel:2023dxt,Aharony:2024fid,Gaberdiel:2024dva,Chakraborty:2025nlb,Eberhardt:2025sbi}. An important aspect of these solvable theories is the computation of their correlation functions. While there exist general prescriptions for computing the correlation functions of symmetric product orbifold CFTs \cite{Dixon:1986qv,Lunin:2000yv,Dei:2019iym}, as well as many exact computations (of mainly 3- and 4-point functions) including \cite{Arutyunov:1997gt,Arutyunov:1997gi,Jevicki:1998bm,Lunin:2001pw,Pakman:2009zz,Pakman:2009ab,Burrington:2012yn,Burrington:2018upk,Roumpedakis:2018tdb,DeBeer:2019oxm,Li:2020zwo,Burrington:2022dii,Burrington:2022rtr,AlvesLima:2022elo,Guo:2023czj,Guo:2024edj,Guo:2025eaf}, computation of generic higher-point functions still remains a challenging problem. The goal of this paper is to compute families of arbitrary higher-point functions and to provide exact formulae.

The necessity of computing higher-point correlation functions of symmetric product orbifold CFTs is twofold. On the one hand, to perform non-trivial tests of the holographic correspondence in some of its most powerful incarnations, namely string theory on AdS$_3$ backgrounds, one has to match correlation functions in the symmetric orbifold CFT, string worldsheet theory, and supergravity \cite{Gaberdiel:2007vu,Dabholkar:2007ey,Taylor:2007hs,Dei:2019osr,Dei:2020zui,Bertle:2020sgd,Dei:2021xgh,Dei:2021yom,Dei:2022pkr,Gaberdiel:2022oeu,Dei:2023ivl,Knighton:2023mhq,Knighton:2024qxd,Yu:2024kxr,Yu:2025qnw}. On the other hand, to reach points in the moduli space of the dual CFT which are suitable to describe black holes, one has to perturb the symmetric orbifold CFT along specific directions, corresponding to exactly marginal (1,1) operators in the CFT. Conformal perturbation theory involves computation of multi-integrals of particular types of higher-point functions, originally considered in \cite{Kadanoff:1978pv,KADANOFF197939,Cardy:1987vr,Kutasov:1988xb,Dijkgraaf:1987jt}, for recent progress see \cite{Behan:2017mwi,Burrington:2023vei} and references therein. The correlation functions computed in this work provide new data for testing the AdS$_3$/CFT$_2$ dualities, but are also relevant for conformal perturbation theory. 

Symmetric product orbifold CFTs \footnote{In this work we use the language of symmetric product orbifolds even though the techniques can be used anytime the group action on the fundamental fields is a permutation -- see e.g. \cite{Belin:2014fna,Belin:2015hwa} for holographic correspondences of this type.} are constructed by copying a {\it seed} CFT $N$ times, and then identifying states that are related by permuting the copies of the seed CFT.  The orbifold theory contains new states on the cylinder where fields are periodic up to the action of the orbifold group: these are the twisted sector states.  The twisted sectors are labeled by the conjugacy classes of the symmetric group $S_{\bold N}$, and we denote the ``conjugacy class of $g$'' as $[g]$.  We note that the conjugacy class $[g]$ is a group action invariant concept (by conjugation), and so is in this sense orbifold invariant.  By the state-operator mapping, there are associated twisted sector operators as well.  The lowest dimension operators in their twist class are called ``bare twists'' and we will denote these with an un-dressed $\sigma_{[g]}$.  When we wish to consider more general twist fields, with possible excitations, we will denote them as $\widehat{\sigma}_{[g]}$.

While the twisted sector operators are labeled by their conjugacy class $[g]$, they may be expanded in terms of effective operators which have twisted boundary conditions which are labeled by individual group elements \cite{Lunin:2000yv,Lunin:2001pw,Pakman:2009zz,Dei:2019iym}.  These non-orbifold invariant twist operators impose boundary conditions up to the action of specific symmetric group element $g\in S_{\bold N}$ acting on the fields.  We denote these by dropping the bracket notation, $\sigma_g$, emphasizing that they correspond to individual group elements rather than conjugacy classes.  Group elements of $S_{\bold N}$ are decomposed in terms of disjoint cycles, and the non-orbifold invariant twist operators corresponding to single-cycle group elements may be considered as fundamental.  This is because correlation functions of twist fields corresponding to group elements with many cycles may be constructed by taking limits of correlation functions constructed from single-cycle twist operators.  We will therefore focus on correlators of single-cycle twist operators in this work.

Being more explicit, single-cycle orbifold-invariant twist operators are obtained from summing over the elements in each conjugacy class
\be\label{tw}
\sigma_{[\hat{w}]}(z)=\frac1{\sqrt{w(N-w)!N!}}\sum_{g\in S_{\bold N}}\sigma_{g(1\cdots w)g^{-1}}\ .
\ee
Above, we have used the shorthand notation $\hat{w}$ to represent one of the cycles in the conjugacy class $[\hat{w}]$, rather than $g$, to emphasize that these are single-cycles ({\bf w}indings).  Without the ``hat'' the $w$ refers to the integer length of the cycle. A generic correlation function of bare twist operators is of the form
\be\label{corr}
\langle\sigma_{[\hat{w}_1]}(z_1)\;\sigma_{[\hat{w}_2]}(z_2)\cdots\sigma_{[\hat{w}_\ell]}(z_\ell)\rangle\ ,
\ee
which may be written as a sum over correlation functions of conjugacy class representative-dependent twist operators
\be\label{corr_ii}
\langle\sigma_{\hat{w}_1}(z_1)\;\sigma_{\hat{w}_2}(z_2)\cdots\sigma_{\hat{w}_\ell}(z_\ell)\rangle\ .
\ee
Thus, the main challenge remains to compute correlation functions of the form \eqref{corr_ii}.  To construct excited twist operator correlators, one may use various techniques \cite{Lunin:2000yv,Lunin:2001pw,Burrington:2012yn,Burrington:2014yia,Burrington:2015mfa,Burrington:2022dii,Burrington:2022rtr}
to dress the calculations of the bare twists.

The correlation functions computed in this work are of the form
\be\label{corr_iii}
\langle\sigma_{\hat{w}_1}(y_1)\;\;\sigma_{\hat{w}_2}(y_2)\;\;\sigma_{\hat{w}_3}(y_3)\;
\underbrace{{\sigma_{\hat{2}}(z_{1})\;\;\cdots\;\;\sigma_{\hat{2}}(z_{k}}}_{k\,{\rm insertions}})\rangle\ \vs{-5}
\ee
where ${\hat{2}}$ refers to some un-specified 2-cycles.  We will focus on the connected part of correlation functions of bare twist operators.  We note that these form the skeleton of the higher-point functions with excited twist operators, particularly those needed in high order conformal perturbation theory. We explicitly provide new exact formulae for $k=1,2$, and we will also comment on how to take limits of the above functions to merge the twist-2 operators into higher twists.

One method to compute correlation functions of twist fields on the sphere in symmetric orbifold CFTs is to find maps to a branched covering surface where the fields twisted under $\sigma_w$ are mapped to a single field with usual periodic boundary conditions \cite{Lunin:2000yv,Lunin:2001pw}. \footnote{A different method to compute correlation functions in symmetric product orbifold CFTs is the stress-energy method of \cite{Dixon:1986qv}. In this work we use the covering space method pioneered in \cite{Lunin:2000yv}, bypassing the need to perform the integrals of the Weyl factor by using \cite{Dei:2019iym}.} The covering space is a Riemann surface whose genus $g$ is determined by the structure of the twists on the original (base) space. Finding covering space maps is in general a difficult problem. We will directly construct new maps which allow us to compute the higher-point functions \eqref{corr_iii}. 

In the large ${\bold N}$ limit (i.e. large central charge ${\bold N}c$ of the symmetric product orbifold CFT), the leading order contribution to the connected correlation functions \eqref{corr} comes from the spherical covering surfaces $g=0$ \cite{Lunin:2000yv} -- see \cite{Burrington:2018upk,Roumpedakis:2018tdb,DeBeer:2019oxm,Li:2020zwo,Burrington:2022dii,Burrington:2022rtr} for a partial list of more recent studies of aspects of large ${\bold N}$ symmetric orbifolds. This limit in turn corresponds holographically to the leading order contribution to the genus expansion of the string worldsheet theory  \cite{Gaberdiel:2018rqv,Eberhardt:2018ouy,Eberhardt:2019ywk}. In this work we will compute connected correlation functions \eqref{corr_iii} with genus 0 covering spaces.

To compute $n$-point functions of bare twist operators, we need certain information about the covering space maps.  The basic procedure for this was worked out in \cite{Lunin:2000yv}, although a more algebraic approach is taken in the work of \cite{Dei:2019iym}.  In \cite{Dei:2019iym}, they take advantage of the fact that a specific fractional mode of the stress tensor acting on a bare twist state gives a null vector (also noted in \cite{Roumpedakis:2018tdb,Burrington:2018upk,DeBeer:2019oxm,Burrington:2022dii,Burrington:2022rtr}).  This may be used to constrain the form of the $n$-point functions of bare twists to a specific function of map parameters, up to a constant.  Requiring the $n$-point function agree with factorization, i.e. the conformal bootstrap, fixes this remaining constant and so fixes the form of the $n$-point function in terms of map parameters.  One then only needs to these map parameters, defined near the ramified points,  from the covering space map.

Near the ramified points at finite locations $t_i$, one defines the expansion
\be
z(t)=z_i+a_i(t-t_i)^{w_i} + \mathcal{O}((t-t_i)^{w_i+1})\  \label{tiExpand}
\ee
where $z$ is the coordinate of the base space, and $t$ is the coordinate on the cover. 
For the point near infinity, we have
\be
z(t)= a_{n-1} t^{w_{n-1}} +  \mathcal{O}(t^{w_{n-1}-1}) \ . \label{tinfExpand}
\ee
In both of these expressions, $w_i$ is the size of the single cycle defining the twist operator.  We will denote
\be
r_i=w_i-1 \label{ramDef}
\ee
which is the ``ramification'' of the point in the map, and denotes the number of ``extra sheets'' that come together at this point.  We also need to define the coefficients of the unramified images of infinity which are given by
\be
z(t)= \frac{C_\rho}{t-t_{\rho}}+ \mathcal{O}((t-t_\rho)^0)\ . \label{InfImage} 
\ee
The $n$-point functions are given by \cite{Dei:2019iym}
\begin{align}
&\langle \sigma_{\hat{w}_0}(z_0,\zb_0) \sigma_{\hat{w}_1}(z_1,\zb_1) \cdots \sigma_{\hat{w}_{n-2}}(z_{n-2},\zb_{n-2}) \sigma_{\hat{w}_{n-1}}(\infty,\bar{\infty}) \rangle \label{npointGen_i} \\[5pt]
&= \prod_{i=0}^{n-2}w_i^{-\frac{c(w_i+1)}{12}} w_{n-1}^{\frac{c(w_{n-1}+1)}{12}} \prod_{j=0}^{n-2} |a_j|^{-\frac{c(w_j-1)}{12 w_j}} |a_{n-1}|^{\frac{c(w_{n-1}-1)}{12 w_n}} \prod_{\rho} |C_{\rho}|^{-\frac{c}{6}}\ ,\nn
\end{align}
where $c$ is the central charge of the seed CFT, and we have assumed $c=\ti{c}$.  Note, in the above we have indexed $i=0 \cdots n-1$ for the $n$ bare twists.  This allows us to denote $(t_0=0,z_0=0)$, and $(t_1=1,z_1=1)$, and we use interchangeably $r_\infty=r_{n-1}$ and $(t_{n-1}=\infty,z_{n-1}=\infty)$. Furthermore, the above is to be read for a specific group element representative.  One must then sum over all preimages of the maps, see \cite{Pakman:2009zz} and \cite{Dei:2019iym}. 

While \eqref{npointGen_i} gives the desired correlation function in terms of map data, one still needs to find the map and extract this data.  It is the purpose of this paper to construct and explore new covering space maps.  

The rest of the paper is organized as follows.  In section \ref{SphereCover} we give a brief introduction to sphere covering spaces and their connection to ordinary differential equations.  In section \ref{GenMaps} we consider the construction of covering space maps.  We start in section \ref{3PtMaps} by considering the hypergeometric differential equation and introduce an interesting limit of the hypergeometric sum that reconstructs the well known map in \cite{Lunin:2000yv}.  In section \ref{nPtMaps} we consider a construction from the literature \cite{Ishkhanyan:2014wma} which writes solutions to Heun's differential equations as a finite sum of hypergeometric functions.  We apply our method for extracting pairs of polynomials from hypergeometric sums to generate the covering space map
\be
z(t)=\frac{f_2(t)}{f_1(t)}=\frac{ \sum\limits_{N=N_{\rm min}}^{N_{\rm max}} b_N\, t^{(N+1)}P^{(N+1),-(n_1+n_3-N)}_{n_3-N-1}(1-2t)}{\sum\limits_{N=N_{\rm min}}^{N_{\rm max}} b_N P^{-(N+1),-(n_1+n_3-N)}_{n_1} (1-2t) }\ , \label{near0.intro}
\ee 
where $n_1$, $n_3$, $\Nmin$, and $\Nmax$ are integers, the coefficients $b_N$ are parameters of the map, and $P_{\gamma}^{\alpha,\beta}$ are the Jacobi polynomials.  We show that the Wronskian has the form
\begin{align}
& W=f_2'f_1-f_2 f_1'= t^{\Nmin} (t-1)^{(n_1+n_3-\Nmax-1)} Q(t)
\end{align}
and so the ramifications of the map at $t=0,1,\infty$ are given by $r_0=\Nmin$, $r_1=(n_1+n_3-\Nmax-1)$, and $r_\infty=n_3-n_1-1$. The identification of ramifications using the Wronskian is reviewed in section \ref{SphereCover}, and the specific case above is derived in section \ref{3PtMaps}.  The polynomial $Q(t)$ has degree $\Delta N=\Nmax-\Nmin$, and the zeros of this polynomial determine the location of a cloud of ramification 1 points in the map: we denote this total ramification by $r_c=\Delta N$.  In section \ref{sec_map_gen} we show that these maps are sufficient to cover all group theoretically allowed $r_0$, $r_1$, $r_\infty$, and $r_c$.  The coefficients $b_N$ and $\lambda b_N$ define the same map and we therefore argue that these coefficients are valued in ${\mathbb{CP}}^{\Delta N}$, which is of the correct dimension to parameterize the $\Delta N$ cross ratios of a $(3+\Delta N)$-point function.  

In section \ref{OPELimSec} we analyze the maps \eqref{near0.intro}.  In section \ref{subsec_opelim} we consider the OPE limits where one of the ramified points in the cloud approaches one of the ramified points at $t=0$, $t=1$, or $t=\infty$, summarized in table \ref{TableOPE}.  We find that the $b_N$ parameterize both the location and the crossing channel of the OPE limits (i.e. which group element representative is taken from amongst the product of the conjugacy classes).  We also consider other OPE limits, including in section \ref{DN2OPEs} where we consider the special case $\Delta N=2$. This is the lowest value of $\Delta N$ where there are multiple ramified points in the cloud and so we can study the OPE limits as these points approach each other, and we find explicitly the cases where they fuse into a ramification 2 point (twist-3), and when they fuse into a ramification 0 point (untwisted).  All examples found in our work give OPE limits as homogeneous polynomials in the $b_N$ which are set to 0, giving them as algebraic varieties in ${\mathbb{CP}}^{\Delta N}$. Finally, in section \ref{CorrSect}, we construct the correlation functions.  We focus on cases where $\Delta N$ is small so that we may analytically evaluate the Wronskian, and use this to find closed form expressions for the 4-point and 5-point functions in the $\Delta N=1,2$ cases.  We end with a discussion and future directions in section \ref{DiscSection}.  We also provide several appendices which give background for Jacobi polynomials (appendix \ref{JacobiIDappx}), proofs of Jacobi polynomial identities used in the main text (appendix \ref{APXproofs}), constraints from group theory that show our maps are general (appendix \ref{appx.radd}), some detailed examples of taking the OPE limits (appendix \ref{Appx.OPE.Examples}), and an algorithmic technique for computing the Wronskian (appendix \ref{WronskGen}).    

\subsection{Brief introduction to sphere covering spaces}
\label{SphereCover}

In this subsection we briefly introduce some necessary components for motivating our covering space maps.  First, a natural place to start is to consider functions that are well defined on the sphere, rather than on a general Riemann surface.  Furthermore, we expect a generic point $z$ to be mapped to a finite number of points $t$ in the covering space, and that all generic points should be mapped to the same number of points on the cover.  This strongly suggests the use of polynomials, and so we consider covering space maps of the form
\be
z(t)=\frac{f_2(t)}{f_1(t)} \label{coverfrac}
\ee  
where $f_1$ and $f_2$ are polynomials that have been fully reduced so that they share no common zeros.  There are special points where the function $z(t)$ is ramified, i.e. in the neighborhood of some $t_i$ the map is of the form \eqref{tiExpand}.  This requirement on the covering map can be rephrased as
\be
\pa z = a_i w_i (t-t_i)^{w_i-1}+\cdots
\ee
where $\pa$ is the derivative with respect to $t$.  We recall the definition of ramification \eqref{ramDef}: $r_i=w_i-1$.  When $r_i=0$ this is an ordinary point where the map is one-to-one in a small neighborhood.  The points where $r_i\ge 1$ are said to be ramified points in the map.  Technically $r_i=-2$ are also ordinary points, and correspond to unramified images of $z=\infty$; any points with $r_i\leq -3$ are also ramified, and correspond to specific cycles in the twist operator at $z=\infty$.  We will consider the case of single-cycle twist operators, and so the point at $z=\infty$ is associated with only one cycle, and we choose to map this to the point $t=\infty$ in the cover.  One may take limits of these maps and place other cycles at the point at infinity, should one choose to.  Thus, for the case at hand, we will consider the case where $r_i\geq 1$ (except for the unramified images of $z=\infty$, given by the simple zeros of $f_1$).  This is the usual presentation, and if forced, we may always choose our map such that multi-cycle twist operators do not get mapped to infinity.

The ramifications are related to the group elements in the symmetric group $S_{\bold N}$, as explained in appendix \ref{appx.radd}.  The ramification of a group element is the minimal number of 2-cycle group elements it takes to construct the group element.  Thus, given a group product that multiplies to the identity, we may count the ramification of group element.  For a given product, the genus of the covering surface is determined by the Riemann-Hurwitz formula
\be\label{RHf}
g=\frac{1}{2} \sum_i r_i -S+1
\ee  
where $S$ is the total number of sheets in the cover, and is simply the total number of distinct indices appearing in the cycles of the group elements in the product under consideration. \footnote{More generally, the Riemann Hurwitz formula can be written in terms of the Euler characteristic of the base space and the covering surface, in which case it reads $\chi^{\uparrow}=S \chi - \sum_i r_i$, where $\chi$ is the Euler characteristic of the base space and $\chi^{\uparrow}$ is the Euler characteristic of the $S$ sheeted cover.  Using the Euler characteristic is more natural for disconnected covering surfaces because the Euler characteristic is additive over the disconnected pieces.} The genus of the cover being 0 puts certain restrictions on what group products one can consider.  These correspond in the orbifold CFT to the leading order in the large ${\bold N}$ limit of the CFT \cite{Lunin:2000yv}. 

Applying the derivative to \eqref{coverfrac} we find
\be
\pa z= \frac{f_2' f_1-f_2 f_1'}{f_1^2}
\ee 
where we have truncated notation $\pa f_i=f_i'$.  The numerator of this expression is the Wronskian of two functions.  This Wronskian must admit zeros of the form $(t-t_i)^{r_i}$, and so one may factor this from the Wronskian.  Doing so for the complete list of ramified points in the map, we see that
\be
W= f_2' f_1-f_2 f_1' = A_0\prod_{i} (t-t_i)^{r_i}\ . \label{wronskpow}
\ee
The Wronskian does not vanish at generic points, and so the space spanned by the values and first derivatives of $f_1$ and $f_2$ at generic points are independent.  The list of ramified points is complete, and so $W$ is fixed up to a constant $A_0$ which depends on the normalization of the functions $f_2$ and $f_1$.  

Seeing the appearance of the Wronskian motivates us to look for guidance from second order differential equations.  We note that given a pair of functions $(f_1,f_2)$, one may always write a linear second order differential equation 
\begin{align}
\pa^2 f(t)-\frac{f_2''(t)f_1(t)-f_2(t)f_1''(t)}{f_2'(t)f_1(t)-f_2(t)f_1'(t)}\pa f(t)+\frac{f_2''(t)f_1'(t)-f_2'(t)f_1''(t)}{f_2'(t)f_1(t)-f_2(t)f_1'(t)}f(t)=0
\end{align}
where the two independent solutions for $f$ are $f_1$ and $f_2$.  The above is simply the Wronskian of three functions $(f,f_1,f_2)$, of which $f$ is kept arbitrary.  We notice the above equation may be written as
\begin{align}\label{diffeqW}
\pa^2 f(t)-\pa\ln(W) \pa f(t)+\frac{f_2''(t)f_1'(t)-f_2'(t)f_1''(t)}{f_2'(t)f_1(t)-f_2(t)f_1'(t)}f(t)=0\ .
\end{align}
For the form of the Wronskian \eqref{wronskpow} the coefficient of $\pa f$ becomes
\be
\pa \ln(W)= \sum_i \frac{r_i}{(t-t_i)}\ . \label{logwronsk}
\ee
Where this function becomes singular are singular points of the differential equation.  Restricting the covering space map to be constructed from polynomials means that the local Frobenius solutions are positive integer power law, and so we expect these to be regular singular points.  Therefore, we recognize this type of differential equation of being of Fuchsian type with regular singular points.  When there are three regular singular points, one may use $sl(2)$ transformations to bring these points to $t=0$, $1$, and $\infty$, and one gets a differential equation of hypergeometric form.  The polynomial ``cousins'' of hypergeometric functions are Jacobi polynomials, and the covering space maps for this case have been constructed in \cite{Lunin:2000yv}.  

For four regular singular points, the story is more complicated: one may not use $sl(2)$ invariance to fix all points.  The CFT interpretation of this is that there is a cross ratio for a 4-point function, and so a non-trivial function of this cross ratio must be calculated.  This type of differential equation is called a Heun differential equation, and has been studied extensively \cite{ronveaux1995}: in this case, the sum \eqref{logwronsk} has only three terms, which correspond to the location of three ramified points in the map at finite positions.  There are polynomial solutions to Heun's equations \cite{Hounkonnou:2013kka,Takemura:2018}, and these have been previously used in the literature to address the problem of 4-point functions \cite{Pakman:2009zz,Dei:2019iym}.  However, we find the presentation for Heun functions as finite sums over hypergeometric functions in \cite{Ishkhanyan:2014wma} particularly useful.

In the presentation of \cite{Ishkhanyan:2014wma}, the sum over hypergeometric functions is found to truncate to a finite sum under certain circumstances.  This finite sum has coefficients that must satisfy a set of algebraic constraints to make them a solution to Heun's differential equation.  We may take advantage of this form of the solution because, as we will show, one may take a limit of hypergeometric functions to generate two Jacobi polynomials.  We show this in the next section exploring the 3-point function. This gives us a method of generating a pair of polynomial solutions to Heun's equation, recalling that one must also impose the algebraic constraints.  Interestingly, without imposing these algebraic constraints, we find that the sums over Jacobi polynomials generated in this way satisfy a more general Fuchsian type of differential equation: one where there are more than three terms in \eqref{logwronsk}.  We show this form of the differential equation by finding the form of the Wronskian for the two sums over Jacobi polynomials.  This Wronskian shows that there are three long cycle twists at $(t=0,z=0)$, $(t=1,z=1)$ and $(t=\infty,z=\infty)$, and a cloud of twist-2 insertions at other points.  The algebraic restrictions in \cite{Ishkhanyan:2014wma} amount to a particular OPE limit where the twist-2 cycles in the cloud ``twist together'' into one long twist operator.  However, without this constraint, the sum provides a more general covering space map.  One may consider other types of OPE limits, as we discuss in section \ref{subsec_opelim}.  

The construction of the covering space map as a ratio of sums over Jacobi polynomials also seems quite natural from a bootstrap perspective.  In a CFT, the building blocks of higher-point functions are 3-point functions.  The higher-point functions are written in terms of conformal blocks (which are generic) along with structure constants (which are the CFT-specific data).  The construction of higher-point covering space maps in terms of the covering space maps for 3-point functions, i.e. Jacobi polynomials, therefore seems quite natural.  We now turn to our construction of the covering space maps.

\section{Maps for \texorpdfstring{$n$}{TEXT}-point functions with three long twists}
\label{GenMaps}

\subsection{3-point function maps from Jacobi polynomials}
\label{3PtMaps}

We first consider the case where there are only three regular singular points: at $t=0$, $t=1$, and $t=\infty$.  Thus, there are only two terms in the sum (\ref{logwronsk}), which is of hypergeometric form
\begin{equation}
f''(t)+\left(\frac{\gamma}{t}+\frac{\alpha+\beta-\gamma+1}{t-1}\right)f'(t)+\frac{\alpha\beta}{t(t-1)}f(t)=0\ . \label{hypergeomdiffeq}
\end{equation}
This is satisfied by the hypergeometric series
\begin{equation}\label{hgs}
f(t)=\,_2F_1\left(\genfrac{}{}{0pt}{}{\alpha,\beta}{\gamma};t\right)= 
\sum_{\ell=0}^\infty \frac{(\alpha)_\ell (\beta)_\ell}{(\gamma)_\ell}\;\frac{t^\ell}{\ell!}\ ,
\end{equation}
where $(\,)_\ell$ denotes the Pochhammer symbol
\begin{equation}
(\alpha)_\ell=\prod_{i=0}^{\ell-1} (\alpha+i)=\frac{\Gamma(\alpha+\ell)}{\Gamma(\alpha)}=
(-1)^\ell \frac{\Gamma(-\alpha+1)}{\Gamma(-\alpha-\ell+1)}\ . \label{pochdef}
\end{equation}
The last expression allows us to more easily consider cases where the argument of gamma functions is near a negative integer. For ease of notation, from now on we shall drop the indices $p$ and $q$ of the hypergeometric series ${}_pF_q$ and exhibit them respectively by the number of terms in the upper and lower levels of the argument: $F\big(\genfrac{}{}{0pt}{}{\alpha_1,\cdots,\alpha_p}{\beta_1,\cdots,\beta_q};t\big)$.

An important property of the Pochhammer symbol is that $(\alpha)_\ell$ becomes 0 when $\alpha$ is a non-positive integer and $\ell$ is sufficiently large: one term will be zero in the product.  More precisely, for $m$ a non-negative integer, $(-m)_\ell=0$ when $\ell\geq m+1$.  However, such 0 values may be regulated by shifting $m$ by $\epsilon_m$, giving
\begin{equation}
(-m+\epsilon_m)_{\ell}= \begin{cases} (-1)^\ell \frac{m!}{(m-\ell)!}+{\mathcal O}(\epsilon_m)\ ,  &\mbox{if $\ell < m+1$} \\[5pt]
(-1)^{m} m! (\epsilon_m) (\ell-m-1)!+{\mathcal O}(\epsilon_m^2)\ ,  &\mbox{if $\ell\geq m+1$} \end{cases} \label{caseslimit}\ .
\end{equation}
The fact that the leading order approximation is either constant or order $\mathcal{O}(\epsilon_m)$ is because at most one term in the product $(-m+\epsilon_m)_{\ell}$ gets close to 0.  This means all Pochhammer symbols, once regulated in the above way, will either be constant or go linearly to zero as the control parameter $\epsilon_m$ goes to zero\footnote{The analogous statement using gamma functions is that all gamma functions near non-positive integers $-m$ behave as $\Gamma(-m+x)=(-1)^m/(m!x)+ {\mathcal O}(1)+\cdots$, i.e. these exhibit only simple poles.  In the Pochhammer symbol $\alpha_\ell=\frac{\Gamma(\alpha+\ell)}{\Gamma(\alpha)}$ the poles in the gamma functions either do not exist, cancel between numerator and denominator, or only the denominator pole exists, leading to a linear 0 in the control parameter.}.  This will allow us to examine the hypergeometric series in an interesting way.

Consider the ``regulated-near-negative-integer'' hypergeometric series with constant coefficient $A$ (the reason for the coefficient will become clear in a moment)
\begin{align}\label{nearinthyp}
A\; F\left(\genfrac{}{}{0pt}{}{-n_1+r_1\epsilon,-n_3+r_3\epsilon}{-n_2+r_2\epsilon};t\right)=
A \sum_{\ell=0}^\infty \frac{(-n_1+r_1\epsilon)_\ell (-n_3+r_3\epsilon)_\ell}{(-n_2+r_2\epsilon)_\ell}\;\frac{t^\ell}{\ell!}
\end{align}
where we pick three non-negative integers $0\leq n_1\leq n_2\leq n_3$.  We note that in the limit as $\epsilon\rightarrow 0$, the above sum is truncated into two ``windows'' for $\ell$ where the regulation parameter $\epsilon$ does not eliminate the term in the sum \eqref{nearinthyp}. The terms that survive the limit are for $0\leq \ell \leq n_1$, and $n_2+1\leq \ell \leq n_3$.  The case $n_3=n_2$ is rather odd, and the second ``window'' doesn't exist because the numerator becomes quadratically 0 at the same term that the denominator becomes linearly 0.  Therefore, we restrict to the cases $0\leq n_1\leq n_2 < n_3$ so that the second window has at least one term.  We note that the regulator $r_3$ never plays a role, while $r_1$ and $r_2$ show up in the form $r_1/r_2$ in the second ``window'' where one numerator Pochhammer symbol and one denominator Pochhammer symbol both become order $\epsilon$.  We suggest the choice $r_2=A$, $r_1=B$, although only the ratio appears.  Using this, along with the limits (\ref{caseslimit}), we find
\be\begin{split}\label{hgsln}
& \lim_{\epsilon\rightarrow 0} A\; F\left(\genfrac{}{}{0pt}{}{-n_1+B\epsilon,-n_3+r_3\epsilon}{-n_2+A\epsilon};t\right) = 
A \frac{n_1!n_3!}{n_2!}\sum_{\ell=0}^{n_1} \frac{(n_2-\ell)!}{(n_1-\ell)!(n_3-\ell)!}\;\frac{(-t)^\ell}{\ell!}  \\[3pt]
&\qquad\qquad\quad+ B(-1)^{n_2-n_1}\frac{n_1!n_3!}{n_2!}
\sum_{\ell=n_2+1}^{n_3}\frac{(\ell-n_1-1)!}{(\ell-n_2-1)!(n_3-\ell)!}\;\frac{(-t)^\ell}{\ell!}\equiv A f_1(t)- Bf_2(t)\ ,
\end{split}\ee
where now all of the Pochhammer symbols have been replaced by factorials.  We see that the hypergeometric sum has separated into two separate polynomials $f_1$ and $f_2$ with two separate coefficients $A$ and $-B$. Thus, any linear relationship that the hypergeometric series satisfies must be satisfied by the two separate sums, so long as the linear relationship does not become singular in the $\epsilon\rightarrow 0$ limit.

The differential equation (\ref{hypergeomdiffeq}) does not become singular in the $\epsilon\rightarrow 0$ limit.  Therefore, we have obtained two distinct polynomials that solve the second order differential equation (\ref{hypergeomdiffeq}) with $\alpha=-n_1,\beta=-n_3,\gamma=-n_2$ and $0\leq n_1\leq n_2 <n_3$, and thus supply a complete set of solutions.  Thus, the function $z(t)=f_2/f_1$ has the form of a covering space map, and the ramified points may be read from the form of the Wronskian -- see \eqref{diffeqW}, \eqref{logwronsk}, and \eqref{hypergeomdiffeq}. At $t=0$ we have ramification $r_0=n_2$, and at $t=1$ we have ramification $r_1=n_1+n_3-n_2-1$. The ramification at infinity is given by $r_{\infty}=n_3-n_1-1$ which is read from the highest order terms in the numerator and denominator of $f_2/f_1$.  Furthermore, the total number of sheets for a generic point $z$ is given by the maximum degree of the polynomials, i.e. $n_3$, the degree of $f_2$.  Thus, plugging in the $r_i$ above and $S=n_3$ in the Riemann-Hurwitz formula \eqref{RHf}, and we have that the genus of the covering surface is $g=0$ as expected.  Furthermore, it is clear that all of these expressions make sense for the range $0\leq n_1 \leq n_2 < n_3$.  However, relaxing the first and last inequality can be useful for certain proofs.  

It is not too hard to identify the polynomials $f_1$ and $f_2$, finding 
\vs{-5}\be\begin{split}
f_1&=\frac{n_1!n_3!}{n_2!}(-1)^{n_1}\frac{(n_2-n_1)!}{n_3!} P^{-(n_2+1),-(n_3+n_1-n_2)}_{n_1}(1-2t) \label{JacPolyFromHyGeo}
\end{split}\ee
and
\begin{align}
f_2(t)&= \frac{n_1!n_3!}{n_2!}(-1)^{n_1}\frac{(n_2-n_1)!}{n_3!} t^{n_2+1}P^{(n_2+1),-(n_1+n_3-n_2)}_{n_3-n_2-1}(1-2t) \ ,
\end{align}
where $P^{\alpha,\beta}_\gamma(x)$ is the Jacobi polynomial ($\alpha, \beta, \gamma$ different from above); see appendix \ref{JacobiIDappx} for definitions and useful identities.  The above functions are easy to show to be the same as the windows in \eqref{hgsln}: one need only shift the indices, and identify the Pochhammer symbols in both the hypergeometric series and the Jacobi polynomials.  This is easiest to do by replacing the Pochhammer symbols with factorials, which one may do in all cases.
{\footnote{As an interesting side note, one may consider a set of generalized hypergeometric series
\begin{equation}
F\left(\genfrac{}{}{0pt}{}{\alpha_1,\alpha_3,\cdots, \alpha_{2W-1}}{\alpha_2,\alpha_4,\cdots,\alpha_{2W-2}};t\right)=\sum_{\ell=0}^\infty \frac{\prod_{w=1}^{W}(\alpha_{2w-1})_\ell}{\prod_{w=2}^{W}(\alpha_{2w-2})_\ell} \frac{t^\ell}{\ell!}\ .
\end{equation}
Consider the $\alpha_i$ near negative integers $-n_i$, appropriately regulated, i.e.
\begin{equation}
F\left(\genfrac{}{}{0pt}{}{-n_1+r_1\epsilon,-n_3+r_3\epsilon,\cdots, 
-n_{(2W-1)}+r_{(2W-1)}\epsilon}{-n_2+r_2\epsilon,-n_4+r_4\epsilon,\cdots,-n_{(2W-2)}+r_{(2W-2)}\epsilon};t\right)\ .
\end{equation}
Above, we require that the $n_i$ satisfy $0\leq n_1$, $n_{2w-1}\leq n_{2w}$, and $n_{2w}<n_{2w+1}$.  The integer $w$ is between $1$ and $W$ labeling the ``windows''.  In the limit that $\epsilon\rightarrow 0$ we find $W$ windows where the coefficients of the generalized hypergeometric sum are non-zero, each with independent coefficients.  These $W$ polynomials satisfy the linear $W^{\rm th}$ order generalized hypergeometric differential equation, providing a complete set of solutions.}}

\subsection{\texorpdfstring{$n$}{TEXT}-point function maps from sums over Jacobi polynomials}
\label{nPtMaps}

We now wish to consider correlation functions beyond the 3-point function.  This suggests having more than two terms in \eqref{logwronsk}. Looking at the case where there are three terms seems to be the next logical step, where the equation becomes Heun's differential equation(s) (see \cite{ronveaux1995} for a monograph on this equation, and its solutions, known as Heun functions).  However, we would like to focus on a construction in the literature \cite{Ishkhanyan:2014wma} where they build Heun functions out of hypergeometric functions.  This will give us a way to build covering space maps with multiple insertions.  

First, Heun's differential equation is
\begin{equation}\label{heundeq}
f''(t)+\left(\frac{\gamma}{t}+\frac{\delta}{t-1}+\frac{\varepsilon}{t-a}\right)f'(t)+\frac{(\alpha\beta t - q)}{t(t-1)(t-a)} f(t)=0\ .
\end{equation}
The extra regular singular point at $t=a$ cannot be fixed with $sl(2)$ transformations, and so must be left with an arbitrary complex location $a$. For us, $\varepsilon$ will be a non-positive integer to give a ramified point in the covering space map. This will lead to ramifications at $t=\{0,1,\infty,a\}$ which we refer to as $\{r_0,r_1,r_\infty,r_a\}$, respectively. The above constants satisfy a relation $\alpha+\beta+1=\gamma+\delta+\varepsilon$.  In addition, there is an extra parameter $q$ above which is not present in the hypergeometric case \eqref{hypergeomdiffeq}.  In the case that $\varepsilon=0$, we recover the hypergeometric equation setting $q=\alpha\beta a$.  Recall that in the hypergeometric series \eqref{hgs} depends symmetrically on $\alpha$ and $\beta$, and so the differential equation \eqref{hypergeomdiffeq} depends only on $\alpha+\beta$ and $\alpha\beta$.

The analysis of \cite{Ishkhanyan:2014wma} builds solutions to Heun's differential equations from a finite sum of hypergeometric functions
\begin{equation}
f(t)=\sum_{n=0}^{n_{\rm max}} c_n\,F\left(\genfrac{}{}{0pt}{}{\alpha,\beta}{\gamma+\varepsilon+n};t\right) \label{geosum}
\end{equation}
with complex coefficients $c_n$.  This takes advantage of the fact that $F(\alpha,\beta;\gamma;t)$ satisfies the differential equation \eqref{hypergeomdiffeq}, and this shares some of the singularity structure of \eqref{heundeq}.  There are restrictions in \cite{Ishkhanyan:2014wma} that allow a recurrence relation to truncate, leading to a finite sum solution.  The relevant term in the recurrence relation reads
\begin{equation} \label{recurUpDown}
{\mathcal P}_n=-\frac{a}{n+\varepsilon+\gamma}(n+\varepsilon)(n+\varepsilon+\gamma-\alpha)(n+\varepsilon+\gamma-\beta)
\end{equation}
and requiring this to vanishes for some $n$ allows the recurrence relation to terminate.  The above can vanish if there is some positive integer $n_{\rm max}$ such that
\begin{align}
\varepsilon=-n_{\rm max}, \qquad \mbox{or} \quad \varepsilon+\gamma-\alpha=
-n_{\rm max}, \qquad \mbox{or} \quad \varepsilon+\gamma-\beta=-n_{\rm max}\ . \label{Pcases}
\end{align}
We consider only the first of these, namely $\varepsilon=-\nmax$ for various reasons.  First, the case $\varepsilon=-\nmax$ is the easiest to analyze because, similar to the last subsection, a limit of the hypergeometric functions appearing in \eqref{geosum} will generate two polynomials with separate coefficients. Furthermore, we will also see that the case $\varepsilon=-\nmax$ generates maps which furnish all group theoretically allowed ramifications for the class of covering maps which \eqref{geosum} can generate.\footnote{One could use the recurrence relations \eqref{recurUpDown} to find a pair of polynomials directly, taking advantage of the other possibilities in \eqref{Pcases}.  This is because both numerator and denominator terms in \eqref{recurUpDown} can become small when parameters are near negative integers.  Having these terms become small for different values of $n$, as $n$ increases can give two windows with distinct coefficients where $c_n$ are finite -- see equation (20) of \cite{Ishkhanyan:2014wma}. The terms in the recurrence relation \eqref{recurUpDown} which become small are also parameters in the hypergeometric functions in \eqref{geosum}, and one would also need evaluate the effect of taking parameters near negative integers in the hypergeometric functions themselves.  We have not explored these avenues due to these complications, as well as the apparent completeness of the case  $\varepsilon=-\nmax$, as discussed below \eqref{Pcases}.}  Therefore, from here on we only consider the case $\varepsilon=-\nmax$.

Now we use the basic construction of the last subsection.  Consider the hypergeometric functions in \eqref{geosum}, this time for each $n$
\begin{equation}
A\; F\left(\genfrac{}{}{0pt}{}{\alpha,\beta}{\gamma-\nmax+n};t\right)=A \sum_{\ell=0}^\infty \frac{(\alpha)_\ell(\beta)_\ell}{(\gamma-\nmax+n)_\ell}\, \frac{t^\ell}{\ell!}\ . \label{hypergeomheun}
\end{equation}
We consider the indices near negative integers
\begin{equation}
\alpha=-n_1+B\epsilon\ ,\qquad \gamma=-n_2+A\epsilon\ , \qquad \beta=-n_3+r_3\epsilon\ , \label{AlBeGaNear}
\end{equation}
and take the limit $\epsilon\rightarrow 0$ to obtain two polynomials: we see that $\varepsilon=-\nmax$ guarantees that $\gamma-\nmax+n$ is also near an integer (note that $\varepsilon$ and $\epsilon$ are different).  The limit is well-defined when none of the coefficients of the hypergeometric series become infinite as $\epsilon\rightarrow 0$.  Therefore we need $(-n_1+B\epsilon)_\ell/(-n_2-\nmax+n+A\epsilon)_\ell$ to be well defined for all $\ell$ in this limit. This requires that the numerator Pochhammer symbol becomes infinitesimal ``first'' (i.e. at a lower value of $\ell$), which gives the constraint $0\leq n_1 \leq n_2+\nmax-n$.  This must be true for all $n$ and so $0\leq n_1\leq n_2$.  In cases where $n_1<n_3\leq n_2+\nmax-n$ is ``out of order'', then the $n^{\rm th}$ term of \eqref{geosum} has no second window.  However, not all such second windows for all $n$ may vanish: we need at least one term proportional to $B$ to get two linearly independent pieces.  To ensure at least one piece proportional to $B$ survives, we must have $n_2< n_3$, and so $0\le n_1\le n_2<n_3$.

Plugging in \eqref{AlBeGaNear} into \eqref{geosum} and then taking the limit $\epsilon\rightarrow 0$ subject to the constraints from the last paragraph then gives one polynomial proportional to $A$ and the other proportional to $-B$, i.e 
\begin{equation}
\lim_{\epsilon\rightarrow 0} A\; \sum_{n=0}^{\nmax} F\left(\genfrac{}{}{0pt}{}{-n_1+B\epsilon,-n_3+r_3\epsilon}{-n_2+A\epsilon-\nmax+n};t\right) =\frac{(-1)^{n_1}(n_1!)(n_2-n_1)!}{(n_2+\nmax)!}\big(Af_1(t)-Bf_2(t)\big) \label{hypergeomheun.fdefs}
\end{equation}
along with the requirement
\vs{-5}\be
0\leq n_1 \leq n_2 < n_3\ . \label{nineq}
\ee
In \eqref{hypergeomheun.fdefs} we have removed a common factor from both windows, which cancel in the covering space map $z=f_2/f_1$.  We find explicitly
\vs{-5}\be\begin{split}\label{jacobisum_genb}
f_1(t) &= \sum_{n=0}^{n_{\rm max}} c_{n} (n_2+n_{\rm max}-n+1)_n (n_2-n_1+1)_{(n_{\rm max}-n)}  \\
& \qquad \qquad \qquad \qquad  \times P^{-(n_2+n_{\rm max}-n+1),-(n_1+n_3-(n_2+n_{\rm max}-n))}_{n_1} (1-2t)\ , \\[2pt]
f_2(t) &= \sum_{n=n_{\min}}^{n_{\rm max}} c_n (n_2+n_{\rm max}-n+1)_n (n_2-n_1+1)_{(n_{\rm max}-n)} t^{(n_2+n_{\rm max}-n+1)}  \\
& \qquad \qquad \qquad \qquad \times 
P^{(n_2+n_{\rm max}-n+1),-(n_1+n_3-(n_2+n_{\rm max}-n))}_{n_3-(n_2+n_{\rm max}-n+1)}(1-2t)\ , \\[3pt]
n_{\rm min}&\equiv{\rm max}\,\Big(0, -\big(n_3-(n_2+\nmax+1)\big)\Big)\ .
\end{split}\ee

In $f_2$ we note that certain occurrences of Jacobi polynomials with negative order would appear if $n_3-(n_2+n_{\rm max}-n+1)\leq -1$.  These are simply an indication that the lower index of the parent hypergeometric function, $n_2+n_{\rm max}-n$, is ``out of order'', i.e. $n_2+n_{\rm max}-n \geq n_3$, and so the second window doesn't exist for that $n$ (but the limit $\epsilon\rightarrow 0$ is still well defined).  We have excluded these terms from the sum by adjusting the lower limit of the sum for $f_2$ above.  Rather than restricting the sum, we find it much more convenient to use the following rule:
\begin{align}
\boxed{\begin{aligned}[t]
     & \mbox{\it Whenever a Jacobi polynomial appears with negative subscript, we set it to 0, i.e.} \label{rule}   \\
     & \kern10em P^{\alpha,\beta}_{-\gamma}(x)\equiv0\ , \qquad\qquad \gamma\in\bz_{>0}\ .
     \end{aligned}}  
\end{align}

For now we have used the construction in \cite{Ishkhanyan:2014wma} as much as we need to.  We can see that the normalization of the constant $c_i$, which seems quite natural in \eqref{geosum} in the sum over Hypergeometric functions, seem somewhat less natural in the sum over Jacobi polynomials. We therefore define new constants which absorb the Pochhammer symbols following the $c_n$'s in \eqref{jacobisum_genb}. Furthermore, we identify
\be
N\equiv n_2+\nmax-n
\ee
as an effective summation variable, where $N$ has a minimum value of $N_{\rm min}=n_2$ and a maximum value of $N_{\rm max}=n_2+\nmax$.  This gives a more convenient form to write the two functions as
\be\begin{split}\label{jacobisum_gen}
f_1(t) &= \sum_{N=N_{\rm min}}^{N_{\rm max}} b_N P^{-(N+1),-(n_1+n_3-N)}_{n_1} (1-2t)\ , \\
f_2(t) &= \sum_{N=N_{\rm min}}^{N_{\rm max}} b_N t^{(N+1)}P^{(N+1),-(n_1+n_3-N)}_{n_3-N-1}(1-2t)\ ,
\end{split}\ee
where
\be\label{bNdef}
b_N\equiv b_{n_2+\nmax-n}=c_{n} (n_2+n_{\rm max}-n+1)_n (n_2-n_1+1)_{(n_{\rm max}-n)} \ .
\ee
While these expressions were derived using the constraints $0\leq n_1 \leq N_{\rm min} < n_3$, we show that the above expressions generate covering space maps with $N_{\rm min}<n_1$ as well. In such cases we consider the sums defining $f_1$ and $f_2$ in \eqref{jacobisum_gen}, and therefore the complex numbers $b_N$, as being fundamental. The identification \eqref{bNdef} is only valid when $\Nmin=n_2\ge n_1$.

We now show that \eqref{jacobisum_gen} generates covering space maps $z=f_2/f_1$ with some ramified points that are easy to identify. To do so, we use two identities
\be\begin{split}
 &P^{-(N+1),-(n_1+n_3-N)}_{n_1}(1-2t)-t^{N+1}P^{N+1,-(n_1+n_3-N)}_{n_3-N-1}(1-2t) \\[5pt]
&\hs{186}  =(-1)^{N}(1-t)^{n_1+n_3-N}P_{N-n_1}^{-(N+1),(n_1+n_3-N)}(1-2t)\ , \label{tm1id}
\end{split}\ee
which, using the identity \eqref{xid1}, we may also write as
\be\begin{split}
& P^{-(n_1+n_3-N),-(N+1)}_{n_1}(2t-1)+(-1)^{n_1+n_3-N}t^{N+1}P^{-(n_1+n_3-N),N+1}_{n_3-N-1}(2t-1) \\[5pt]
& \hs{218} =(1-t)^{n_1+n_3-N}P_{N-n_1}^{(n_1+n_3-N),-(N+1)}(2t-1)\ , \label{tm1idSWAP} 
\end{split}\ee
which are proved as \eqref{tm1idAPX} and \eqref{tm1idAPXSWAP} in appendix \ref{APXproofs}.  Importantly, the above identities hold for {\it all} integers $-\infty<N<\infty$ subject to the rule \eqref{rule}.  We will also need the identity
\be \label{idInfty}
P^{\alpha,\beta}_{\gamma}(1-2t)=t^\gamma\, P^{-(2\gamma+\alpha+\beta+1),\beta}_\gamma \Big(1-\frac{2}{t}\Big)\ ,
\ee 
which we prove in \eqref{idInftyAPX} and is true for all integers $-\infty <\gamma < \infty$, using the rule \eqref{rule}.

Using these identities, we may write the covering space map in three equivalent ways.  First, we have
\be
z(t)= \frac{f_2(t)}{f_1(t)}= \frac{ \sum\limits_{N=N_{\rm min}}^{N_{\rm max}} b_N t^{(N+1)}P^{(N+1),-(n_1+n_3-N)}_{n_3-N-1}(1-2t)}{\sum\limits_{N=N_{\rm min}}^{N_{\rm max}} b_N P^{-(N+1),-(n_1+n_3-N)}_{n_1} (1-2t) }\label{near0}
\ee 
which is adapted to be expanded near the point $t=0, z=0$.  We refer to this as the ``near 0'' form of the map.  We may use \eqref{tm1id} to to write 
\be
1-z(t)=\frac{f_1(t)-f_2(t)}{f_1(t)} =
\frac{ \sum\limits_{N=N_{\rm min}}^{N_{\rm max}} b_N (1-t)^{n_1+n_3-N}P_{N-n_1}^{(n_1+n_3-N),-(N+1)}(2t-1) }
{\sum\limits_{N=N_{\rm min}}^{N_{\rm max}} b_N P^{-(n_1+n_3-N),-(N+1)}_{n_1} (2t-1) }  \label{near1}
\ee
which is adapted to be expanded near the point $t=1, z=1$.  We refer to this as the ``near 1'' form of the map.  We finally may use the identity \eqref{idInfty} to write
\be
\frac{1}{z(t)} = \left(\frac{1}{t}\right)^{n_3-n_1}
\frac{\sum\limits_{N=N_{\rm min}}^{N_{\rm max}} b_N P^{(n_3-n_1),-(n_1+n_3-N)}_{n_1} \left(1-\frac{2}{t}\right) }
{\sum\limits_{N=N_{\rm min}}^{N_{\rm max}} b_N P^{-(n_3-n_1),-(n_1+n_3-N)}_{n_3-N-1}\left(1-\frac{2}{t}\right)}\label{nearinfty}
\ee
which is adapted to the point at $t=\infty, z=\infty$.  We will refer to this as the ``near $\infty$'' form of the map.  These three forms of the map make it easy to identify the ramifications at each point, namely
\be
r_0=\Nmin\ , \qquad r_1= n_1+n_3-\Nmax-1\ , \qquad r_\infty=n_3-n_1-1 \ . \label{r01infty}
\ee
We require all of these ramifications to be greater than or equal to 1, such that these points are ramified.  We note that $\Nmin$ may in fact be less than $n_1$.  In such a case some of the Jacobi polynomials are 0 in the numerator of the ``near 1'' form of the map \eqref{near1}, given the rule \eqref{rule}.

From equations \eqref{near0}, \eqref{near1}, and \eqref{nearinfty}, it is clear that the Wronskian has the general form
\begin{align}
& W=f_2'f_1-f_2 f_1'= t^{\Nmin} (t-1)^{(n_1+n_3-\Nmax-1)} Q(t) \label{wronskform}\ .
\end{align}
The power of $t^{\Nmin}$ in \eqref{wronskform} is guaranteed by the common factor of $t^{\Nmin+1}$ appearing in the numerator of \eqref{near0}. Note that the Wronskian is identically equal to $W=(f_2-f_1)'f_1-(f_2-f_1) f_1'$, and so the power of $(t-1)$ in \eqref{wronskform} is guaranteed by the common factor of $(t-1)^{n_1+n_3-\Nmax}$ in the numerator of \eqref{near1}. Finally, we see that the polynomial $Q(t)$ must have the form
\be \label{Qform}
Q(t)=\sum_{i=0}^{\Delta N} A_{i}\, t^{\Delta N-i}\ ,\qquad\qquad\qquad \Delta N \equiv \Nmax-\Nmin\ ,
\ee
by matching degrees of polynomials in \eqref{wronskform}: $f_1$ is of degree $n_1$ and $f_2$ is of degree $n_3$. Above, $\Delta N$ is the range of the sums appearing in $f_1$ and $f_2$ -- see \eqref{jacobisum_gen}. The coefficients $A_i$ in $Q(t)$ are quadratic homogeneous polynomials of the coefficients $b_{N}$.  Three combinations of the $A_i$ may be found in the general case, which we discuss in section \ref{subsec_opelim}.  Interestingly, these give all the information needed to find $Q$ in the $\Delta N=0,1,2$ cases.  For cases $\Delta N\geq 3$, one may iteratively generate the $A_i$, as explained in appendix \ref{WronskGen}, and if $\Delta N$ is finite, this is finitely many steps.

Generically, the polynomial $Q(t)$ has $\Delta N$ distinct zeros, and the locations of these zeros define the location of points with ramification 1.  These maps, therefore, correspond to having a cloud of twist-2 operators surrounding the three long twists at $(t=0,z=0)$, $(t=1,z=1)$, and $(t=\infty,z=\infty)$.  The total ramification of this cloud of operators is $r_c=\Delta N$.  It is important to realize that these are a \emph{cloud} of operators, not a single operator, and so they may have interesting OPE limits amongst themselves. 

Note that the simultaneous scaling of the coefficients $b_N\rightarrow \lambda b_N$ give rise to the same covering space map.  There are $\Delta N+1$ such $b_N$ coefficients, and so the space of maps is clearly $\Delta N$ dimensional, exactly the dimension spanned by the cross-ratios in a $(3+\Delta N)$-point function, and matching the number of ramification one points in the cloud $r_c=\Delta N$. The scaling symmetry suggests that these coefficients are valued in $\mathbb{CP}^{\Delta N}$.  We will see that the map parameters $b_N$ parameterize more than just the cross-ratios, but also control which type of OPE limit is produced when two ramified points come together (i.e. which group product is taken amongst the many possibilities in the conjugacy class product).  We will find in section \ref{subsec_opelim} that our conditions for OPE limits are given by homogeneous polynomials in the $b_N$ set to 0.  These are natural subspaces of $\mathbb{CP}^{\Delta N}$: algebraic varieties.

\subsection{Generality of the maps}\label{sec_map_gen}

In this subsection, we consider whether the maps \eqref{near0}, or equivalently \eqref{near1}, or \eqref{nearinfty} are general enough to give any $r_0$, $r_1$, $r_\infty$, and $r_c=\Delta N$.  We will find that they are by showing that all maps corresponding to group theoretically allowed ramifications can be constructed for some choice of $n_1$, $n_3$, $\Nmin$ and $\Nmax$.   

We begin by pointing out some restrictions that the maps \eqref{near0}, \eqref{near1}, or \eqref{nearinfty} must obey so that they are not pathological. First, in \eqref{near0}, it is clear that we must have at least some terms with $n_3-N-1\geq 0$, otherwise $z(t)=0$.  Therefore $N_{\rm min}<n_3$.  Similarly, by examining \eqref{near1} we see that we need some terms with $N-n_1\geq 0$, and so $\Nmax\geq n_1$.  This guarantees that there exists some $N$ that have $n_1\leq N < n_3$.  This is the special window where \eqref{tm1id} and \eqref{tm1idSWAP} have all three Jacobi polynomials present, and corresponds exactly to case \ref{case3} in appendix \ref{APXproofs} where these identities are proven. All in all, we find the set of constraints
\be\label{goodmaps}
n_3-n_1\ge2\ ,\qquad\; \Nmax\ge\Nmin\ ,\qquad\; 1\le\Nmin<n_3\ ,\qquad\; 0\le n_1\le\Nmax<n_1+n_3-1\ ,
\ee
such that the maps are not pathological, and the points $(t=0,z=0)$, $(t=1,z=1)$, and $(t=\infty,z=\infty)$ are ramified. 

We now consider how the ramifications of the map \eqref{near0} can be restricted by group theoretic considerations.  First we recall the ramifications
\begin{align}\label{rams}
r_0= \Nmin\ , \qquad\;\; r_1= n_1+n_3-\Nmax-1\ , \qquad\;\; r_{\infty}= n_3-n_1-1\ , \qquad\;\; r_{c}=\Delta N\ ,
\end{align}
where $r_c$ is the total ramification from the cloud.  We define the ramification of a group element by $r_{g}=\sum_k r_k=\sum_k(w_k-1)$ where $r_k=w_k-1$ are the ramifications of each cycle appearing in the group element when it is expressed as a set of disjoint cycles, and $w_k$ is just the length of each cycle.  We have the following statement that
\begin{align}
\boxed{\begin{aligned}[t]
     & \mbox{if} \qquad \prod_i g_i = e, \qquad \mbox{then} \qquad  \sum_{i\neq j} r_{g_i} \geq r_{g_j} \qquad \mbox{for each $g_j$ in the product} \label{RamSubAdd}  
     \end{aligned}}  
\end{align}
where $e$ is the identity element and $g_i$ are cycles.  We refer to this as ramification subadditivity and prove it in appendix \ref{appx.radd}.  

Applying restriction \eqref{RamSubAdd} to each of the ramifications associated with individual cycles for our maps, we have   
\be\label{cloudConstraint}
r_0\leq r_1+r_\infty+r_c\ , \qquad\qquad\qquad r_1\leq r_0 +r_\infty+r_c\ , \qquad\qquad\qquad r_\infty\leq r_0 +r_1+r_c\ .
\ee
One must {\it not} enforce $r_c\leq r_0 +r_1+r_\infty$ because the cloud is composed of distinct ramified points which correspond to distinct cycles.  These can merge and lower the total ramification in the cloud independently. Inserting \eqref{rams} in \eqref{cloudConstraint} we find
\be\label{cloudConstraintN}
\Nmin \leq n_3-1\ , \qquad\qquad\qquad n_1\leq \Nmax\ ,\qquad\qquad\qquad 0\leq n_1\ .  
\ee
We combine \eqref{cloudConstraintN} with the restrictions that the points at $0$, $1$, and $\infty$ are ramified $r_0\ge1$, $r_1\ge1$, $r_\infty\ge1$, along with $r_c\ge0$, and find exactly the constraints \eqref{goodmaps}.  

We can consider this in another way by inverting \eqref{rams}, finding
\begin{align}\label{nofr}
n_1=\frac{r_0+r_1-r_\infty+r_c}{2}\ , \quad\;\; \Nmin=r_0\ ,\quad\;\;  n_3=\frac{r_0+r_1+r_\infty+r_c}{2}+1\ , \quad\;\; \Delta N=r_c\ .
\end{align} 
The equation for $n_3$ is exactly the Riemann Hurwitz formula \eqref{RHf} with genus $g=0$, and insists that the sum of the ramifications is an even number. This is also guaranteed by group theory because the twists must multiply to the identity, which is an even element of the group $S_{\bold N}$. That $n_3$ is an integer also guarantees that $n_1$ is an integer because flipping the sign of $r_\infty$ differs by $2r_\infty$ in the numerator of the expression for $n_1$. The constraints from \eqref{cloudConstraint} guarantee that equations \eqref{nofr} define integers which satisfy \eqref{goodmaps} when $r_0\ge1$, $r_1\ge1$, $r_\infty\ge1$, and $r_c\ge0$.  

Therefore, the constraints imposed by insisting that the maps are well defined \eqref{goodmaps} are the same constraints one gets from the above group theoretic considerations.  Thus, any group product $\prod_i g_i=e$ composed of three long cycles and an arbitrarily large number of 2-cycles is captured by one of our covering space maps \eqref{near0}, or equivalently \eqref{near1}, or \eqref{nearinfty}.  To construct the map, one starts with ramifications $r_0$, $r_1$, and $r_\infty$, and the desired number of ramification 1 points in the cloud $r_c=\Delta N$: these must be specified in a way that is consistent with group theory.  Using \eqref{nofr} one obtains the parameters of the map which fix ramifications $n_1$, $n_3$, $\Nmin$, and $\Nmax$, and these obey the constraints \eqref{goodmaps} automatically.  The $b_N$, valued in $\mathbb{CP}^{\Delta N}$, parameterize the location of the $\Delta N$ twist-2 operators, or equivalently the $\Delta N$ cross ratios present in a $(3+\Delta N)$-point function.  If, as was suggested in \cite{Dei:2019iym}, all maps to sphere covering spaces are connected by analytic continuation, i.e. transport of twist operators around each other, then our maps \eqref{near0} represent a complete set up to transport. 

We return briefly to discussing the approach of \cite{Ishkhanyan:2014wma}, and ask whether the maps $\eqref{near0}$ gives a general set of maps of for the general four-point function of single-cycle operators.  Following the above discussion for the general maps, we find that the ramifications in this case are
\begin{align} \label{ramsa}
& r_0= \Nmin\, ,\qquad\;\; r_1= n_1+n_3-\Nmax-1 \ , \qquad\;\; r_{\infty}=n_3-n_1-1 \ ,  \qquad\;\;  r_a= \Delta N \ ,
\end{align}
where now $r_a$ is the ramification of the additional fourth point (we use the subscript $a$ to emphasize this is not a cloud of ramified points, but a single ramified point at $t=a$, following the notation of \cite{Ishkhanyan:2014wma}, and equation \eqref{heundeq}).  One may also arrive at these ramifications by following the discussion of \cite{Ishkhanyan:2014wma} and reading the relevant coefficients in the Heun equation: one can read $(\ln(W))'$ by comparing  \eqref{heundeq} to \eqref{diffeqW}.

One must again impose ramification subadditivity \eqref{RamSubAdd}, which gives 
\be\begin{split}
& r_0\leq r_1+r_\infty+r_a\ , \qquad\qquad r_1\leq r_0 +r_\infty+r_a\ , \\
& r_\infty\leq r_0 +r_1+r_a\ , \qquad\qquad  r_a\leq r_0 +r_1+r_\infty\ ,
\end{split}\ee
and substituting \eqref{ramsa}, we find
\be\Nmin \leq n_3-1\ , \qquad\qquad n_1\leq \Nmax\ , \qquad\qquad 0\leq n_1\ , \qquad\qquad  1\leq n_3\ .
\ee
We can generate another constraint by considering the following.  Given four ramifications, $q_1$, $q_2$, $q_3$, and $q_4$, it must be that $q_1+q_2 \geq q_3+q_4$ or $q_1+q_2\leq q_3+q_4$.  In either circumstance, we map the two points with the larger sum of ramifications to $t=0$ and $\infty$, and map one of the others to $t=1$ leaving the fourth ramified point at $t=a$.  Thus, without loss of generality,
\be
r_0+r_\infty \geq r_1+r_a
\ee   
which gives
\be
\Nmin \geq n_1\ .
\ee
This constraint reduces us to the case where the sum of hypergeometric equations \eqref{geosum} with the near integer values \eqref{AlBeGaNear} admits a well defined $\epsilon\rightarrow 0$ limit, and directly generates the polynomials appearing in $\eqref{near0}$, i.e. the case where $\Nmin=n_2$ and $n_1\leq n_2 < n_3$.  Thus, this single case appears to be sufficient to generate maps with arbitrary ramifications for four single-cycle operators, and so up to transport, generates the complete set of maps. 

We now recall the additional algebraic constraints in \cite{Ishkhanyan:2014wma}.   Their approach is to find recurrence relations between the coefficients $c_n$ in \eqref{jacobisum_genb} (equivalently the $b_N$ in \eqref{jacobisum_gen}), which ultimately come from the Heun equation.  These determine the $c_{n}$ in terms of $c_0$, $a$, and $q$ -- see \eqref{heundeq} and \eqref{geosum}.  In \cite{Ishkhanyan:2014wma}, $c_0$ is usually set to 1, which we may also do by scaling.  The rest of the $c_i$ only depend on the parameters of the recurrence relation.  The parameters that define the recurrence relation are the parameter $a$, which gives the location of the new ramified point, and $q$, the parameter in the Heun equation, and the integers $n$, $n_1$, and $n_3$.  Writing the recurrence relation in matrix form gives a matrix which must have determinant 0, giving an algebraic relationship relating $a$ and $q$, i.e. this constraint gives an effective $q(a)$.  In this way there is only one complex parameter remaining: $a$, the location of the new ramified point.  For the polynomials that we have extracted, this means that there is only one remaining parameter, once the algebraic constraints of \cite{Ishkhanyan:2014wma} are imposed.  This is expected to parameterize the single cross-ratio for a four point function of four single-cycle twists.

One may consider reparameterizing the algebraic constraints from \cite{Ishkhanyan:2014wma}.  For example, one could solve for $a$ in terms of $c_1$ to keep dependence on map parameters, and then the determinant relationship would give $q(a(c_1))$.  Thus, all of the quantities could ultimately be related to the single map parameter $c_1$ via algebraic constraints.  However, given the discussion above, we have a physical interpretation for these algebraic constraints: we view these as the OPE limit where the $\Delta N$ twist-2 operators fuse into a single-cycle twist with ramification $\Delta N$, i.e. a $(\Delta N+1)$-cycle operator.  Thus, the general map \eqref{near0} along with the algebraic constraints from \cite{Ishkhanyan:2014wma}, which we interpret as an OPE limit, give the general map for four single-cycle twist operators where the covering surface is a sphere, up to transport.  

We will see one example of an OPE limit where ramified points in the cloud approach each other, specifically for the case $\Delta N=2$ in section \ref{DN2OPEs}.  There, we will see that the algebraic constraint that fuses the two twist-2 operators into a single twist-3 operator is again given by a homogeneous polynomial in the $b_N$ set to 0, in this case a cubic, and so is an algebraic variety subspace of $\mathbb{CP}^{\Delta N=2}$.  In section \ref{OPELimSec} we consider what constraints on the $b_N$ can generate OPE limits.  This will help us identify singularity structures that appear in the correlation functions, which we address in section \ref{CorrSect}.

\section{OPE Limits}
\label{OPELimSec}

We start our analysis by considering the OPE limit when one of the ramification 1 points in the cloud approaches one of the points at $(t=0,z=0)$, $(t=1,z=1)$, $(t=\infty, z=\infty)$.  We use the forms of the maps \eqref{near0}, \eqref{near1}, and \eqref{nearinfty} which are adapted to the points $(t=0,z=0)$, $(t=1,z=1)$, $(t=\infty, z=\infty)$, respectively.  We consider how to identify other types of OPE limits as well.  We then consider the Wronskian in the general case, and are able to construct certain coefficients of the Wronskian which are written in terms of the OPE limit polynomials.  All of this helps us to construct the $n$-point functions in section \ref{CorrSect} by identifying singularity structures.  We specifically concentrate on the $\Delta N=1,2$ cases, where we can find $Q(t)$ exactly, and construct the $n$-point functions in closed form.  

\subsection{OPE limits using \texorpdfstring{$b_N$}{TEXT} coefficients and small \texorpdfstring{$\Delta N$}{TEXT} Wronskians}\label{subsec_opelim}

In this subsection we construct the OPE limits where one of the ramification 1 points in the cloud approaches the points at $t=0$, $t=1$, and $t=\infty$, which are structurally similar, and write the limits as restrictions on the $b_N$.  The  twist 2 operators in the cloud may either ``twist up'' the operator they approach when they share one copy index, or ``twist down'' the operator they approach when they share two copy indices.  We will show that the $b_N$ can parameterize both types of OPE limits, each as different linear constraints on the $b_N$.  

We may address this rather generically, and so we consider the general structure of the three forms of our maps \eqref{near0}, \eqref{near1}, and \eqref{nearinfty}.  The three functions $z(t)$, $1-z(t)$, and $1/z(t)$ given in \eqref{near0}, \eqref{near1}, and \eqref{nearinfty} will be represented generically as some function $Z$.  These should be thought of as functions of either $t$, $(1-t)$, or $1/t$, respectively, and we use the generic variable $T$ to represent these three possibilities in the three respective cases.  Thus, all three versions of the map are of the basic structure
\be
Z(T)= \frac{T^{r_q+1}P_2(T)}{P_1(T)}
\ee
where $r_q$ is the ramification of the point for which the map is adapted ($T=0$), and $P_2$ and $P_1$ are polynomials with lowest order terms which are constants.  To generate a ``twist up'' at the point of concern, we simply take $P_2(T=0)=0$, which sets the constant part of this polynomial to 0.  This will be a linear constraint on the $b_N$.  Under this constraint, we may factor out a $T$, writing $P_2(T)= T \ti{P}_2(T)$ where $\ti{P}_2(T)$ is the remaining polynomial.  This gives 
\be
Z= \frac{T^{r_q+1}P_2(T)}{P_1(T)} \xrightarrow[P_2(T=0)=0]{} Z= \frac{T^{(r_q+1)+1}\ti{P}_2(T)}{P_1(T)}\ .
\ee
We see that the linear constraint $P_2(T=0)=0$ leaves one fewer degree of freedom amongst the $b_N$ (and so there is one fewer ramified point in the cloud), but increases the ramification at $T=0$, identifying it as the twist up.  To twist down, we simply impose $P_1(T=0)=0$, which is a different linear constraint on the $b_N$.  Under this constraint, we similarly factor $P_1(T)=T\ti{P}_1(T)$, and so we find
\be
Z= \frac{T^{r_q+1}P_2(T)}{P_1(T)} \xrightarrow[P_1(T=0)=0]{}  Z= \frac{T^{(r_q-1)+1}{P}_2(T)}{\ti{P}_1(T)}\ .
\ee
In this case, the linear constraint leaves one fewer degree of freedom amongst the $b_N$ (so there is one fewer ramified point in the cloud), and also decreases the ramification at $T=0$, identifying it as the twist down.  Furthermore, it should be noted that because a power of $T$ has been cancelled, the map has one fewer sheet.  In this case, the total ramification has been lowered by 2, but the number of sheets has been lowered by 1, leaving result of the Riemann-Hurwitz formula for the genus $g=0$ unaffected.\footnote{One may also consider simply decreasing the ramification by 2, giving $g=-1$, which is also technically true: this is the genus for two disconnected spheres, interpreting the genus through the Euler characteristic $\chi=(2-2g)$, and realizing that the Euler characteristic is additive.  This has the added benefit of emphasizing that indeed the other copy of the seed CFT is present.  It is simply inert under the group elements chosen to represent their conjugacy classes for the operators in the correlator, and so this inert copy ``lives'' on its own sphere.}     

There are three forms of the map with two possible OPE limits each, giving six OPE limits we may find in this way.  Following the above prescription, we arrive at the following identifications
\be\label{limt0}
\lim_{t\rightarrow 0} f_1(t) = \frac{(-1)^{n_1}\Nmin!}{(n_1)!(\Nmax-n_1)!} g_{(0,\downarrow)}\ ,\quad \lim_{t\rightarrow 0} \frac{f_2(t)}{t^{\Nmin+1}}= \frac{(n_3)!}{(\Nmin+1)!(n_3-\Nmin-1)!} g_{(0,\uparrow)}\ ,
\ee
and 
\begin{align}\label{limt1}
&\hs{-16} \lim_{t\rightarrow 1}  f_1(t)= \frac{(n_1+n_3-\Nmax-1)!}{(n_1)!(n_3-\Nmin-1)!} g_{(1,\downarrow)}\ , \\[8pt]
& \kern10em \lim_{t\rightarrow 1} \frac{f_2(t)-f_1(t)}{(t-1)^{n_1+n_3-\Nmax}}= \frac{(-1)^{n_3-\Nmax-1} (n_3)!}{(n_1+n_3-\Nmax)!(\Nmax-n_1)!} g_{(1,\uparrow)}\ ,\nn 
\end{align}
and 
\be\label{limtinf}
\lim_{t\rightarrow \infty} \frac{f_2(t)}{t^{n_3}} =\frac{(-1)^{n_3-\Nmin-1}(n_3-n_1-1)!}{(\Nmax-n_1)!(n_3-\Nmin-1)!}g_{(\infty,\downarrow)}\ , \quad \lim_{t\rightarrow \infty} \frac{f_1(t)}{t^{n_1}}= \frac{ (n_3)!}{(n_1)!(n_3-n_1)!} g_{(\infty,\uparrow)} \ , \\
\ee
where we have named the linear constraint polynomials of the coefficients $b_N$ 
\begin{align}
&g_{(0,\uparrow)}=b_{\Nmin}\ , && g_{(0,\downarrow)}=\sum\limits_{N=N_{\rm min}}^{N_{\rm max}}(N-n_1+1)_{(\Nmax-N)}(\Nmin+1)_{(N-\Nmin)} b_N\ , \nn \\
&g_{(1,\uparrow)}=b_{\Nmax}\ , &&g_{(1,\downarrow)}=\sum\limits_{N=N_{\rm min}}^{N_{\rm max}}(n_1+n_3-\Nmax)_{(\Nmax-N)}(n_3-N)_{(N-\Nmin)} b_N\ ,\label{gdefs}\\
&g_{(\infty,\uparrow)}=\sum\limits_{N=\Nmin}^{\Nmax} b_N\ , && g_{(\infty,\downarrow)}=\sum\limits_{N=N_{\rm min}}^{N_{\rm max}} (-1)^{(N-\Nmin)}(N-n_1+1)_{(\Nmax-N)} (n_3-N)_{(N-\Nmin)} b_N  \ . \nn 
\end{align}
The definitions of the constraint polynomials $g_{(t_i,\updownarrow)}$ have been chosen such that the sums over $N$ are unconstrained between $\Nmin$ and $\Nmax$, although some of the Pochhammer symbols may be 0 (but not all of them).  The cases when these Pochhammer symbols are zero directly correspond to cases where rule \eqref{rule} applies, setting some Jacobi polynomials to 0 in one of the three forms of the map \eqref{near0}, \eqref{near1}, or \eqref{nearinfty}.   

In addition, one may solve the linear constraints as some set of linear functions $b_N(\{B_{M}\})$ of new variables $B_M$, where there is one fewer $B_M$ than there are $b_N$ (and so the bounds of the sum on $M$ are smaller, i.e. $\Delta M = \Delta N-1$).  Written in this way, the new functions reproduce the form of our maps exactly after taking the OPE limit, using the $B_M$.  To find these linear functions $b_N(\{B_{M}\})$,  we start with the known final form of the covering space map after the OPE limit has been taken, written in terms of the $B_M$.  We shift the sums appropriately, and find Jacobi polynomial identities that shift indices in appropriate ways.  This allows us to directly find the functions $b_N(\{B_{M}\})$.  We give example calculations for the twist up and twist down OPE limits as one of the operators approaches $t=0$ in appendix \ref{Appx.OPE.Examples}.  We simply summarize the other OPE constraints in table \ref{TableOPE} below.  The solutions $b_N(\{B_M\})$ always solve the linear homogeneous equation in the $b_N$ by making this equation a telescoping sum in the $B_M$ which telescopes to 0, which is straightforward to verify.      

\begin{table}[ht!]
\sbox0{{ 
\scriptsize
\hspace*{-1cm} \begin{tabular}{|c|c|c|c|c|c|}
\hline
Approach  & Twist & Algebraic  & $b_N(B_N)$ & Identity & Equivalent \\
Point & up/down & Restriction & Solution & Used & shift \\
\hline
$\phantom{\Bigg( \Bigg)}$ \hspace*{-0.8cm} $t=0$ & up & $g_{(0,\uparrow)}=b_{\Nmin}=0$ & $b_N=B_N$ for $N\neq \Nmin$ & NA & $\Nmin\rightarrow \Nmin+1$\\
\hline
    $\phantom{\Bigg( \Bigg)}$                        &     &                                                                    &  $b_N=\frac{1}{n_3}\left((N+1) B_{N-1}  - (N-n_1) B_{N-2}\right)$     &   & $n_i\rightarrow n_i-1$ \\
 $\phantom{\Bigg( \Bigg)}$ \hspace*{-0.8cm} $t=0$ & down & \hspace*{-0.1cm}  $g_{(0,\downarrow)}=\sum\limits_{N=N_{\rm min}}^{N_{\rm max}}(N-n_1+1)_{(\Nmax-N)}$ &  $\Nmin\leq N\leq \Nmax$, & \eqref{PidSumIndexShift}  & $\Nmax \rightarrow \Nmax-2$  \vspace*{-0.25cm}  \\
 $\phantom{\Bigg( \Bigg)}$                  &      &     \hspace*{3cm} $\times(\Nmin+1)_{(N-\Nmin)} b_N =0$                                                              &     $B_{\Nmin-2}=0,\quad$ $B_{\Nmax-1}=0$                                          &                                    & $\Nmin \rightarrow \Nmin-1$ \\[20pt]
\hline
$\phantom{\Bigg( \Bigg)}$ \hspace*{-0.8cm} $t=1$ & up & $g_{(1,\uparrow)}=b_{\Nmax}=0$ & $b_N=B_N$ for $N\neq \Nmax$ & NA & $\Nmax\rightarrow \Nmax-1$ \\
\hline
$\phantom{\Bigg( \Bigg)}$                        &     &                                                                    &  $b_N=\frac{1}{n_3}\left((n_1+n_3-N) B_{N-1}  - (n_3-N-1) B_{N}\right)$     &   & $n_i\rightarrow n_i-1$ \\
$\phantom{\Bigg( \Bigg)}$ \hspace*{-0.8cm} $t=1$  & down &\hspace*{-0.1cm} $g_{(1,\downarrow)}=\sum\limits_{N=N_{\rm min}}^{N_{\rm max}}(n_1+n_3-\Nmax)_{(\Nmax-N)}$ &  $\Nmin\leq N\leq \Nmax$  & \eqref{PidSumIndexShift.Switch} & $\Nmax\rightarrow \Nmax-1$   \vspace*{-0.25cm} \\
$\phantom{\Bigg( \Bigg)}$ \hspace*{-0.8cm}       &      &    \hspace*{3cm}$\times (n_3-N)_{(N-\Nmin)} b_N=0$                                                                        &          $B_{\Nmin-1}=0,\quad$ $B_{\Nmax}=0$          &                                 & \\[20pt]
\hline
$\phantom{\Bigg( \Bigg)}$ \hspace*{-0.8cm}            &    &                                 & $b_N=(B_N-B_{N-1})$ &             &                                                    \\
$\phantom{\Bigg( \Bigg)}$ \hspace*{-0.8cm} $t=\infty$ & up &  $g_{(\infty,\uparrow)}=\sum\limits_{N=\Nmin}^{\Nmax} b_N=0$  & $\Nmin\leq N \leq \Nmax$  & \eqref{tPid1.inf} & $\Nmax\rightarrow \Nmax-1$ \\
$\phantom{\Bigg( \Bigg)}$ \hspace*{-0.8cm}            &    &                                & $B_{\Nmin-1}=0,\qquad B_{\Nmax}=0$ &  \eqref{idforinfty1}    & $n_1\rightarrow n_1-1$ \\
\hline
$\phantom{\Bigg( \Bigg)}$                        &     &                                                                                                                             &  $b_N=-\frac{1}{n_3}\left((n_3-N-1)B_N+(N-n_1)B_{N-1}\right)$     &\eqref{idforinfty3}   &   \\
$\phantom{\Bigg( \Bigg)}$ \hspace*{-0.8cm} $t=\infty$  & down & \hspace*{-0.1 cm} $g_{(\infty,\downarrow)}=\sum\limits_{N=N_{\rm min}}^{N_{\rm max}} (-1)^{(N-\Nmin)} (n_3-N)_{(N-\Nmin)}$ &  $\Nmin\leq N\leq \Nmax$  & \eqref{idforinfty2} &$\Nmax\rightarrow \Nmax-1$ \vspace*{-0.25cm} \\
$\phantom{\Bigg( \Bigg)}$ & &     \hspace*{3cm}$\times(N-n_1+1)_{(\Nmax-N)} b_N=0$     & $B_{\Nmin-1}=0,\quad B_{\Nmax=0}$  & & \\[5pt]
$\phantom{\Bigg( \Bigg)}$                                &  &                                                                           &                               &        &  $n_3\rightarrow n_3-1$  \\
\hline
\end{tabular}}}
\centering
\rotatebox{90}{\begin{minipage}{ 0.95\textheight}
  \usebox0
  
  \caption{Table of OPE limits with algebraic constraints.  We name each of the above linear constraints for later use.\label{TableOPE}}
\end{minipage}}
\end{table}
\newpage

We can see that the OPE limits in table \ref{TableOPE} seem quite natural geometrically.  We have claimed that the maps \eqref{near0} correspond to the most general set with three long twists operators and a cloud of $\Delta N$ twist-2 operators, and that the parameters of the maps $b_N$ are valued in ${\mathbb{CP}}^{\Delta N}$.  One way of testing such a claim would be to consider OPE limits.  One can immediately see that a twist-2 operator approaching a long single-cycle twist has two non-trivial OPE limits: where the twist two increases the ramification of the long twist, or where it decreases the ramification of the long twist.  In either case, one is left with a cloud of $\Delta N-1$ twist-2 operators, and three long twists.  By our claim, this should correspond to a map of the same form \eqref{near0} with some coefficients $B_M$ which take values in ${\mathbb{CP}}^{\Delta N-1}$.  This suggests that the OPE limits are given by embedding ${\mathbb{CP}}^{\Delta N-1}$ inside of ${\mathbb{CP}}^{\Delta N}$.  The natural way of accomplishing this embedding is with linear homogenous polynomials in the $b_N$ which are set to 0.  This is exactly the implementation of the OPE limits in table \ref{TableOPE}.    

There are also OPE limits where the group product is trivial, i.e. no indices are shared between the twist at the point of concern and the twist-2 that approaches the operator at this point.  Let us consider the case where the point of concern is $t=0$, and so we consider the form of the map \eqref{near0}.  This type of ``inert'' OPE limit would be found by insisting that a zero of $Q(t)$, determining a ramified point, is shared by the polynomial $\sum_{N=N_{\rm min}}^{N_{\rm max}} b_N t^{(N-\Nmin)}P^{(N+1),-(n_1+n_3-N)}_{n_3-N-1}(1-2t)$ which defines a place where $z=0$ but that $t\neq 0$.   This would be determined by the resultant
\be
{\rm Res}\left(Q(t),\sum_{N=N_{\rm min}}^{N_{\rm max}} b_N t^{(N-\Nmin)}P^{(N+1),-(n_1+n_3-N)}_{n_3-N-1}(1-2t)\right)=0
\ee
which is again a homogenous polynomial constraint on the $b_N$ set to 0, given that the coefficients of $Q(t)$ are quadratic homogenous polynomials in the $b_N$.  This makes this constraint another algebraic variety subspace of $\mathbb{CP}^{\Delta N}$.  The fact that two distinct ramified points on the cover are both mapped to $z=0$ means that the cycles defining the group element at $z=0$ are in fact disjoint: different copies of the CFT are twisted together, but not into each other.  We have not explored these types of OPE limits further.

We can also use the above considerations to extract the leading order coefficients of $Q(t)$ as $t\rightarrow 0,1,\infty$ by realizing that these only depend on the leading order behavior of $f_1$ and $f_2$ in these limits, and these have been computed in \eqref{limt0}-\eqref{limtinf}.  We find
\begin{align}
& A_0 = \lim_{t\rightarrow \infty} \frac{W}{t^{n_1+n_3-1}}=\lim_{t\rightarrow \infty} \frac{Q(t)}{t^{\Delta N}}=(n_3-n_1)\left(\lim_{t\rightarrow \infty}   \frac{f_2(t)}{t^{n_3}}\right)\left( \lim_{t\rightarrow \infty}\frac{f_1(t)}{t^{n_1}}\right)\nn\\[3pt]
& \phantom{A_0}  = \frac{(-1)^{(n_3-\Nmin-1)} (n_3)!}{n_1!(\Nmax-n_1)!(n_3-\Nmin-1)!}\, g_{(\infty,\downarrow)}\,g_{(\infty,\uparrow)}\ , \label{A0gen}
\end{align}
\begin{align}
& A_{\Delta N} = \lim_{t\rightarrow 0} \frac{W}{t^{\Nmin} (t-1)^{n_3-\Nmax-1}}= Q(0)  \nn \\[3pt]
& \phantom{A_{\Delta N}} =(-1)^{n_1+n_3-\Nmax-1} (\Nmin+1) \left(\lim_{t\rightarrow 0} \frac{f_2(t)}{t^{\Nmin+1}}\right)\left(\lim_{t\rightarrow 0} f_1(t)\right)  \nn \\[3pt]
& \phantom{A_{\Delta N}} =\frac{(-1)^{(n_3-\Nmax-1)} (n_3)!}{n_1!(\Nmax-n_1)!(n_3-\Nmin-1)!}\, g_{(0,\downarrow)}\, g_{(0,\uparrow)}\ , \label{ADeltaNgen}
\end{align}
and
\begin{align}
& A_{\Sigma}\equiv \sum_{i=0}^{\Delta N} A_i =\lim_{t\rightarrow 1} \frac{W}{t^{\Nmin}(t-1)^{n_1+n_3-\Nmax-1}}=Q(1)\nn \\[3pt]
& \phantom{A_{\Sigma}} = (n_1+n_3-\Nmax) \left(\lim_{t\rightarrow 1}\frac{f_2(t)-f_1(t)}{(t-1)^{n_1+n_2-\Nmax}}\right)\left(\lim_{t\rightarrow 1} f_1(t)\right) \nn \\[3pt]
& \phantom{A_{\Sigma}} = \frac{(-1)^{n_3-\Nmax-1} (n_3)!}{(n_1)!(n_3-\Nmin-1)!(\Nmax-n_1)!}g_{(1,\downarrow)}g_{(1,\uparrow)} \ . \label{ASiggen}
\end{align}
To help understand the above, consider the identification of $A_0$.  Here we have used the fact that only the leading order coefficients of $f_1$ and $f_2$ in the $t\rightarrow \infty$ limit need to be computed: the $n_3-n_1$ comes about from taking the derivative of the top term in $f_2$ and $f_1$, which have degrees $n_3$ and $n_1$ respectively.  The identification of $A_{\Delta N}$ is arrived at by realizing $f_2' f_1$ has lowest power $t^{\Nmin}$, but $f_2f_1'$ has lowest power $t^{\Nmin+1}$ since the leading order in $f_1$ near $t=0$ is a constant ($f_2f_1'$ is identically 0 if $f_1$ is constant).  Thus, only the $f_2'f_1$ term contributes to the leading order term in the Wronskian in the $t\rightarrow 0$ limit.  This also explains the factor of $\Nmin+1$ on the second line of \eqref{ADeltaNgen}, since $f_2$ goes to $0$ as $t^{\Nmin+1}$.  Similar logic applies to $A_{\Sigma}$.  

The above solves for the Wronskian analytically in the cases $\Delta N=0,1,2$.  We note that the $\Delta N=0$ case is trivial: $W=t^{\Nmin} (t-1)^{n_1+n_3-\Nmin}A_0$ and $\Nmin=\Nmax=n_2$, giving the answer quoted, for example, in \cite{Lunin:2000yv}.  

In the case $\Delta N=1$, we have
\be
Q(t)= A_0 t + A_1
\ee   
where $A_0$ and $A_1$ are given in \eqref{A0gen} and \eqref{ADeltaNgen}, respectively, and $\Nmax=\Nmin+1$.  In this case we have found a map with three long twists and a single twist-2 insertion.  These polynomials are of Heun type, which have been considered previously \cite{Pakman:2009zz,Dei:2019iym} for use as covering space maps, however, using recursion relations to solve for coefficients.  These recursion relations may be feasibly solved for finite size twists.  Here we have the solution in closed form, at least for a case where three of the twists are large, and one is a twist-2 insertion.  Further, for us the location of the new twist insertion on the cover is known analytically in terms of the map parameters: $t_2=-A_1/A_0$.

Finally, in the $\Delta N=2$ case, we have $Q(t)= A_0 t^2 + A_1 t + A_2$, and $A_1$ is also found analytically, 
\be
A_1=A_{\Sigma}-A_0-A_2 \ . \label{A1WhenDN2}
\ee
Therefore, for $\Delta N=2$ we have
\be
Q(t)=A_0 t^2 + (A_{\Sigma}-A_0-A_2) t + A_2 \label{DN2Qanalytic}
\ee
where $A_0$, $A_2$, and $A_{\Sigma}$ are respectively given by \eqref{A0gen}, \eqref{ADeltaNgen}, and \eqref{ASiggen}, with $\Nmax=\Nmin+2$.  

A general method for finding $Q(t)$ for $\Delta N \geq 3$ is discussed in appendix \ref{WronskGen}.  This algorithmic approach finds the polynomial $Q$ in $\Delta N$ steps, and so is feasible when $\Delta N$ is not too large.  

The form of $Q(t)$ in \eqref{DN2Qanalytic} makes the discriminant of $Q$ in the $\Delta N=2$ case easy to write
\be
{\rm Disc}\,(A_0 t^2 + (A_{\Sigma}-A_0-A_2) t + A_2)= A_0^2+A_2^2+A_{\Sigma}^2-2A_0A_2-2A_0A_{\Sigma}-2A_2A_{\Sigma}
\ee
which has an obvious interchange symmetry amongst $A_0$, $A_2$, and $A_{\Sigma}$.

\subsection{\texorpdfstring{$\Delta N=2$}{TEXT} special OPE limits}\label{DN2OPEs}

In the case that $\Delta N=2$, there are two twist-2 operators in the cloud.  These cloud operators can approach the points at $(t=0,z=0)$, $(t=1,z=1)$, or $(t=\infty,z=\infty)$, as discussed in the previous subsection.  However, now we have the possibility that the two twist-2 operators in the cloud can approach each other.  These may ``twist down'' into an untwisted operator, or ``twist up'' into a twist-3 operator.  This can be easily addressed by insisting that the two zeros of $Q(t)=\sum_{i=0}^2 A_i t^{2-i}$ are coincident, which we explore by calculating the discriminant of $Q$ 
\begin{align}
{\rm Disc}\,(Q(t))&=A_1^2-4A_0A_2= A_0^2+A_2^2+A_{\Sigma}^2-2A_0A_2-2A_0A_{\Sigma}-2A_2A_{\Sigma} \label{Q2Discrim} \\
&=\left(\frac{n_3! }{n_1!(\Nmin+2-n_1)!(n_3-\Nmin-1)!}\right)^2 g_{(c,\downarrow)}\, g_{(c,\uparrow)} \nn 
\end{align}
where $A_0$, $A_2$, and $A_{\Sigma}$ are given by \eqref{A0gen}, \eqref{ADeltaNgen}, and \eqref{ASiggen} for the case $\Delta N=2$, and where $g_{(c,\downarrow)}$ and $g_{(c,\uparrow)}$ define a factorization of the discriminant.  We give these factors momentarily.  Insisting that the discriminant vanishes so that the zeros of $Q(t)$ are coincident now breaks into two separate cases: $g_{(c,\downarrow)}=0$; or $g_{(c,\uparrow)}=0$.  The first case reads
\begin{align}\label{cloud2twistdown}
g_{(c,\downarrow)}&\equiv(\Nmin-n_1+2)(n_1+n_3-\Nmin-1)b_{\Nmin} +(\Nmin-n_1+2)(n_3-\Nmin-1)b_{(\Nmin+1)}\nn \\[4pt]
&   +(\Nmin+2)(n_3-\Nmin-1)b_{(\Nmin+2)}=0  
\end{align}
and can be shown to be the ``twist down'' case.  We show this in appendix \ref{DN2TwistDownAppx}. 

The other possibility is an OPE limit where the operators in the cloud ``twist up'' into a twist-3, i.e. a ramification 2 point, and this is given by the other factor in \eqref{Q2Discrim}, 
\begin{align}
g_{(c,\uparrow)}&\equiv4(\Nmin-n_1+1)(n_1+n_3-\Nmin-2)b_{\Nmin}^2 b_{(\Nmin+2)} \label{cloud2twistup} \\
& -(\Nmin-n_1+2)(n_1+n_3-\Nmin-1)b_{\Nmin}b_{(\Nmin+1)}^2 \nn \\
&  +4\big((\Nmin + 1)(n_1+n_3-\Nmin- 2) - n_1n_3\big)b_{\Nmin}b_{(\Nmin+1)}b_{(\Nmin+2)}  \nn \\
&  + 4(\Nmin + 1)(n_3-\Nmin - 2)b_{\Nmin}b_{(\Nmin+2)}^2 \nn \\
& -(n_3-\Nmin-1)(\Nmin-n_1+2)b_{(\Nmin+1)}^3 \nn \\
&   -(\Nmin+2)(n_3-\Nmin-1)b_{(\Nmin+1)}b_{(\Nmin+2)}^2=0\nn
\end{align}
which is a homogenous cubic constraint on the $b_N$.  We note that this restriction is an algebraic variety inside of $\mathbb{CP}^{\Delta N=2}$, similar to all other OPE limits found.  The above constraint applied to the map \eqref{near0} furnishes a map that has three long twists and a single twist-3 operator.  

Enforcing either constraint so that the discriminant vanishes, the position on the cover that the two ramification 1 points approach is given by
\begin{align}
t_3 & =-\frac{A_1}{2 A_0} = \frac{ g_{(0,\uparrow)}g_{(0,\downarrow)}+g_{(\infty,\uparrow)}g_{(\infty,\downarrow)}-g_{(1,\uparrow)}g_{(1,\downarrow)}}{2 g_{(\infty, \uparrow)} g_{(\infty,\downarrow)}}\ .\label{a2location}
\end{align}

Thus, there are two OPE limits where the two points in the cloud on the cover approach each other.  One case is the requirement \eqref{cloud2twistdown} in which the two twist-2 operators merge to give an untwisted operator at \eqref{a2location}.  The map simplifies to
\be
z(t)=\frac{f_2}{f_1}= \frac{t^{\Nmin+1} P_{n_3-\Nmin-2}^{(\Nmin+1),-(n_1+n_3-\Nmin-2)}(1-2t)}{P^{-(\Nmin+1),-(n_1+n_3-\Nmin-2)}_{n_1-1}(1-2t)} \label{twistDownCloudMap}
\ee
which we show in appendix \ref{Appx.OPE.Examples}.  This is just a case where we replace $n_i'=n_i-1$, $\Nmin'= \Nmin$, and $\Nmax'=\Nmax-2$ so that $\Nmax'=\Nmin'=n_2$ and there is no sum.  Furthermore, these replacements leave the ramifications of the points at $t=0,1,\infty$ unaffected.  We may simplify \eqref{a2location} in this case, realizing that \eqref{cloud2twistdown} is linear. We solve \eqref{cloud2twistdown} for $b_{(\Nmin+1)}$ and substitute into \eqref{a2location}, finding
\be
t_{\downarrow}= \frac{b_{\Nmin}(\Nmin-n_1+2)}{b_{\Nmin}(\Nmin-n_1+2)+b_{(\Nmin+2)}(n_3-\Nmin-1)} \ . \label{tmMain}
\ee
This agrees with \eqref{tmAppx} which is the zero of the linear \eqref{LinCancCloud}.  This linear cancels between the numerator and denominator polynomials, identifying it as a twist down -- see in appendix \ref{DN2TwistDownAppx}.

Note that the original 5-point function, i.e three long twists and a pair of twist-2 insertions, has two cross ratios.  One might be concerned that the OPE limit \eqref{cloud2twistdown} is only a linear relationship between the $b_N$, and so should decrease the dimension of the space of maps only by one.  However, it is important to note that $t_{\downarrow}$ is a marked point on the cover where the ramified points approach each other.  At this point we expect a full OPE expansion.  Even the bare twists have such an expansion, given by fractional modes of the stress tensor acting at the corresponding position in the base space, or equivalently modes of the covering space stress tensor (along with terms arising from the Schwarzian) acting at the point $t=t_{\downarrow}$ \cite{Burrington:2018upk}.  If the twists are excited twists, rather than bare twists, other fractional modes of fields can also appear, for example modes of the superconformal currents \cite{DeBeer:2019oxm}.  Therefore, the above OPE limit still results in a four point function: the fourth operator is in the untwisted sector, and so does not show up in the covering space map directly.  Furthermore, the expression \eqref{tmMain} is scaling invariant under $b_N\rightarrow \lambda b_N$, and so is determined by a point in ${\mathbb{CP}}^1$, as should be expected: a linear algebraic variety \eqref{cloud2twistdown} inside of $\mathbb{CP}^2$ is $\mathbb{CP}^1$.  Of course $\mathbb{CP}^1$ is just the sphere, and so the marked point $t_\downarrow$ takes values on the covering space sphere.  

In the twist up case, when \eqref{cloud2twistup} is enforced, the two ramification 1 points merge into a ramification 2 point located at \eqref{a2location}, which is also $b_N$ scaling invariant.  If we take this in a limiting way, the 5-point function becomes singular, as expected: this is a contact singularity.  However, to get the correct 4-point function, we recognize that the two zeros are coincident, and  so the Wronskian is given by
\be
W= t^{\Nmin} (t-1)^{n_1+n_3-(\Nmin+2)-1} A_0 (t-t_3)^2
\ee 
where $t_3$ is \eqref{a2location} with \eqref{cloud2twistup} imposed.  In this case, there is only one additional ramified point, other than $t=0,1,\infty$, which contributes to the 4-point function calculation.  This is distinct from the case where we impose \eqref{cloud2twistup} in a limiting way, which would correspond to the singular limit of a 5-point function, where the zeros of $Q(t)$ are close, but distinct -- see the discussion surrounding \eqref{a3eq}.

\section{Correlation functions}
\label{CorrSect}

In the previous sections, we have attempted to be as general as possible analyzing the maps \eqref{near0}.  We will continue this for the time being while considering the $n$-point function calculation.  

First, the $n$-point functions are given by \cite{Dei:2019iym}
\begin{align}
&\langle \sigma_{\hat{w}_0}(z_0,\zb_0) \sigma_{\hat{w}_1}(z_1,\zb_1) \cdots \sigma_{\hat{w}_{n-2}}(z_{n-2},\zb_{n-2}) \sigma_{\hat{w}_{n-1}}(\infty,\bar{\infty}) \rangle \label{npointGen} \\[5pt]
&= \prod_{i=0}^{n-2}w_i^{-\frac{c(w_i+1)}{12}} w_{n-1}^{\frac{c(w_{n-1}+1)}{12}} \prod_{j=0}^{n-2} |a_j|^{-\frac{c(w_j-1)}{12 w_j}} |a_{n-1}|^{\frac{c(w_{n-1}-1)}{12 w_n}} \prod_{\rho} |C_{\rho}|^{-\frac{c}{6}}\ ,\nn
\end{align}
which is to be read for a specific group element representative.  In the above $a_i$ are defined to be the leading order coefficients of the expansion of $z(t)$ near a ramified point, as in \eqref{tiExpand}, the $w_i=r_i+1$ are the lengths of the single-cycle twists, the $C_\rho$ are the coefficients describing the unramified images of $z=\infty$, see \eqref{InfImage}, and $c$ is the central charge of the seed CFT (and we have assumed $c=\ti{c}$).  One must then sum over all preimages of the maps, see \cite{Pakman:2009zz} and \cite{Dei:2019iym}.  Recall that we have indexed $i=0, \cdots, n-1$ for the $n$ bare twists.  This allows us to denote $(t_0=0,z_0=0)$, and $(t_1=1,z_1=1)$, and we use interchangeably $r_\infty=r_{n-1}$ and $(t_{n-1}=\infty,z_{n-1}=\infty)$. 

The $a_i$ may be computed by noting that
\be\begin{split}
& z(t) = z_j + a_j (t-t_j)^{w_j} + \cdots\ , \\
& \pa z(t)= w_j a_{j} (t-t_j)^{r_j}+\cdots = \frac{W}{f_1^2}=A_0\, \frac{\prod_{i=0}^{n-2} (t-t_i)^{r_i}}{f_1^2}\ ,
\end{split}\ee
where $A_0$ is the leading order coefficient appearing in $Q(t)$.  We can therefore identify
\be
a_j= A_0 \frac{\prod\limits_{i\neq j} (t_j-t_i)^{r_i}}{w_j(f_1(t_j))^2}\ . \label{ajGen}\vs{-7}
\ee

The ramified points at finite locations are given by $t=0$, $t=1$, and the $\Delta N$ zeros of $Q(t)$, i.e. $t_i$ for $i=2,\cdots, \Delta N+1$ where $\Delta N+1=n-2$.  This gives us
\begin{align}
&a_0= (-1)^{r_1} \frac{Q(0)}{(\Nmin+1)(f_1(0))^2}=\frac{(-1)^{n_1} (n_1)!(n_3)!(\Nmax-n_1)!}{(\Nmin+1)((\Nmin)!)^2(n_3-\Nmin-1)!} \frac{g_{(0,\uparrow)}}{g_{(0,\downarrow)}}\ , \label{a0Find} \\[7pt]
&a_1=\frac{Q(1)}{(n_1+n_3-\Nmax)(f_1(1))^2}= \frac{(-1)^{n_3-\Nmax-1}(n_1)!(n_3)!(n_3-\Nmin-1)!}{(n_1+n_3-\Nmax)((n_1+n_3-\Nmax-1)!)^2(\Nmax-n_1)!} \frac{g_{(1,\uparrow)}}{g_{(1,\downarrow)}}\ .\nn 
\end{align}
We note that the generic expression $\eqref{ajGen}$, and the specific expressions \eqref{a0Find} depend on the leading coefficients of the polynomial $Q(t)$, i.e. $A_{0}$, $A_{\Delta N}$, and $A_{\Sigma}$ which have been found in the general case in \eqref{A0gen}, \eqref{ADeltaNgen}, and \eqref{ASiggen}, respectively. 

For the other points, we recognize that we do not need to calculate the $a_i$ individually, but rather the product of the $a_i$: these all have the same ramification of 1.  Individually, they are given by
\be
a_i= \frac{t_i^{\Nmin} (t_i-1)^{n_1+n_3-\Nmax-1} \lim\limits_{t\rightarrow t_i} \frac{Q(t)}{(t-t_i)}}{2(f_1(t_i))^2}\ , \qquad \qquad i\neq 0,1, n-1\ .
\ee
The above $i$ only runs over the ramified positions labeled by $2\leq i\leq n-2$, and these are associated with the cloud of twist-2 operators.  These therefore refer to the zeros of $Q(t)$.  We recognize that 
\be
Q(t)=A_0 \prod_{i=2}^{\Delta N+1} (t-t_i)
\ee
and so we may write the product over the $a_i$ as
\be
\prod_{i=2}^{\Delta N+1} a_i  = \frac{\left(\frac{g_{(0,\uparrow)}g_{(0,\downarrow)}}{g_{(\infty,\uparrow)}g_{(\infty,\downarrow)}}\right)^{\Nmin} \left(\frac{g_{(1,\downarrow)}g_{(1,\uparrow)}}{g_{(\infty,\downarrow)}g_{(\infty,\uparrow)}}\right)^{n_1+n_3-\Nmax-1} \frac{(-1)^{\frac12\Delta N(\Delta N-1)}{\rm Disc}(Q)}{A_0^{\Delta N-2}}}{2^{\Delta N} \left(\frac{{\rm Res}(Q,f_1)}{A_0^{n_1^{\;}}}\right)^2} \ . \label{prodai}\vs{-5}
\ee
Above we have written the answer in terms of the resultant ${\rm Res}(Q,f_1)/A_0^{n_1}$, where $A_0$ has been given in \eqref{A0gen}.  In what follows, we will be considering $Q$ with small degree, and so it is more convenient to evaluate this as ${\rm Res}(Q,f_1)/A_0^{n_1}=\prod_{i=2}^{n-2} f(t_i)$ so that there are just a few evaluations of the known function $f_1$ at the location of the zeros of $Q$.  We will approach this in a systematic way  momentarily.

We next consider the point at infinity, writing
\be\begin{split}
& z(t)= a_{n-1} t^{w_{n-1}}+\mathcal{O}(t^{w_{n-1}-1})\ ,\\
& \pa z(t)= w_{n-1} a_{n-1}t^{r_{n-1}} +\cdots =A_0 \frac{\prod_{i} (t-t_i)^{r_i}}{w_j(f_1(t))^2}\ .
\end{split}\ee
Remembering that $w_{n-1}-1=r_{n-1}=n_1-n_3-1$, and that the Wronskian diverges like $t^{n_3+n_1-1}$, we identify
\be
a_{n-1}=\frac{A_0}{(n_3-n_1)} \frac{1}{\left(\lim\limits_{t\rightarrow \infty} \frac{f_1}{t^{n_1}}\right)^2}=\frac{(-1)^{n_3-\Nmin-1}(n_3-n_1)\big((n_3-n_1-1)!\big)^2 (n_1)!}{(n_3)!(\Nmax-n_1)!(n_3-\Nmin-1)!}\, \frac{g_{(\infty,\downarrow)}}{g_{(\infty,\uparrow)}} \ .\label{anm1Find}\vs{-5}
\ee
One may also read this directly from the ratio $z=f_2/f_1$ in the limit $t\rightarrow \infty$.

The result \eqref{anm1Find}, up to a constant, is the ``twist down'' $g_{(\infty,\downarrow)}$ divided by the ``twist up'' $g_{(\infty,\uparrow)}$ .  Interestingly, this restores some of the symmetry between the ramified points at finite locations, and the ramified point at infinity by noting the extra minus sign in the powers of $a_{n-1}$ appearing in \eqref{npointGen}.  This is also noted in \cite{Dei:2019iym} and \cite{Eberhardt:2019ywk} by a redefinition of the coefficient $a_{n-1}$ for the point at infinity, although we have noted it rather directly here by identifying the polynomials in the $b_N$ that compose these coefficients, and how these correspond to different types of OPE limits. 

Finally, we need the product over the $C_\rho$.  These are the unramified images of infinity and are given by the zeros of $f_1$, which we have denoted $t_\rho$ where $\rho$ runs from $1$ to $n_1$. We see that
\be\begin{split}
&z(t)= \frac{C_\rho}{(t-t_{\rho})} +\cdots\ ,\\
&\pa z  = \frac{-C_{\rho}}{(t-t_{\rho})^2}+\cdots= \frac{W}{(f_1(t))^2}= \frac{t^{\Nmin} (t-1)^{n_1+n_3-\Nmax-1} Q(t)}{(f_1(t))^2}\ ,
\end{split}\ee
and so
\be
C_{\rho}= - \frac{t_{\rho}^{\Nmin} (t_{\rho}-1)^{n_1+n_3-\Nmax-1} Q(t_{\rho})}{\left(\lim\limits_{t\rightarrow t_{\rho}}\frac{f_1(t)}{(t-t_\rho)}\right)^2}\ .\vs{-5}
\ee
We note that
\begin{align}
f_1(t)= \frac{(n_3)!}{n_1!(n_3-n_1)!}\, g_{(\infty,\uparrow)} \prod_{\rho=1}^{n_1} (t-t_{\rho})\ .
\end{align}
With this, we have
{\footnotesize
\begin{align} \label{prodC}
& (-1)^{n_1}\prod_{\rho=1}^{n_1} C_{\rho} \\ 
&{\footnotesize =\frac{\left(\frac{(\Nmin)!(n_3-n_1)!g_{(0,\downarrow)}}{(n_3)!(\Nmax-n_1)! g_{(\infty,\uparrow)}}\right)^{\Nmin} \left(\frac{(-1)^{n_1}(n_3-n_1)!(n_1+n_1-\Nmax-1)! g_{(1,\downarrow)}}{(n_3)!(n_3-\Nmin-1)!g_{(\infty,\uparrow)}}\right)^{n_1+n_3-\Nmax-1}\left( \frac{(-1)^{n_1\Delta N} {\rm Res}(Q,f_1)}{\left(\frac{(n_3)!}{n_1!(n_3-n_1)!} g_{(\infty,\uparrow)} \right)^{\Delta N}}\right) }{\left(\frac{{\rm Disc}(f_1)}{\left(\frac{(n_3)!}{n_1!(n_3-n_1)!} g_{(\infty,\uparrow)} \right)^{n_1-2}}\right)^2}}\ . \nn 
\end{align}}
\\\noindent Each piece of the $n$-point function in \eqref{npointGen}, namely \eqref{a0Find}, \eqref{prodai}, and \eqref{prodC}, includes terms which are the polynomials in the $b_N$ which give OPE constraints discussed in section \ref{OPELimSec}, helping identify the singularity structure.

In this last expression, we have a formula that depends on ${\rm Disc}(f_1)$, which is a homogeneous polynomial in the $b_N$ (possibly factorizable).  We expect this discriminant to be expressible using ${\rm Res}(Q,f_1)$ for the following reasons.  First, if ${\rm Disc}(f_1)=0$, then $f_1$ has repeated roots.  If this is the case, then the repeated zeros must also be zeros of the Wronskian.  This type of zero must not be part of the prefactor $t^{\Nmin} (t-1)^{n_1+n_2-\Nmax-1}$ of $W$ since these are generic, and independent of the values of the $b_N$.  Therefore, if ${\rm Disc}(f_1)=0$ for some specific values of $b_N$, then $Q$ has a zero at the repeated root, and so ${\rm Res}(Q,f_1)=0$ at those same values of $b_N$.  This implies that every zero of ${{\rm Disc}(f_1)}$ is contained in the zeros of ${\rm Res}(Q,f_1)$.  This means that we expect the polynomials in the $b_N$ that compose ${\rm Disc}(f_1)$ must be contained in the polynomials in the $b_N$ that compose ${\rm Res}(Q,f_1)$.  In addition, if for some values of $b_N$ the polynomials $f_1$ and $f_2$ share a zero at some $t_i$, the Wronskian must also be zero at this $t_i$.  These are exactly the ``twist down'' OPE limits we have seen in section \ref{OPELimSec}, but will also involve the additional ``twist down'' OPEs that occur when distinct members of the cloud come together in an OPE limit.  We also expect the ``twist up'' to appear, given that the discriminant and resultant depend on the leading coefficient of the polynomials involved.  We will have one example of such a limit below in the $\Delta N=2$ case.  

Thus, we expect ${\rm Res}(Q,f_1)$ to contain ${\rm Disc}(f_1)$ along with additional factors of the ``twist down'' and ``twist up'' OPE constraint polynomials.  In what follows, we simply reverse engineer ${\rm Disc}(f_1)$  for the specific cases at hand, $\Delta N=1,2$.  We do so by calculating specific examples, identifying polynomials in the $b_N$, and fixing powers and coefficients. As mentioned above,  ${\rm Res}(Q,f_1)/A_0^{n_1}= \prod_{i=2}^{n-1} f_1(t_i)$ will be more efficient for us because the latter product will contain only $\Delta N$ terms, which will be small for the examples considered.  Given the discussion above, we must find the Wronskian in the special cases, and then reverse engineer ${\rm Disc}(f_1)$ in terms of ${\rm Res}(Q,f_1)$, and this will furnish the $n$-point functions in closed form.

We start with the case $\Delta N=1$.  We have
\be
Q(t)=A_0 t + A_1
\ee
where the coefficients are given by the generic formulas \eqref{A0gen} and \eqref{ADeltaNgen}.  The location of the new ramified point is given by
\begin{align}
t=t_2&=-\frac{A_1}{A_0}= \frac{g_{(0,\uparrow)}g_{(0,\downarrow)}}{g_{(\infty,\uparrow)}g_{(\infty,\downarrow)} }\\[5pt]
&=\frac{b_{\Nmin}\big((\Nmin-n_1+1)b_{\Nmin}+(\Nmin+1)b_{\Nmin+1}\big)}{(b_{\Nmin}+b_{\Nmin+1})
\big((n_3-\Nmin)b_{\Nmin}- (\Nmin-n_1+1) b_{\Nmin+1}\big)}\ .\nn
\end{align}
One may check the above by considering the case $-\varepsilon=\nmax=1$ in \cite{Ishkhanyan:2014wma}, constructing the $2\times 2$ matrix equation (20) in that work.  Solving this system of 2 equations for $q$ and $t_2$ (they call $t_2=a$ in their work) in terms of $c_0$ and $c_1$ gives the same answer as the above for $t_2$, keeping in mind one must change notation from $c_i$ to the $b_{N}$ as in \eqref{bNdef}.  Reverse engineering the discriminant, we find
{\small
\begin{align}
{\rm Disc}(f_1)&= \frac{f_1(t_2)(g_{(\infty,\uparrow)})^{n_1-1} (g_{(\infty,\downarrow)})^{n_1}}{g_{(0,\downarrow)}g_{(1,\downarrow)}}   \label{RI1} \\[3pt]
&\times \prod_{j=1}^{n_1} j^{j+3-2n_1} \prod_{j=1}^{n_1-1}\big(j-(\Nmin+1)\big)^{j-1} \prod_{j=2}^{n_1}\big(j-(n_1+n_3-\Nmin)\big)^{j-2} \prod_{j=1}^{n_1}(j-n_3-1)^{n_1-j} \ . \nn 
\end{align}}

\noindent One may check that the above gives the correct answer in the limiting cases $b_{\Nmin}=0$ and $b_{\Nmin+1}=0$, when these limits are allowed.  In these cases $f_1$ reduces to a single Jacobi polynomial. These discriminants are known \cite{Szego} and are in fact part of how we have reverse engineered the answer.

In the case $\Delta N=2$ we have\vs{-5}
\be
Q(t)=A_0 t^2 + A_1 t + A_2
\ee  
where the $A_i$ are given by \eqref{A0gen}, \eqref{ADeltaNgen}, and $A_1$ is given in this special case by \eqref{A1WhenDN2}, using \eqref{ASiggen}.  The two zeros of $Q$ are given by
\be
t_{\pm}= \frac{-A_1 \pm \sqrt{A_1^2-4 A_0 A_2}}{2A_0} \ . \label{tpmdef}
\ee
We again reverse engineer the discriminant of $f_1$ finding
\begin{align}
&{\rm Disc}(f_1) = \frac{f_1(t_+)f_1(t_-) (g_{(\infty,\uparrow)})^{n_1-1} (g_{(\infty,\downarrow)})^{n_1}}{g_{(0,\downarrow)} g_{(1,\downarrow)}g_{(c,\downarrow)}}\label{RI2}  \\[3pt]
& \qquad \times \prod_{j=1}^{n_1} j^{j+4-2n_1} \prod_{j=1}^{n_1-2} \big(j-(\Nmin+1)\big)^{j-1} \prod_{j=3}^{n_1} \big(j-(n_1+n_3-\Nmin)\big)^{j-3} \prod_{j=1}^{n_1} (j-n_3-1)^{n_1-j} \ .\nn
\end{align}

Equations \eqref{RI1} and \eqref{RI2} therefore give the last remaining ingredients to express the $n$-point function \eqref{npointGen} in closed form for $\Delta N=1,2$.

In the case that $\Delta N=2$ and \eqref{cloud2twistup} is enforced, there is only a single $a_i$ associated with the point $t=t_3$, rather than the two $a_i$.  This is because \eqref{cloud2twistup} enforces that $Q$ has the form $Q=A_0(t-t_3)^2$ with $t_3$ given in \eqref{a2location}.  Thus, rather than $\prod_{j=2}^{3} a_j$ for the two zeros of $Q$, there is only one new $a_i$ to calculate 
\be \label{a3eq}
a_3 = \frac{t_3^{\Nmin} (t_3-1)^{n_1+n_3-\Nmax-1}A_0}{2\big(f_1(t_3)\big)^2}
\ee
which we can use in \eqref{npointGen}, keeping in mind that all equations in this case must be read with an extra algebraic constraint \eqref{cloud2twistup}.   

\section{Discussion}
\label{DiscSection}

In this paper we have considered covering space maps where both the covering space and base space are spheres.  We have shown that the maps \eqref{near0}, or equivalently \eqref{near1}, or \eqref{nearinfty}, have arbitrarily large ramifications at $(t=0, z=0)$, $(t=1, z=1)$ and $(t=\infty, z=\infty)$, and a cloud of ramification 1 points by computing the Wronskian $W$ in \eqref{wronskform}.  The location of the ramification 1 points are given by the zeros of the polynomial $Q(t)$ in \eqref{Qform}, which is part of $W$. We have shown that the class of maps \eqref{near0} cover all group theoretically allowed ramifications in this class.  The integers in the map $n_1$, $n_3$, $\Nmin$, and $\Nmax$ can be found directly from these ramifications through \eqref{nofr}, and give non-pathological maps satisfying \eqref{goodmaps}.  

There are $\Delta N+1$ coefficients defining the maps \eqref{near0}, which we have called $b_N$.  We argued they are valued in $\mathbb{CP}^{\Delta N}$ due to an invariance of the maps under the scaling $b_N\rightarrow \lambda b_N$.  Therefore, the space of maps is $\Delta N$ dimensional, the correct dimension to parameterize the $\Delta N$ cross ratios for a $(3+\Delta N)$-point function.  The map parameters $b_N$ control the cross ratios, but also control which group product channel is taken when considering an OPE limit, i.e. when the ramified points approach each other.  We considered a set of these OPE limits for the general map \eqref{near0}, which are summarized in table \ref{TableOPE}.  We also considered other OPE limits in section \ref{DN2OPEs} and in appendix \ref{Appx.OPE.Examples}.  These OPE limits are all seen to be given by homogenous polynomials in the coefficients $b_N$, and so are algebraic variety subspaces of $\mathbb{CP}^{\Delta N}$.

To compute $n$-point functions, one generally needs to be able to compute certain coefficients in the maps.  First, we need to evaluate the Wronskian \eqref{wronskform}, which we have shown can be computed in closed form algorithmically in $\Delta N$ steps (see appendix \ref{WronskGen}), finding the polynomial $Q$ analytically.  We have done this explicitly up to $\Delta N=2$.  Next, we need the discriminant of $Q$, which furnishes polynomials in $b_N$ that encode OPE limits of operators in the cloud approaching each other.  One also needs to compute the resultant ${\rm Res}(Q(t),f_1(t))$, and for small $\Delta N$ this is simply evaluation of $f_1$ at the zeros of $Q$.  Finally one also needs to compute the discriminant of $f_1$, which we argue can be written in terms of the polynomials in $b_N$ which make up the resultant ${\rm Res}(Q(t),f_1(t))$, and other OPE limit polynomials already found.   Thus, our previous identification of the OPE limits helps express the coefficients that we use to construct the correlators, and in such a way that the singularity structure is clear.  Concentrating on the cases $\Delta N=0,1,2$, we were able to write closed form answers for the Wronskian, and the constants necessary to construct the $4$-point and $5$-point functions in these cases.  

One of our primary motivations for this work is with an eye towards holography.  A canonical example is the D1-D5 CFT.  To move to the point on the moduli space that is well described by classical supergravity, one must deform the theory along a specific direction in the moduli space.  In the D1-D5 CFT, the pertinent exactly marginal operator is in the twist-2 sector  \cite{Seiberg:1999xz,Larsen:1999uk,David:1999ec}.  Thus, to track the parameters of the theory, and ultimately observables, amounts to using conformal perturbation theory to high order.  These calculations would involve a large number of twist-2 operators to be inserted, for which our maps have direct relevance. 

These considerations also begin to outline our future directions.  We would like to use our maps to further explore conformal perturbation theory applied to candidate holographic orbifold CFTs.  This includes programmes of finding the change of the dimensions of operators and the change of structure constants -- see \cite{Burrington:2023vei} and references therein.  The problem of studying the changes to the structure constants has been limited to low twist operators.  Here, we have constructed maps for correlation functions with large twists in closed form, and obtained in some cases closed form solutions for the $n$-point correlation functions.  

Importantly, the closed form expressions are given in terms of the covering space map parameters $b_N$. One must sum over all images of the map to construct the full correlation function.  For a given point in the base space, $z_0=f_2(t)/f_1(t)$ has many solutions in $t$, even for ramified points \cite{Pakman:2009zz,Pakman:2009ab,Dei:2019iym}.  However, at least part of this sum is encoded in the map parameters.  Consider one of the simplest cases, a 4-point function with a single twist-2 insertion, which needs to be integrated over to compute a perturbation to the structure constant.   In this case we have the location of the ramified point as 
\be \label{t2Disc}
t=t_2=-\frac{A_1}{A_0}= \frac{g_{(0,\uparrow)}g_{(0,\downarrow)}}{g_{(\infty,\uparrow)}g_{(\infty,\downarrow)} }=\frac{b_{\Nmin}\big((\Nmin-n_1+1)b_{\Nmin}+(\Nmin+1)b_{\Nmin+1}\big)}{(b_{\Nmin}+b_{\Nmin+1})\big((n_3-\Nmin)b_{\Nmin}- (\Nmin-n_1+1) b_{\Nmin+1}\big)} \ .
\ee
Given a specific value for the local coordinate $r=b_{\Nmin+1}/b_{\Nmin}$ on $\mathbb{CP}^{1}$, we see that we can read a value $\zeta=f_2(t_2(r))/f_1(t_2(r))$: i.e. this specific value of $r$ defines the cross ratio $\zeta$.  However, the equation $\zeta=f_2(t_2(r))/f_1(t_2(r))$ for a fixed value of $\zeta$ has many solutions, and corresponds to distinct maps.  The number of these maps are called the connected Hurwitz number $H$.  One may check that the above computation agrees with \cite{Liu:2006psh} (also used in \cite{Dei:2019iym}) which gives the number of such maps in this case $H\!\!=\!\!\!\min\limits_{i=0,\cdots,3}\! \Big(w_i(S+1-w_i)\Big)$ with $S=n_3$, $w_0=\Nmin+1$, $w_1=n_1+n_3-\Nmax$, $w_2=2$, and $w_3=n_3-n_1$; we have checked  this for many values of $n_i$ and $\Nmin$.      

In conformal perturbation theory one has to integrate over the position of the deformation operator, i.e. over all possible values of $\zeta$.  This would be integrating over a set of $H$ patches of $\mathbb{CP}^{1}$, and each patch represents one cover of the base space (the cross ratio ranges over the base space sphere).  It seems it might be more efficient to simply change to the coordinates of $\mathbb{CP}^{1}$ and integrate over this coordinate, and this would account for these $H$ transport equivalent maps.  One would have to be careful about how to regulate the integrals, given the non-trivial form of the maps $z=f_2/f_1$ and the nontrivial form of $t_2$ in \eqref{t2Disc}: hard disk regulators in the original $z$ plane would have complicated pre-images in the $t$-plane cover, as well as in the $\mathbb{CP}^1$ parameterized by $b_{\Nmin}$ and $b_{\Nmin+1}$, and so a careful treatment would be necessary.  Similarly, the twist-3 deformation operators in \cite{Belin:2020nmp} would presumably be integrated over the 1 complex dimensional algebraic variety inside of $\mathbb{CP}^{2}$ defined by \eqref{cloud2twistup} with similar complications.  We hope to explore these problems in future work.

\section*{Acknowledgements}

We are thankful to the Mainz Institute for Theoretical Physics for hospitality where part of this work was done, particularly during the workshop “Exact Results and Holographic Correspondences”, and we thank the participants of this workshop for many interesting conversations.  BAB is thankful for funding support from Hofstra University including startup funds and faculty research and development grants, and for support from the Scholars program at KITP, which is supported in part by grants NSF PHY-1748958 and PHY-2309135 to the Kavli Institute for Theoretical Physics (KITP), where some of this work was completed.

\appendix

\section{Collected identities for Jacobi Polynomials}\label{JacobiIDappx}

The Jacobi polynomials are defined through the series expansion
\begin{align}
P_\gamma^{\alpha,\beta}(x)=\sum_{\ell=0}^\gamma \frac{(\gamma+\alpha+\beta+1)_\ell (\alpha+\ell+1)_{\gamma-\ell}}{\ell!(\gamma-\ell)!}\left(\frac{x-1}{2}\right)^\ell \label{JacobiSumForm}
\end{align}
where $(\kappa)_{\delta}=(\kappa)(\kappa+1) \cdots (\kappa+\delta-1)$ is Pochhammer's symbol (i.e. the ``rising factorial'' with $\delta$ terms).  This can be written in terms of the gamma function
\begin{equation}
(\kappa)_{\delta}=\frac{\Gamma(\kappa+\delta)}{\Gamma(\kappa)}
\end{equation}
keeping in mind that regulation may be necessary for negative integer values.  There is also the identity
\begin{equation}
(-\kappa)_{\delta}=(-1)^\delta (\kappa-\delta+1)_\delta\ .
\end{equation}

Some particularly useful values of Jacobi polynomials are (assuming $\gamma\geq 0$, and that $\alpha, \beta, \gamma$ are integers)
\begin{align}
P_\gamma^{\alpha, \beta}(1)= \frac{(\alpha+1)_\gamma}{\gamma!}= \begin{cases} \frac{(\alpha+\gamma)!}{\alpha! \gamma!}\qquad  &\text{if $\alpha\geq0$} \\[4pt]
(-1)^\gamma \frac{ (-\alpha-1)!}{(-\alpha-\gamma-1)! \gamma!} \qquad &\text{if $\alpha\leq-1$ and $\alpha+\gamma+1\leq 0$} \label{valueat1i}\\[4pt]
0 \qquad &\text{otherwise}
\end{cases}\ .
\end{align}
Using the reflection symmetry \eqref{xid1} below, one finds
\begin{align}
P_\gamma^{\alpha, \beta}(-1)=(-1)^\gamma P_\gamma^{\beta,\alpha}(1)=
\begin{cases} (-1)^\gamma\frac{(\beta+\gamma)!}{\beta! \gamma!}\qquad  &\text{if $\beta\geq0$} \\[4pt]
 \frac{ (-\beta-1)!}{(-\beta-\gamma-1)! \gamma!} \qquad &\text{if $\beta\leq-1$ and $\beta+\gamma+1\leq 0$} \label{valueat1}\\[4pt]
0 \qquad &\text{otherwise}
\end{cases}\ .
\end{align}
The Jacobi polynomials can also be written using the Rodrigues formula
\begin{equation}
P^{\alpha,\beta}_{\gamma}(x)= \frac{(-1)^\gamma}{2^\gamma \gamma!}(1-x)^{-\alpha}(1+x)^{-\beta} \left(\frac{d}{dx}\right)^\gamma\left((1-x)^{\alpha+\gamma}(1+x)^{\beta+\gamma}\right)\ .
\end{equation}

The Jacobi polynomials have the following recursion relations and symmetries, most of which can be proved relatively quickly using either the series expression or Rodrigues formula above
\begin{align}
&P^{\alpha,\beta}_{\gamma}(-x)=(-1)^\gamma P^{\beta,\alpha}_{\gamma}(x)\ , \label{xid1} \\[6pt]
&P^{\alpha,\beta-1}_{\gamma}(x)-P^{\alpha-1,\beta}_{\gamma}(x) = P^{\alpha,\beta}_{\gamma-1}(x) \ , \label{xid2} \\[6pt]
&\frac{(1-x)}{2}P^{\alpha+1,\beta}_{\gamma}(x) +\frac{(1+x)}{2}P^{\alpha,\beta+1}_{\gamma}(x) =P^{\alpha,\beta}_{\gamma}(x)\ , \label{xid3} \\[6pt]
&(2\gamma+\alpha+\beta+1)P^{\alpha,\beta}_{\gamma}(x) =(\gamma+\alpha+\beta+1)P^{\alpha,\beta+1}_{\gamma}(x) +(\gamma+\alpha)P^{\alpha,\beta+1}_{\gamma-1}(x)\ , \label{xid4}\\[9pt]
&(2\gamma+\alpha+\beta+1)P^{\alpha,\beta}_{\gamma}(x) =(\gamma+\alpha+\beta+1)P^{\alpha+1,\beta}_{\gamma}(x) -(\gamma+\beta)P^{\alpha+1,\beta}_{\gamma-1}(x)\ , \label{xid5}\\[6pt]
&(2\gamma+\alpha+\beta+2)\frac{(1+x)}{2}P^{\alpha,\beta+1}_{\gamma}(x)= (\gamma+1)P^{\alpha,\beta}_{\gamma+1}(x)+(\gamma+\beta+1)P^{\alpha,\beta}_{\gamma}(x)\ , \label{xid6}\\[6pt]
&(2\gamma+\alpha+\beta+2)\frac{(1-x)}{2}P^{\alpha+1,\beta}_{\gamma}(x)= -(\gamma+1)P^{\alpha,\beta}_{\gamma+1}(x)+(\gamma+\alpha+1)P^{\alpha,\beta}_{\gamma}(x)\ . \label{xid7}
\end{align}
The following identity connects Jacobi polynomials of different order, but equivalent indices:
\begin{align}
& P_{\gamma+1}^{\alpha,\beta}(x)=\left(A_{\gamma}^{\alpha,\beta} x + B_{\gamma}^{\alpha,\beta}\right)P_{\gamma}^{\alpha,\beta}(x) -C_{\gamma}^{\alpha,\beta}P_{\gamma-1}^{\alpha,\beta}(x)\ ,\\[8pt]
& A_{\gamma}^{\alpha,\beta} = \frac{(2\gamma+\alpha+\beta+1)(2\gamma+\alpha+\beta+2)}{2(\gamma+1)(\gamma+\alpha+\beta+1)}\ ,\quad  
B_{\gamma}^{\alpha,\beta} = \frac{(\alpha^2-\beta^2)(2\gamma+\alpha+\beta+1)} {2(\gamma+1)(\gamma+\alpha+\beta+1)(2\gamma+\alpha+\beta)}\ ,\nn \\[5pt]
& C_{\gamma}^{\alpha,\beta} = \frac{(\gamma+\alpha)(\gamma+\beta)(2\gamma+\alpha+\beta+2)} {(\gamma+1)(\gamma+\alpha+\beta+1)(2\gamma+\alpha+\beta)}\ .\nn
\end{align}
In addition, the Jacobi Polynomials have the following first derivative
\begin{equation}
2\pa_x P^{\alpha,\beta}_{\gamma}(x)=(\gamma+\alpha+\beta+1)P^{\alpha+1,\beta+1}_{\gamma-1}(x)\ ,
\end{equation}
and satisfy the differential equation
\be
\pa_{x}^2 P_{\gamma}^{\alpha,\beta}(x)+\left(\frac{\alpha+1}{x-1}+\frac{\beta+1}{x+1}\right)\pa_{x} P_{\gamma}^{\alpha,\beta}(x) +\frac{\gamma(\gamma+\alpha+\beta+1)}{2}\left(\frac{-1}{x-1}+\frac{1}{x+1}\right)P_{\gamma}^{\alpha,\beta}(x)=0.
\ee

For more complete treatment of Jacobi Polynomials, see \cite{NIST} (this entry in the bibliography includes a link to an online version which is frequently updated).

\section{Proofs for specific identities used}\label{APXproofs}

\subsection{Identities used to adapt maps to $t=1$ and $t=\infty$}
We consider the identity
\begin{align}
(-1)^{n_1} P^{-(N+1),-(n_1+n_3-N)}_{n_1}(1-2t)
&-(-1)^{n_1}t^{N+1}P^{N+1,-(n_1+n_3-N)}_{n_3-N-1}(1-2t)\label{tm1idAPX} \\[5pt]
&  =(-1)^{N-n_1}(1-t)^{n_1+n_3-N}P_{N-n_1}^{-(N+1),(n_1+n_3-N)}(1-2t) \nn
\end{align}
used in the main text \eqref{tm1id}.  Here and throughout we will assume that $n_1$ and $n_3$ are integers satisfying $0\leq n_1<n_3$, and we will be able to show that \eqref{tm1idAPX} is true for all integers $N$, using \eqref{rule}, i.e. $P_{-n}^{\alpha,\beta}\equiv0$ for $n\geq 1$.  The above is written as a power series in $t$, and we will often find it convenient to write it as a power series in $(t-1)$ which we can do using the identity $P^{\alpha,\beta}_{\gamma}(x)=(-1)^\gamma P^{\beta,\alpha}_{\gamma}(-x)$, giving an equivalent form of \eqref{tm1idAPX} as 
\begin{align}
P^{-(n_1+n_3-N),-(N+1)}_{n_1}(2t-1)
&+(-1)^{n_1+n_3-N}t^{N+1}P^{-(n_1+n_3-N),N+1}_{n_3-N-1}(2t-1)\label{tm1idAPXSWAP} \\[5pt]
& =(1-t)^{n_1+n_3-N}P_{N-n_1}^{(n_1+n_3-N),-(N+1)}(2t-1)\ .\nn
\end{align}
We explore the equivalent identities \eqref{tm1idAPX} and \eqref{tm1idAPXSWAP} in three cases:
\begin{enumerate}
\item $N\ge n_3$ such that $P^{-(n_1+n_3-N),N+1}_{n_3-N-1}$ is set to 0 in \eqref{tm1idAPXSWAP} (similarly in \eqref{tm1idAPX})\label{case1}
\begin{enumerate}
\item $N\ge n_1+n_3$\label{case1a}
\item $n_3\le N< n_1+n_3$\label{case1b}
\end{enumerate}
\item $N<n_1$ such that $P_{N-n_1}^{(n_1+n_3-N),-(N+1)}$ is set to 0 in \eqref{tm1idAPXSWAP} (similarly in \eqref{tm1idAPX}) \label{case2}
\begin{enumerate}
\item $0\le N+1\le n_1$\label{case2a}
\item $N+1< 0$\label{case2b}
\end{enumerate}
\item $n_1\le N<n_3$ such that all Jacobi polynomials are present in \eqref{tm1idAPX} and \eqref{tm1idAPXSWAP}\label{case3}
\end{enumerate}
The main cases above are motivated by the absence of certain Jacobi polynomials.  The sub-cases are motivated by the powers of $t$ and $(1-t)$ appearing in the identity \eqref{tm1idAPX} or \eqref{tm1idAPXSWAP}: when these powers become negative, the Jacobi polynomials multiplying them will be seen to truncate, making the expressions evaluate to polynomials.

\begin{figure}
\begin{center}
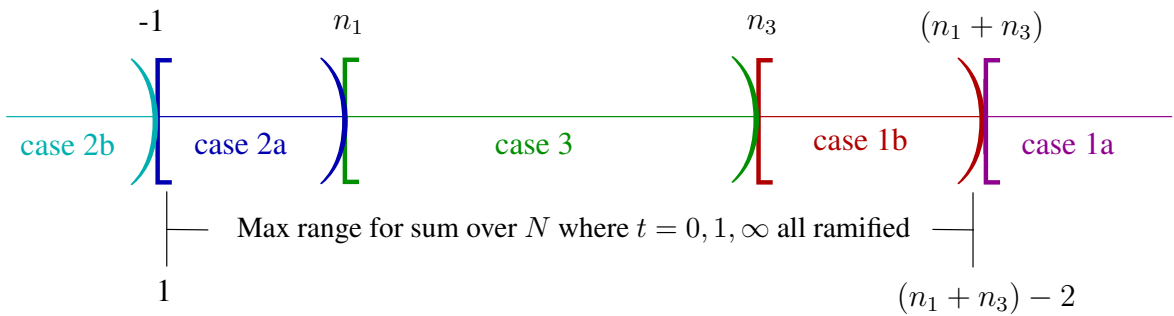
\end{center}
\caption{The ranges of $N$ for the different cases \ref{case1}-\ref{case3}.    
}
\end{figure}

We can prove the cases \ref{case1}-\ref{case3} using a set of standard manipulations which is worth pointing out.  First, if a Jacobi polynomial has a negative subscript it is omitted \eqref{rule}.  Next, in any given Jacobi polynomial the Pochhammer symbols only involve integer arguments, and some of these Pochhammer symbols are zero: a Pochhammer symbol $(-m)_n$ is 0 when $n\geq m+1$ (with $m,n$ integers and where $m$ non-negative).  This will truncate the sums defining certain Jacobi polynomials.  After doing so, we will often shift summation indices, which does not affect whether a given Pochhammer symbol is 0 or not.  After this, the remaining Pochhammer symbols are all non-zero with integer arguments, and we express such Pochhammer symbols in a ``normal form'' by writing them in terms of factorials, noting that
\begin{equation}
(m+1)_n= \frac{(m+n)!}{m!}\ , \qquad\qquad\qquad (-m)_n=(-1)^n(m-n+1)_n=(-1)^n\frac{m!}{(m-n)!}\ .
\end{equation}
Here we have assumed that $m$ and $n$ are non-negative in the first equation and that $0\leq n\leq m$ in the second equation.  In the case where there are only two Jacobi polynomials, these make identities \eqref{tm1idAPX} and \eqref{tm1idAPXSWAP} simple to check: the bounds of the sums and the individual terms become identically equal.  We show one such example (case \ref{case1b}) where all the above steps are executed.  The other cases where only two Jacobi polynomials are present, i.e. cases \ref{case1a}, \ref{case2a}, and \ref{case2b} follow the same steps, or with a simpler set of steps.  Case \ref{case3} is sufficiently different that we include its proof separately, although the above manipulations still prove to be useful.       

We now address case \ref{case1b} as our example where only two Jacobi polynomials are present.  In this case $N\geq n_3$, which eliminates the Jacobi polynomials with subscript $N-n_3-1$, making \eqref{tm1idAPXSWAP} become
\begin{align}
& P^{-(n_1+n_3-N),-(N+1)}_{n_1}(2t-1)=(1-t)^{n_1+n_3-N}P_{N-n_1}^{(n_1+n_3-N),-(N+1)}(2t-1)\ . \label{case1EQ}
\end{align}
Plugging in the expansion \eqref{JacobiSumForm} for the left hand side of \eqref{case1EQ}, we find
\begin{align}
P^{-(n_1+n_3-N),-(N+1)}_{n_1}(2t-1) & =\sum_{\ell=0}^{n_1} \frac{(-n_3)_\ell (-(n_1+n_3-N)+\ell+1)_{(n_1-\ell)}}{\ell!(n_1-\ell)!} (t-1)^\ell\ \label{tm1idLHSor} 
\end{align}
We note that the Pochhammer symbol $(N-(n_1+n_3)+\ell+1)_{n_1-\ell}$ is 0 unless $\ell$ is sufficiently large: $\ell$ must be large enough to make the smallest term in the product greater than 0.  This gives that $\ell\geq (n_1+n_3)-N$, and we note that $0\leq (n_1+n_3)-N \leq n_1$ for case \ref{case1b}.  This gives
\be\begin{split}
&P^{-(n_1+n_3-N),-(N+1)}_{n_1}(2t-1)  \\
&=\sum_{\ell=n_1+n_3-N}^{n_1} \frac{(-n_3)_\ell (-(n_1+n_3-N)+\ell+1)_{(n_1-\ell)}}{\ell!(n_1-\ell)!} (t-1)^\ell   \\
&=(t-1)^{(n_1+n_3-N)}\sum_{\ell=0}^{N-n_3} \frac{(-n_3)_{(\ell+n_1+n_3-N)} (\ell+1)_{(N-n_3-\ell)}}{(n_1+n_3-N+\ell)!(N-n_3-\ell)!} (t-1)^\ell   \\
&=(1-t)^{(n_1+n_3-N)}\sum_{\ell=0}^{N-n_3} \frac{(-1)^{\ell}n_3! (N-n_3)!}{(N-n_1-\ell)!\ell!(n_1+n_3-N+\ell)!(N-n_3-\ell)!} (t-1)^\ell \label{1bLHS}  
\end{split}\ee
where in the third line we shift the summation index and the last line write the Pochhammer symbols (which are all non-zero) in terms of factorials and absorb some factors of $-1$ into the prefactor.  Plugging in the expansion \eqref{JacobiSumForm} for the right hand side of \eqref{case1EQ}, we find
\be\begin{split}\label{1brhs}
& (1-t)^{n_1+n_3-N}P_{N-n_1}^{(n_1+n_3-N),-(N+1)}(2t-1)  \\[3pt]
& \qquad\qquad = (1-t)^{n_1+n_3-N}\sum_{\ell=0}^{N-n_1} \frac{(n_3-N)_\ell (n_1+n_3-N+\ell+1)_{(N-n_1-\ell)}}{\ell!(N-n_1-\ell)!} (t-1)^\ell\ . 
\end{split}\ee
We note that the Pochhammer symbol $(n_3-N)_\ell=0$ if $\ell\geq\ell_{\rm max}= N-n_3+1$, which truncates the sum to
\be\begin{split}
& (1-t)^{n_1+n_3-N}P_{N-n_1}^{(n_1+n_3-N),-(N+1)}(2t-1) \\
& \qquad = (1-t)^{n_1+n_3-N}\sum_{\ell=0}^{N-n_3} \frac{(n_3-N)_\ell (n_1+n_3-N+\ell+1)_{(N-n_1-\ell)}}{\ell!(N-n_1-\ell)!} (t-1)^\ell \\
& \qquad = (1-t)^{(n_1+n_3-N)}\sum_{\ell=0}^{N-n_3} \frac{(-1)^{\ell}n_3! (N-n_3)!}{(N-n_1-\ell)!\ell!(n_1+n_3-N+\ell)!(N-n_3-\ell)!} (t-1)^\ell \label{1brhs2}
\end{split}\ee
where in the last line we write these non-zero Pochhammer symbols in terms of factorials.  This matches \eqref{1bLHS} to \eqref{1brhs2}, concluding case \ref{case1b}.  Cases \ref{case1a}, \ref{case2a}, and \ref{case2b} follow the same steps.

We now address case \ref{case3} where $n_1\leq N < n_3$ and all Jacobi polynomials are present in \eqref{tm1idAPX}.  We start with the hypergeometric identity \cite[section 9.13]{Gradshteyn:1943cpj} multiplied by a constant $A$
\be\begin{split}
A\,F\left(\genfrac{}{}{0pt}{}{\alpha,\beta}{\gamma};t\right)&=A\,\frac{\Gamma(\gamma)\Gamma(\gamma-\alpha-\beta)}{\Gamma(\gamma-\alpha)\Gamma(\gamma-\beta)}\,F\left(\genfrac{}{}{0pt}{}{\alpha,\beta}{\alpha+\beta-\gamma+1};1-t
\right) \\[8pt]
& +(1-t)^{\gamma-\alpha-\beta}\frac{\Gamma(\gamma)\Gamma(\alpha+\beta-\gamma)}{\Gamma(\alpha)\Gamma(\beta)}\,A\,F\left(\genfrac{}{}{0pt}{}{\gamma-\alpha,\gamma-\beta}{\gamma-\alpha-\beta+1};1-t\right)\ .\label{hypID1}
\end{split}\ee
We take as before
\be\begin{split}
&\alpha=-n_1+B\epsilon\ , \qquad\qquad\qquad\qquad\qquad\qquad\;\;\,
\gamma=-N + A\epsilon\ , \qquad\qquad \beta=-n_3+r\epsilon\ ,\\
&\gamma-\alpha= -N+n_1+(A-B)\epsilon\ ,\qquad\qquad\qquad \gamma-\beta=-N+n_3+(A-r)\epsilon\ ,\\
&\gamma-\alpha-\beta=-N+n_1+n_3 +(A-B-r)\epsilon\ ,
\end{split}\ee
where $B$ and $r$ are also constants. We note that with the ordering $n_1\leq N<n_3$, only certain combinations of $\alpha, \beta, \gamma$ are close to negative integers.  Further, the hypergeometric function on the left hand side of \eqref{hypID1} has two windows of $\ell$ that survive the limit, as does the first hypergeometric function on the right hand side of \eqref{hypID1}.  However, the second geometric function on the right hand side of \eqref{hypID1} has only one surviving window for $\ell$.  Taking the $\epsilon\rightarrow 0$ limit of the identity we find
\begin{align}\label{hypID2}
&A\,\sum_{\ell=0}^{n_1} \frac{(-n_1)_\ell (-n_3)_\ell}{(-N)_\ell\ell!}\, t^\ell+ B\! \sum_{\ell=N+1}^{n_3} \frac{(-1)^{n_1-N} n_1!(\ell-n_1-1)!(-n_3)_\ell}{N! (\ell-N-1)! \ell!}\,t^\ell  \\
&=\frac{(-1)^{n_1} (n_3-N)_{n_1}}{(N-n_1+1)_{n_1}}\, (A-B)\sum_{\ell=0}^{n_1} \frac{(-n_1)_\ell (-n_3)_\ell}{(-n_1-n_3+N+1)_\ell \ell!}\, (1-t)^\ell  \nn \\
&\; + \frac{(-1)^{n_1} (n_3-N)_{n_1}}{(N-n_1+1)_{n_1}}\,\frac{(A-B)B}{(A-B-r)} \sum_{\ell=n_1+n_3-N}^{n_3}\frac{(-1)^{n_3-N}n_1!(\ell-n_1-1)!(-n_3)_\ell}{(n_1+n_3-N-1)!(\ell-n_1-n_3+N)!\ell!}\, (1-t)^\ell \nn \\
&\; - \frac{n_1! n_3!}{N!(n_1+n_3-N)!}\,\frac{Br}{(A-B-r)}\,(1-t)^{n_1+n_3-N} \sum_{\ell=0}^{N-n_1} \frac{(-N+n_1)_\ell(-N+n_3)_\ell}{(-N+n_3+n_1+1)_\ell \ell!}\,(1-t)^\ell\ .\nn
\end{align}
Interestingly, the last two sums on the right hand side of the above expression are the same function with different coefficients.  This can be shown by examining the second to last line of the above equation
\begin{align}
&\frac{(-1)^{n_1} (n_3-N)_{n_1}}{(N-n_1+1)_{n_1}}\!\! \sum_{\ell=n_1+n_3-N}^{n_3}\! \frac{(-1)^{n_3-N}n_1!(\ell-n_1-1)!(-n_3)_\ell}{(n_1+n_3-N-1)!(\ell-n_1-n_3+N)!\ell!}\, (1-t)^\ell \label{idprove1} \\
&=\frac{(-1)^{n_1}n_1! (n_3-N)_{n_1}}{(N-n_1+1)_{n_1}}\, \sum_{\ell=0}^{N-n_1} \frac{(-1)^{n_3-N}(\ell+n_3-N-1)!(-n_3)_{(\ell+n_1+n_3-N)}}{(n_1+n_3-N-1)!\ell!(\ell+n_1+n_3-N)!}\, (1-t)^{\ell+n_1+n_3-N} \nn \\
&=\frac{n_1!(n_1+n_3-N-1)! (N-n_1)!}{(n_3-N-1)! N!}\,(1-t)^{n_1+n_3-N} \nn \\
& \qquad \qquad \times\sum_{\ell=0}^{N-n_1} \frac{(-1)^\ell (\ell+n_3-N-1)!n_3!}{(N-n_1-\ell)!(n_1+n_3-N-1)!\ell!(\ell+n_1+n_3-N)!}\,(1-t)^\ell \nn \\
&=\frac{n_1!(N-n_1)!}{(n_3-N-1)! N!}\,(1-t)^{n_1+n_3-N}\sum_{\ell=0}^{N-n_1} \frac{(-1)^\ell (\ell+n_3-N-1)!n_3!}{(N-n_1-\ell)!\ell!(\ell+n_1+n_3-N)!}\,(1-t)^\ell\ .\nn
\end{align}
Above we have shifted the sum, and used identities to write all Pochhammer symbols in terms of factorials.  Doing a similar replacement in the last line of \eqref{hypID2} we find
\begin{align}
& \frac{n_1! n_3!}{N!(n_1+n_3-N)!} (1-t)^{n_1+n_3-N}\sum_{\ell=0}^{N-n_1} \frac{(-N+n_1)_\ell(-N+n_3)_\ell}{(-N+n_3+n_1+1)_\ell \ell!}(1-t)^\ell \label{idprove2}\\
& = \frac{n_1! n_3!}{N!(n_1+n_3-N)!}(1-t)^{n_1+n_3-N}\!\!
\sum_{\ell=0}^{N-n_1} \frac{(-1)^\ell(N-n_1)! (n_3-N+\ell-1)! (n_1+n_3-N)!}{(N-n_1-\ell)!(n_3-N-1)!(n_1+n_3-N+\ell)! \ell!}(1-t)^\ell \nn \\
& = \frac{n_1! n_3!}{N!}(1-t)^{n_1+n_3-N}\sum_{\ell=0}^{N-n_1} \frac{(-1)^\ell(N-n_1)! (n_3-N+\ell-1)! }{(N-n_1-\ell)!(n_3-N-1)!(n_1+n_3-N+\ell)! \ell!}(1-t)^\ell\ .\nn
\end{align}
The coefficients of $(1-t)^\ell$ in the last lines of \eqref{idprove1} and \eqref{idprove2} are seen to be equivalent.  This allows us to rewrite the identity \eqref{hypID2} as
\begin{align}
&A\sum_{\ell=0}^{n_1} \frac{(-n_1)_\ell (-n_3)_\ell}{(-N)_\ell\ell!}\, t^\ell+ B\!\! \sum_{\ell=N+1}^{n_3} \frac{(-1)^{n_1-N} n_1!(\ell-n_1-1)!(-n_3)_\ell}{N! (\ell-N-1)! \ell!}\,t^\ell\label{hypID2_ii}  \\
&=\frac{(-1)^{n_1} (n_3-N)_{n_1}}{(N-n_1+1)_{n_1}}\, (A-B)\sum_{\ell=0}^{n_1} \frac{(-n_1)_\ell (-n_3)_\ell}{(-n_1-n_3+N+1)_\ell \ell!}\, (1-t)^\ell  \nn \\
&+ \frac{n_1! n_3!}{N!(n_1+n_3-N)!}\,B\, (1-t)^{n_1+n_3-N} \sum_{\ell=0}^{N-n_1} \frac{(-N+n_1)_\ell(-N+n_3)_\ell}{(-N+n_3+n_1+1)_\ell \ell!}\,(1-t)^\ell\nn
\end{align} 
where we see that the regulator $r$ plays no role on the right hand side, which we should have anticipated because it plays no role on the left hand side. 

All the sums in the above equation are related to Jacobi polynomials:
\begin{align}
& \sum_{\ell=0}^{n_1} \frac{(-n_1)_\ell (-n_3)_\ell}{(-N)_\ell\ell!}\,t^\ell=\frac{(-1)^{n_1} n_1!(N-n_1)!}{N!}\, P_{n_1}^{-(N+1),-(n_1+n_3-N)}(1-2t)\ ,\qquad\qquad\qquad\label{jpoly1}
\end{align}
\begin{align}
& \sum_{\ell=N+1}^{n_3} \frac{(-1)^{n_1-N} n_1!(\ell-n_1-1)!(-n_3)_\ell}{N! (\ell-N-1)! \ell!}\,t^\ell = -\frac{(-1)^{n_1}n_1!(N-n_1)!}{N!}\, P_{n_3-N-1}^{(N+1),-(n_1+n_3-N)}(1-2t)\ ,\label{jpoly2} 
\end{align}
\be\begin{split}
& \frac{(-1)^{n_1} (n_3-N)_{n_1}}{(N-n_1+1)_{n_1}} \sum_{\ell=0}^{n_1} \frac{(-n_1)_\ell (-n_3)_\ell}{(-n_1-n_3+N+1)_\ell \ell!}\, (1-t)^\ell  \\
&\kern15em = \frac{n_1!(N-n_1)!}{N! }\, P_{n_1}^{-(n_1+n_3-N),-(N+1)}(2t-1)\ ,\qquad\;\;\label{jpoly3}
\end{split}\ee
\be\begin{split}
& \frac{n_1! n_3!}{N!(n_1+n_3-N)!}\sum_{\ell=0}^{N-n_1} \frac{(-N+n_1)_\ell(-N+n_3)_\ell}{(-N+n_3+n_1+1)_\ell \ell!}(1-t)^\ell \\
&\kern15em = \frac{n_1! (N-n_1)!}{N!}P_{N-n_1}^{n_1+n_3-N,-(N+1)}(2t-1)\ .\qquad\quad\label{jpoly4}
\end{split}\ee
These statements can be proven similarly to the last cases: truncating sums, and then replacing Pochhamer symbols by factorials.

Plugging in \eqref{jpoly1}-\eqref{jpoly4} into the identity \eqref{hypID2_ii} and canceling the common factor of ${n_1!(N-n_1)!}/{N!}$ gives
\be\begin{split}
&A(-1)^{n_1}P_{n_1}^{-(N+1),-(n_1+n_3-N)}(1-2t)-
B(-1)^{n_1}t^{N+1}P_{n_3-N-1}^{(N+1),-(n_1+n_3-N)}(1-2t)\\[5pt]
=\;&(A-B)P_{n_1}^{-(n_1+n_3-N),-(N+1)}(2t-1)+B (1-t)^{n_1+n_3-N} P_{N-n_1}^{n_1+n_3-N,-(N+1)}(2t-1).
\end{split}\ee 
We see that the above represents two identities, given that $A$ and $B$ are independent. We see that the $B=0$ case is the familiar identity $P_{\gamma}^{\alpha,\beta}(x)=(-1)^\gamma P^{\beta,\alpha}_{\gamma}(-x)$ identity.  The case of interest for us is $A=B=1$ and gives
\be\begin{split}
&(-1)^{n_1}P_{n_1}^{-(N+1),-(n_1+n_3-N)}(1-2t)- (-1)^{n_1}t^{N+1}P_{n_3-N-1}^{(N+1),-(n_1+n_3-N)}(1-2t)\\[5pt]
=\; &(1-t)^{n_1+n_3-N} P_{N-n_1}^{n_1+n_3-N,-(N+1)}(2t-1)=(-1)^{(N-n_1)}P_{N-n_1}^{-(N+1),n_1+n_3-N}(1-2t)
\end{split}\ee
concluding case \ref{case3}.

Effectively, \eqref{tm1idAPX} works by swapping the long twist operator at $(t=0,z=0)$ with the long twist operator at $(t=1,z=1)$.  We wish to also consider swapping the operators at $(t=0,z=0)$ and $(t=\infty, z=\infty)$.  This is accomplished by showing that
\be \label{idInftyAPX}
P^{\alpha,\beta}_{\gamma}(1-2t)=t^\gamma P^{-(2\gamma+\alpha+\beta+1),\beta}_\gamma \left(1-\frac{2}{t}\right)
\ee 
which appears as \eqref{idInfty} in the main text.  We prove this by examining the right hand side:
\begin{align}
& t^\gamma P^{-(2\gamma+\alpha+\beta+1),\beta}_\gamma \left(1-\frac{2}{t}\right) \\
& =(-1)^\gamma(-t)^\gamma \sum_{\ell=0}^\gamma \frac{(\gamma-(2\gamma+\alpha+\beta+1)+\beta+1)_\ell (-(2\gamma+\alpha+\beta+1)+\ell+1)_{(\gamma-\ell)}}{\ell!(\gamma-\ell)!} \left(-\frac{1}{t}\right)^\ell \nn \\
& =(-1)^\gamma \sum_{\ell=0}^\gamma \frac{\big(-(\gamma+\alpha)\big)_\ell \big(-(2\gamma+\alpha+\beta+1\big)+\ell+1)_{(\gamma-\ell)}}{\ell!(\gamma-\ell)!}\,(-t)^{\gamma-\ell} \nn \\
& =(-1)^\gamma \sum_{\ell=0}^\gamma \frac{\big(-(\gamma+\alpha)\big)_{(\gamma-\ell)} \big(-(\gamma+\alpha+\beta)-\ell\big)_{\ell}}{\ell!(\gamma-\ell)!}\,(-t)^{\ell}\nn
\end{align}
where in the last line we have changed the order of the sum, swapping $\ell\leftrightarrow \gamma-\ell$.  We end by reversing the order of the Pochhammer symbols, i.e. using $(a)_m=(-1)^m(1-m-a)_m$, and so
\begin{align}
& t^\gamma P^{-(2\gamma+\alpha+\beta+1),\beta}_\gamma \left(1-\frac{2}{t}\right) = \sum_{\ell=0}^\gamma \frac{(\alpha+1+\ell)_{(\gamma-\ell)} (\gamma+\alpha+\beta+1)_{\ell}}{\ell!(\gamma-\ell)!}(-t)^{\ell}=P^{\alpha,\beta}_{\gamma}(1-2t)
\end{align}
proving the assertion.

\subsection{Identities used for OPE limits}\label{app_OPEid}

We start with by commenting on the identities used in table \ref{TableOPE} to show that the OPE limit gives a map of the same form.  In this table, one should use the form of the map \eqref{near0}, \eqref{near1}, or \eqref{nearinfty}, depending on the approach point.  To verify that the map is of the same form, one may plug in $b_{N}(\{B_M\})$ in the table into the appropriate form of the map.  By shifting the summation indices in the numerator and denominator, the sums can be written as sums where the coefficients are $B_N$, rather than $B_{N-1}$ or other shifts.  Once this is accomplished, a combination of Jacobi polynomials appears as the coefficient of $B_N$, and this combination of Jacobi polynomials can be addressed using the stated identities to simplify them to one Jacobi polynomial.  If table \ref{TableOPE} has only one identity listed, then this identity is used both in the numerator and denominator; if table \ref{TableOPE} has two identities listed, then the upper identity is used in the numerator and the lower identity is used in the denominator.  This final form of the map matches the general form \eqref{near0}, \eqref{near1}, or \eqref{nearinfty}, appropriate to the point, with $b_N\rightarrow B_N$ and shifts to $n_1$, $n_3$, $\Nmin$, and $\Nmax$ labeled as the ``equivalent shift'' column in the table. These shifts accomplish the change to the ramifications at the points $(t=0,z=0)$, $(t=1,z=1)$ and $(t=\infty, z=\infty)$ appropriate for the OPE. We now turn to proving the identities in table \ref{TableOPE}  by using the basic identities in appendix \ref{JacobiIDappx}. 

We start by proving the identities used when the approach point is $t=0$. First, we show
\be
tP^{\alpha,\beta}_{\gamma}=P^{\alpha-1,\beta-1}_{\gamma+1}-P^{\alpha-1,\beta}_{\gamma+1}\ . \label{tPid1.0}
\ee
This identity is easy to establish, using \eqref{JacobiSumForm}, which we use on the right hand side, finding
\begin{align}
P^{\alpha-1,\beta-1}_{\gamma+1}-P^{\alpha-1,\beta}_{\gamma+1} &= \sum_{\ell=0}^{\gamma+1}\left( \frac{(\gamma+\alpha+\beta)_\ell(\alpha+\ell)_{\gamma+1-\ell}}{\ell! (\gamma+1-\ell)!} - \frac{(\gamma+\alpha+\beta+1)_\ell(\alpha+\ell)_{\gamma+1-\ell}}{\ell! (\gamma+1-\ell)!}\right) (-t)^\ell\ .
\end{align}  
We note that the $\ell=0$ term vanishes, and so we may start the sum at $\ell=1$.  Shifting the sum indices down, we find
\begin{align}
& P^{\alpha-1,\beta-1}_{\gamma+1}-P^{\alpha-1,\beta}_{\gamma+1} \\
&= \sum_{\ell=0}^{\gamma}\left( \frac{(\gamma+\alpha+\beta)_{(\ell+1)}(\alpha+\ell+1)_{\gamma-\ell}}{(\ell+1)! (\gamma-\ell)!} - \frac{(\gamma+\alpha+\beta+1)_{(\ell+1)}(\alpha+\ell+1)_{\gamma-\ell}}{(\ell+1)! (\gamma-\ell)!}\right) (-t)^{\ell+1} \nn\\
&= \sum_{\ell=0}^{\gamma}\frac{(\gamma+\alpha+\beta+1)_{(\ell)}(\alpha+\ell+1)_{\gamma-\ell}}{(\ell)! (\gamma-\ell)!}\left(\frac{(\gamma+\alpha+\beta)}{\ell+1} - \frac{(\gamma+\alpha+\beta+\ell+1)}{\ell+1}\right)(-t)^{\ell+1} \nn \\
&= \sum_{\ell=0}^{\gamma}\frac{(\gamma+\alpha+\beta+1)_{(\ell)}(\alpha+\ell+1)_{\gamma-\ell}}{(\ell)! (\gamma-\ell)!}(-1)(-t)^{\ell+1}= t\,P_{\gamma}^{\alpha,\beta}(1-2t)\nn
\end{align}
establishing the identity \eqref{tPid1.0}.  Shifting $\gamma$ we find
\be
tP^{\alpha,\beta}_{\gamma-1}(1-2t)=P^{\alpha-1,\beta-1}_{\gamma}(1-2t)-P^{\alpha-1,\beta}_{\gamma}(1-2t)\ . \label{tPid1}
\ee
We similarly use identity \eqref{xid7}, but written using the $t$ variable i.e. $x=1-2t$ and shifting $\gamma$ down by one and shifting $\alpha$ down by one. This gives
\be\begin{split}
& (2\gamma+\alpha+\beta-1)\,t\,P^{\alpha,\beta}_{\gamma-1}(1-2t)=-\gamma P^{\alpha-1,\beta}_\gamma+(\gamma+\alpha-1)P^{\alpha-1,\beta}_{\gamma-1}(1-2t) \\[5pt]
&\quad\, = -\gamma P^{\alpha-1,\beta}_\gamma(1-2t)+(\gamma+\alpha-1)\big(P^{\alpha-1,\beta-1}_\gamma(1-2t)-P^{\alpha-2,\beta}_\gamma(1-2t)\big)\label{tPid2}
\end{split}\ee
where in the second line we have used identity \eqref{xid2}.  Now, we take $\gamma$ times \eqref{tPid1} and subtract it from \eqref{tPid2}, finding
\be
(\gamma+\alpha+\beta-1)\,t\,P^{\alpha,\beta}_{\gamma-1}(1-2t) = (\alpha-1)P^{\alpha-1,\beta-1}_{\gamma}(1-2t)-(\gamma+\alpha-1)P^{\alpha-2,\beta}_\gamma(1-2t) \label{PidSumIndexShift}
\ee
which we use in showing \eqref{PidSumIndexShift.Main}.  It should be noted that the above identity is still valid for $\gamma=0$, recognizing that $P^{\alpha,\beta}_1=1$ and defining $P^{\alpha,\beta}_{-1}=0$.  It continues to be trivially true for $\gamma\leq -1$ as well, given rule \eqref{rule}.  This identity is the only one used in the approaches to $t=0$ and so concludes these cases.     

Next, we consider the approaches to the point $t=1$. We redefine $t\rightarrow (1-t)$ in \eqref{PidSumIndexShift} to arrive at the related identity
\be
(\gamma+\alpha+\beta-1)(1-t)P^{\alpha,\beta}_{\gamma-1}(2t-1) = (\alpha-1)P^{\alpha-1,\beta-1}_{\gamma}(2t-1) -(\gamma+\alpha-1)P^{\alpha-2,\beta}_\gamma(2t-1) \label{PidSumIndexShift.Switch}
\ee
which is the only identity needed for the OPE limit when the approach point is $t=1$, concluding this case.        

We may substitute $t\rightarrow 1/t$  into \eqref{tPid1.0} and get
\be
\frac{1}{t}P^{\alpha,\beta}_{\gamma-1}\left(1-\frac2t\right)=P^{\alpha-1,\beta-1}_{\gamma}\left(1-\frac2t\right)-P^{\alpha-1,\beta}_\gamma\left(1-\frac2t\right) \label{tPid1.inf}
\ee
which is used for the twist up OPE at $t=\infty$ for the numerator sum.  The basic identity
\be
P^{\alpha,\beta-1}_{\gamma}(x)-P^{\alpha-1,\beta}_{\gamma}(x) = P^{\alpha,\beta}_{\gamma-1}(x) \label{idforinfty1}
\ee
is used for the twist up OPE near $t=\infty$ for the denominator sum. We take \eqref{tPid1} with $t\rightarrow 1/t$ to give 
\be
(\gamma+\alpha+\beta-1)\frac{1}{t}P^{\alpha,\beta}_{\gamma-1}\left(1-\frac2t\right)  = (\alpha-1)P^{\alpha-1,\beta-1}_{\gamma}\left(1-\frac2t\right)-(\gamma+\alpha-1)P^{\alpha-2,\beta}_\gamma\left(1-\frac2t\right) \label{PidSumIndexShift.Inv}
\ee
which is used for the twist down OPE near $t=\infty$ for the numerator sum.  We next start with \eqref{xid4} and \eqref{xid2} with some indices shifted as
\be
P^{\alpha,\beta}_{\gamma}(x)=P^{\alpha-1,\beta+1}_{\gamma}(x) + P^{\alpha,\beta+1}_{\gamma-1}(x)\ .
\ee
We take $\gamma+\alpha$ times the above equation, subtract it from \eqref{xid4}, and evaluate at $x=1-2/t$.  This gives
\be
(\gamma+\beta+1)P_\gamma^{\alpha,\beta}\left(1-\frac2t\right)=(\gamma+\alpha+\beta+1)P_\gamma^{\alpha,\beta+1}\left(1-\frac2t\right)-(\gamma+\alpha)P_{\gamma-1}^{\alpha-1,\beta+1}\left(1-\frac2t\right) \label{idforinfty3} 
\ee
which is the identity needed for the twist down OPE near $t=\infty$ in the numerator sum.  We next begin with \eqref{xid7} with some shifts to the indices.
\begin{align}\label{xid7_ii}
(2\gamma+\alpha+\beta+1)\,\frac{(1-x)}{2}\,P^{\alpha+1,\beta+1}_{\gamma-1}(x)= -\gamma P^{\alpha,\beta+1}_{\gamma}(x)+(\gamma+\alpha)P^{\alpha,\beta+1}_{\gamma-1}(x)
\end{align}
and compare with \eqref{xid4}. We eliminate the first Jacobi polynomial on the right hand side by multiplying \eqref{xid7_ii} by $(\gamma+\alpha+\beta+1)$, multiplying \eqref{xid4} by $\gamma$, and then adding.  Doing so, a common factor of $(2\gamma+\alpha+\beta+1)$ cancels and we find
\begin{align}
(\gamma+\alpha+\beta+1)\,\frac{(1-x)}{2}\,P_{\gamma-1}^{\alpha+1,\beta+1}(x)=-\gamma P_{\gamma}^{\alpha,\beta}(x)+(\gamma+\alpha)P_{\gamma-1}^{\alpha,\beta+1}(x)\ .
\end{align}
Specializing to $x=1-2/t$ we obtain
\begin{align}
(\gamma+\alpha+\beta+1)\,\frac{1}{t}\,P_{\gamma-1}^{\alpha+1,\beta+1}\left(1-\frac2t\right)=-\gamma P_{\gamma}^{\alpha,\beta}\left(1-\frac2t\right)+(\gamma+\alpha)P_{\gamma-1}^{\alpha,\beta+1}\left(1-\frac2t\right)  \label{idforinfty2}
\end{align}
which is the identity needed for the twist down OPE limit near $t=\infty$ for the denominator polynomial.

\section{OPE limit examples}\label{Appx.OPE.Examples}

\subsection{OPE limits for approaches to \texorpdfstring{$t=0$}{TEXT}}

Here we will construct the OPE limits explicitly for the case where one of the ramified points in the cloud approaches the point at $t=0$.  We begin by considering the ``near 0" maps \eqref{near0}
\be
z(t)=\frac{ \sum\limits_{N=N_{\rm min}}^{N_{\rm max}} b_N t^{(N+1)}P^{(N+1),-(n_1+n_3-N)}_{n_3-N-1}(1-2t)}{\sum\limits_{N=N_{\rm min}}^{N_{\rm max}} b_N P^{-(N+1),-(n_1+n_3-N)}_{n_1} (1-2t) }\ .\label{near0copy}\vs{-5}
\ee
Factoring out a $t^{\Nmin+1}$ from the sum, we see that the remaining polynomial has a constant term coming only from $N=\Nmin$.  Therefore, the situation where the ramification at $t=0$ increases is given by
\be
b_{\Nmin}=0\ . \label{t0twistupOPE}
\ee  
This is a homogeneous polynomial in the $b_N$, making it a well defined algebraic variety subspace of the ${\mathbb{CP}}^{\Delta N}$ defined by the $b_N$.  Enforcing this changes the map to
\begin{align}
z(t)=\frac{ \sum\limits_{N=N_{\rm min}+1}^{N_{\rm max}} b_N t^{(N+1)}P^{(N+1),-(n_1+n_3-N)}_{n_3-N-1}(1-2t)}{\sum\limits_{N=N_{\rm min}+1}^{N_{\rm max}} b_N P^{-(N+1),-(n_1+n_3-N)}_{n_1} (1-2t) }\ .\label{near0copynew}
\end{align}
Thus, $\Delta N$ has decreased by one, making the cloud smaller by one twist-2 operator, but increasing the ramification at $t=0$, i.e. $r_0=\Nmin+1$.  This, therefore, represents an OPE limit where one of the twist-2 operators in the cloud approaches the operator at $(t=0,z=0)$ and increases the length of the cycle by one.  This is a ``twist up'' part of the operator product expansion, and corresponds to the ramification preserving product between simple cycles -- see appendix \ref{appx.radd}.  There is no further processing to be done: \eqref{near0copynew} is already in the correct form. To generate \eqref{near0copynew} one takes \eqref{near0copy} and replaces $\Nmin\rightarrow \Nmin+1$.  

We note that if $\Nmin=n_3-1$, then the numerator of \eqref{near0copynew} would become identically zero.  This means that one may not ``twist up'' the operator at the origin.  If one does this, the ramification subadditivity constraints become impossible to satisfy, i.e. there is no group product with those ramifications that can multiply to the identity.  This does not mean that the operators on the base space cannot approach each other: it simply means that the exchange channels produced in the OPE limit only contain the ``twist down'' or ``ramification lowering'' products, which we now discuss.  

So, how do we identify the ``twist down'' OPE?  This can be accomplished by requiring that the constant term in the denominator of \eqref{near0copy} is 0.  This allows one to factor a $t$ and cancel one power in the numerator, lowering the ramification at the origin by 1.  This is easy to write as a condition on the $b_N$ by setting the denominator of \eqref{near0copy} evaluated at $t=0$ to 0, i.e.
\be
\sum_{N=N_{\rm min}}^{N_{\rm max}} b_N P^{-(N+1),-(n_1+n_3-N)}_{n_1} (1)
= \sum_{N=N_{\rm min}}^{N_{\rm max}} b_N \frac{(-N)_{n_1}}{n_1!}=\frac{(-1)^{n_1}\Nmin!}{(n_1)!(\Nmax-n_1)!}\, g_{(0,\downarrow)}=0 \label{t0twistdownOPE}
\ee     
which is also a homogeneous polynomial in the $b_N$ and so defines an algebraic variety subspace of ${\mathbb{CP}}^{\Delta N}$.  We note that many of the above Pochhammer symbols may be 0, given that $-N$ is negative, while $-N+n_1-1$ may be non-negative: these correspond exactly to the cases where the Jacobi polynomials in the numerator of \eqref{near1} are 0 by rule \eqref{rule}.  However, not all such terms are zero given that some $N$ are in the range $n_1\leq N<n_3$, and so the above does represent a constraint.  

We do have a solution to the constraint \eqref{t0twistdownOPE}, written in table \ref{TableOPE}.  However, it might be concerning exactly how one arrives at this solution as being the correct one that gives a new map in the same form.  We may generate the solution in table \ref{TableOPE} by knowing the answer.  This situation we are examining is where the twist at the origin has had the ramification decreased by 1, i.e. $\Nmin\rightarrow \Nmin-1$.  We see that this is possible if one of the twists in the cloud had approached the operator at $(t=0,z=0)$ and a ``ramification decreasing'' product has been taken, lowering $\Delta N$ by 1 as well, and so $\Nmax\rightarrow \Nmax-2$.  However, we require that the ramifications at $t=1$ and $t=\infty$ remain unchanged.  This can be accomplished by $n_1\rightarrow n_1-1$ and $n_3\rightarrow n_3-1$.  Therefore, we expect the condition \eqref{t0twistdownOPE} to transform the map to the form
\be
z(t)=\frac{ \sum\limits_{N=N_{\rm min}-1}^{N_{\rm max}-2} B_N t^{(N+1)}P^{(N+1),-(n_1+n_3-2-N)}_{n_3-1-N-1}(1-2t)}{\sum\limits_{N=N_{\rm min}-1}^{N_{\rm max}-2} B_N P^{-(N+1),-(n_1+n_3-2-N)}_{n_1-1} (1-2t) } \label{near0copyMod}
\ee
for some set of constants $B_N$, which we now set about identifying.  Note that there is one fewer non-zero $B_N$ than there are $b_N$, and this should span the same space as the $b_N$ with the constraint \eqref{t0twistdownOPE}.  For the ease of notation we define
\be\label{endB0}
B_N \equiv 0 \qquad \mbox{for} \qquad N\leq \Nmin-2\qquad{\rm and}\qquad N\geq \Nmax-1\ .
\ee
This will allow us to extend the sums in what follows.  

We first consider the denominator of \eqref{near0copyMod} and substitute in the identity
\begin{align}
(\gamma+\alpha+\beta-1)tP^{\alpha,\beta}_{\gamma-1}(1-2t) & = (\alpha-1)P^{\alpha-1,\beta-1}_{\gamma}(1-2t) -(\gamma+\alpha-1)P^{\alpha-2,\beta}_\gamma(1-2t) \label{PidSumIndexShift.Main}
\end{align}
which is proved in appendix \ref{APXproofs} -- see \eqref{PidSumIndexShift}. It may not be obvious why identity \eqref{PidSumIndexShift.Main} is the correct one to use.  This identity is found by seeking an identity where shifts in indices $\alpha$, $\beta$, $\gamma$ can be compensated for by shifts in the summation index $N$ in \eqref{near0copyMod}, making them the Jacobi polynomial on the right hand side the same after shifting summation indices. This gives us an idea of how to manipulate basic Jacobi polynomial identities to find one that is useful.

The identity \eqref{PidSumIndexShift.Main} is true for all integer $\gamma$ using rule \eqref{rule}.  Inserting \eqref{PidSumIndexShift.Main} in the denominator of \eqref{near0copyMod} gives\vs{-8}
\be\begin{split}
& \sum_{N=N_{\rm min}-1}^{N_{\rm max}-2} B_N P^{-(N+1),-(n_1+n_3-2-N)}_{n_1-1} (1-2t) \\
&= \frac{1}{n_3\,t} \sum_{N=N_{\rm min}-1}^{N_{\rm max}-2} B_N (N+2) P^{-(N+2),-(n_1+n_3-1-N)}_{n_1} (1-2t) \\
& -\frac{1}{n_3\,t} \sum_{N=N_{\rm min}-1}^{N_{\rm max}-2} B_N (N-n_1+2)  P^{-(N+3),-(n_1+n_3-2-N)}_{n_1} (1-2t)\ .
\end{split}\ee
We make the functions look identical by shifting summation indices separately:
\begin{align}
& \sum_{N=N_{\rm min}-1}^{N_{\rm max}-2} B_N P^{-(N+1),-(n_1+n_3-2-N)}_{n_1-1} (1-2t)  \nn \\
&= \frac{1}{n_3\,  t} \sum_{N=N_{\rm min}}^{N_{\rm max}-1} B_{N-1} (N+1) P^{-(N+1),-(n_1+n_3-N)}_{n_1} (1-2t) \nn\\
& -\frac{1}{n_3\, t} \sum_{N=N_{\rm min}+1}^{N_{\rm max}} B_{N-2} (N-n_1)  P^{-(N+1),-(n_1+n_3-N)}_{n_1} (1-2t)\ . 
\end{align}
Now, given our definitions \eqref{endB0}, we may extend both of the sums, writing
\begin{align}
& \sum_{N=N_{\rm min}-1}^{N_{\rm max}-2} B_N P^{-(N+1),-(n_1+n_3-2-N)}_{n_1-1} (1-2t)  \nn \\
&= \frac{1}{n_3\, t} \sum_{N=N_{\rm min}}^{N_{\rm max}} \big((N+1) B_{N-1}  - (N-n_1) B_{N-2} \big)P^{-(N+1),-(n_1+n_3-N)}_{n_1} (1-2t)    
\end{align}
which is of the form of the denominator of \eqref{near0}, dressed with an additional factor.  This allows us to identify
\be
b_N=\frac{1}{n_3}\big((N+1) B_{N-1}  - (N-n_1) B_{N-2}\big) \label{bid0}
\ee
for $\Nmin\leq N\leq \Nmax$.  We have written the $\Delta N+1$ constants $b_N$ as linear functions of the $\Delta N$ constants $B_N$, and so there must be a liner relationship between them, which is precisely the relationship \eqref{t0twistdownOPE}.  This is easy to show by plugging in the above constraint.  We find
\begin{align}
&\sum_{N=N_{\rm min}}^{N_{\rm max}} b_N \frac{(-N)_{n_1}}{n_1!} =\frac{1}{n_3} \sum_{N=N_{\rm min}}^{N_{\rm max}} \big((N+1) B_{N-1}  - (N-n_1) B_{N-2}\big) \frac{(-N)_{n_1}}{n_1!}\ .\label{telescopeSum0}
\end{align}
The constant $B_{\ti{N}}$ appears in the sum above when $N=\ti{N}+1$ or when $N=\ti{N}+2$.  The total coefficient of $B_{\ti{N}}$ is\vs{-5}
\be\begin{split}
&\frac{1}{n_1!n_3}\left((\ti{N}+2)(-(\ti{N}+1))_{n_1} -(\ti{N}-n_1+2)(-(\ti{N}+2))_{n_1}\right)  \\[2pt]
=\,&\frac{(-1)^{n_1}}{n_1!n_3}\left((\ti{N}+2)(\ti N-n_1+2)_{n_1} -(\ti{N}-n_1+2)(\ti N-n_1+3)_{n_1}\right) \\[2pt]
=\,&\frac{(-1)^{n_1}}{n_1!n_3}\left((\ti{N}-n_1+2)_{n_1+1} -(\ti{N}-n_2+2)_{n_1+1}\right)=0\ .
\end{split}\ee  
Thus, the sum \eqref{t0twistdownOPE} becomes a telescoping sum with 0 end contributions when \eqref{bid0} is implemented, making the identification \eqref{bid0} equivalent to the constraint \eqref{t0twistdownOPE}.  This can also be seen by taking the summation bounds in \eqref{telescopeSum0} and replacing them with a sum over $N$ that goes from $-\infty$ to $\infty$ and shifting the two sum indices.  This is possible because most of the $B_N$ are 0 in this sum, automatically truncating it. 

Thus, we find that \eqref{bid0} implements the linear relationship \eqref{t0twistdownOPE}.  Plugging \eqref{bid0} into the denominator of \eqref{near0copy} gives\vs{-5}
\be\label{near0copy_den}
\sum_{N=N_{\rm min}}^{N_{\rm max}}\!\! b_N P^{-(N+1),-(n_1+n_3-N)}_{n_1} (1-2t)
= t\!\!\! \sum_{N=\Nmin-1}^{\Nmax-2}\!\!\! B_N P^{-(N+1),-(n_1-1+n_3-1-N)}_{n_1-1}(1-2t)\ .
\ee

We may also check that \eqref{bid0} transforms the numerator in the appropriate way.  Plugging in \eqref{bid0} into the numerator of \eqref{near0} we find
\be\begin{split}
&\sum_{N=N_{\rm min}}^{N_{\rm max}} b_N t^{(N+1)}P^{(N+1),-(n_1+n_3-N)}_{n_3-N-1}(1-2t)  \\
=\frac{1}{n_3} &\sum_{N=N_{\rm min}}^{N_{\rm max}} (N+1)\,B_{N-1}\, t^{(N+1)}P^{(N+1),-(n_1+n_3-N)}_{n_3-N-1}(1-2t)  \\
 -\frac{1}{n_3} &\sum_{N=N_{\rm min}}^{N_{\rm max}} (N-n_1)\,B_{N-2}\, t^{(N+1)}P^{(N+1),-(n_1+n_3-N)}_{n_3-N-1}(1-2t)  \\
=\frac{1}{n_3} &\sum_{N=N_{\rm min}-1}^{N_{\rm max}-1} (N+2)\,B_{N}\, t^{(N+2)}P^{(N+2),-(n_1+n_3-1-N)}_{n_3-1-N}(1-2t)  \\
 -\frac{1}{n_3} &\sum_{N=N_{\rm min}-2}^{N_{\rm max}-2} (N-n_1+2)\,B_{N}\, t^{(N+3)}P^{(N+3),-(n_1+n_3-2-N)}_{n_3-1-N-2}(1-2t)
\end{split}\ee
where in the last step we have shifted the sum indices such that $B_N$ appears as a coefficient in both.  In the last equality, we may drop the top term in the first sum, and the bottom term in the second sum, given \eqref{endB0}.  This allows us to combine the sums into
\begin{align}\label{numBN}
&\sum_{N=N_{\rm min}}^{N_{\rm max}} b_N\, t^{(N+1)}P^{(N+1),-(n_1+n_3-N)}_{n_3-N-1}(1-2t) =\frac{1}{n_3} \sum_{N=N_{\rm min}-1}^{N_{\rm max}-2}\!t^{(N+2)} B_{N}  \\[3pt] &\qquad\times \left((N+2) P^{(N+2),-(n_1+n_3-1-N)}_{n_3-1-N}(1-2t) -(N-n_1+2)\,t\, P^{(N+3),-(n_1+n_3-2-N)}_{n_3-1-N-2}(1-2t)\right)\ . \nn
\end{align}
We again use \eqref{PidSumIndexShift.Main} with $\alpha=N+3$, $\gamma=n_3-N-2$ and $\beta=-(n_1+n_3-2-N)$ and obtain 
\be\label{near0copy_num}
\sum_{N=N_{\rm min}}^{N_{\rm max}} b_N t^{(N+1)}P^{(N+1),-(n_1+n_3-N)}_{n_3-N-1}(1-2t)=\!\!\!\!\sum_{N=N_{\rm min}-1}^{N_{\rm max}-2}\!\!\!t^{(N+2)} B_{N} P_{n_3-1-N-1}^{N+1,-(n_1-1+n_3-1-N)}(1-2t)\ .
\ee

All in all, plugging in  \eqref{near0copy_den} and \eqref{near0copy_num} into \eqref{near0} we find
\be\begin{split}
z(t)&=\frac{ \sum\limits_{N=N_{\rm min}}^{N_{\rm max}} b_N t^{(N+1)}P^{(N+1),-(n_1+n_3-N)}_{n_3-N-1}(1-2t)}{\sum_{N=N_{\rm min}}^{N_{\rm max}} b_N P^{-(N+1),-(n_1+n_3-N)}_{n_1} (1-2t) }  \\[3pt]
&= \frac{\sum\limits_{N=N_{\rm min}-1}^{N_{\rm max}-2}t^{(N+2)} B_{N} P_{n_3-1-N-1}^{N+1,-(n_1-1+n_3-1-N)}(1-2t)}{ t \sum_{N=\Nmin-1}^{\Nmax-2} B_N P^{-(N+1),-(n_1-1+n_3-1-N)}_{n_1-1}(1-2t)}  \\[3pt]
&= \frac{\sum\limits_{N=N_{\rm min}-1}^{N_{\rm max}-2}t^{(N+1)} B_{N} P_{n_3-1-N-1}^{N+1,-(n_1-1+n_3-1-N)}(1-2t)}{ \sum_{N=\Nmin-1}^{\Nmax-2} B_N P^{-(N+1),-(n_1-1+n_3-1-N)}_{n_1-1}(1-2t)}
\end{split}\ee 
which has the exact same form as our covering space maps \eqref{near0copy} with $n_1\rightarrow n_1-1$, $n_3\rightarrow n_3-1$, $\Nmin\rightarrow \Nmin-1$, and $\Nmax\rightarrow \Nmax-2$.  This implements a ``twist down'' at the origin from a twist-2 in the cloud approaching the origin, i.e. $r_0\rightarrow r_0-1$ and $r_c\rightarrow r_c-1$.  However, the ramifications $r_1$ and $r_\infty$ remain unchanged. We note that in this operation the covering space has lost a sheet, directly implemented by $n_3\rightarrow n_3-1$, which is the total number of sheets.  Thus, the general solution for a point $z_0=z(t)$ will have one fewer solution after \eqref{bid0} or equally \eqref{t0twistdownOPE} is implemented.  This loss of a sheet is seen as the direct cancelation of common polynomials in the numerator and denominator; $t$ in this case.  The order of the polynomial canceled in the reduction of the polynomials is exactly the number of sheets lost.  

We have therefore identified both kinds of OPE limits that give non-trivial group products when a twist-2 operator approaches the operator at the origin.        

As pointed out in the main text, the other forms of the map \eqref{near1} and \eqref{nearinfty} are similar in structure, and so one may read off the constraints analogous to \eqref{t0twistupOPE} and \eqref{t0twistdownOPE}.  These are given in table \ref{TableOPE}.  The identities used above, as well as those needed for $t=1$ and $t=\infty$ OPE limits, are proved in section \ref{app_OPEid}.  The appropriate identities to use are always found by insisting that the final form of the Jacobi polynomials are the same after shifting the index $N$, helping us identify the relevant identity needed.  One may, of course, simply plug in $b_N(\{B_M\})$ in table \ref{TableOPE} into the relevant form of the map, shift the sum index so that $B_N$ is the coefficient in the sums, and then use the quoted identity (which should be obvious how to use, given the coefficient of $B_N$).  This will result in a map of the same form \eqref{near0} or \eqref{near1} or \eqref{nearinfty} depending on the point of approach.

\subsection{\texorpdfstring{$\Delta N=2$}{TEXT} cloud twist down OPE}
\label{DN2TwistDownAppx}

We consider the OPE constraint for the $\Delta N=2$ case \eqref{cloud2twistdown}, which we claim is a ``twist down'' OPE limit.   We show this here.  First, we take $f_1$ with a convenient prefactor
\begin{align}
&\hspace*{-0.5cm}(\Nmin-n_1+2)(n_3-\Nmin-1)f_1 \nn \\
& \qquad =(\Nmin-n_1+2)(n_3-\Nmin-1)\sum_{N=N_{\rm min}}^{\Nmin+2} b_N P^{-(N+1),-(n_1+n_3-N)}_{n_1} (1-2t) \ . 
\end{align}
We solve \eqref{cloud2twistdown} for $b_{\Nmin+1}$ and substitute into the above, finding
\begin{align}
&\hspace*{-0.5cm}(\Nmin-n_1+2)(n_3-\Nmin-1)f_1 \nn \\
&=b_{\Nmin}(\Nmin-n_1+2)\Big((n_3-\Nmin-1)P^{-(\Nmin+1),-(n_1+n_3-\Nmin)}_{n_1} (1-2t) \nn \\
& \qquad \qquad  -(n_1+n_3-\Nmin-1)P^{-(\Nmin+2),-(n_1+n_3-\Nmin-1)}_{n_1} (1-2t)\Big) \label{cloud2twistdownmid1} \\
&+b_{(\Nmin+2)}(n_3-\Nmin-1)\Big((\Nmin-n_1+2)P^{-(\Nmin+3),-(n_1+n_3-\Nmin-2)}_{n_1} (1-2t) \nn \\
& \qquad \qquad  - (\Nmin+2)P^{-(\Nmin+2),-(n_1+n_3-\Nmin-1)}_{n_1} (1-2t)\Big) \ . \nn
\end{align}
Next consider the identity \eqref{PidSumIndexShift.Switch} and use the $(-1)^\gamma P_{\gamma}^{\alpha,\beta}(-x)=P^{\beta,\alpha}_\gamma(x)$ identity to write it as 
\be
(\gamma+\alpha+\beta-1)(1-t)P_{\gamma-1}^{\beta,\alpha}(1-2t)=-(\alpha-1)P_\gamma^{\beta-1,\alpha-1}(1-2t)+(\gamma+\alpha-1)P_\gamma^{\beta,\alpha-2}(1-2t)\ .
\ee
Evaluating this at $\alpha=-(n_1+n_3-\Nmin-2)$, $\beta=-(\Nmin+1)$, and $\gamma=n_1$, we find
\be\begin{split}\label{cloud2twistdownmid1_i}
&-n_3(1-t)P_{n_1-1}^{-(\Nmin+1),-(n_1+n_3-\Nmin-2)}(1-2t)  \\
& \qquad\qquad\qquad\qquad\qquad\qquad =(n_1+n_3-\Nmin-1)P_{n_1}^{-(\Nmin+2),-(n_1+n_3-\Nmin-1)}(1-2t)  \\
& \qquad \qquad\qquad\qquad\qquad\qquad -(n_3-\Nmin-1)P_{n_1}^{-(\Nmin+1),-(n_1+n_3-\Nmin)}(1-2t) 
\end{split}\ee
and immediately recognize the coefficient of $b_{\Nmin}$ in \eqref{cloud2twistdownmid1}.  We similarly use identity \eqref{PidSumIndexShift} with $\alpha=-(\Nmin+1)$, $\beta=-(n_1+n_3-\Nmin)+2$, and $\gamma=n_1$ and find
\be\begin{split}\label{cloud2twistdownmid1_ii}
&-n_3tP^{-(\Nmin+1),-(n_1+n_3-\Nmin-2)}_{n_1-1}(1-2t)  \\
&\qquad\qquad\qquad\qquad\qquad\qquad = -(\Nmin+2)P^{-(\Nmin+2),-(n_1+n_3-\Nmin-1)}_{n_1}(1-2t)  \\
& \qquad\qquad\qquad\qquad\qquad\qquad\quad+(\Nmin-n_1+2)P^{-(\Nmin+3),-(n_1+n_3-\Nmin-2)}_{n_1}(1-2t) 
\end{split}\ee
which allows us to identify the coefficient of $b_{(\Nmin+2)}$ in \eqref{cloud2twistdownmid1}.  Using \eqref{cloud2twistdownmid1_i} and \eqref{cloud2twistdownmid1_ii}, we find that $f_1$ \eqref{cloud2twistdownmid1} becomes
\begin{align}
(\Nmin-n_1+2)(n_3-\Nmin-1)f_1 &=-n_3 P^{-(\Nmin+1),-(n_1+n_3-\Nmin-2)}_{n_1-1}(1-2t) \label{cloud2twistdownmid1fin} \\
& \times(b_{\Nmin}(\Nmin-n_1+2)(t-1)+b_{\Nmin+2}(n_3-\Nmin-1)t). \nn
\end{align}
We note that this single Jacobi polynomial is the appropriate one for the denominator of \eqref{near0copy} with $n_i\rightarrow n_i-1$, $\Nmin\rightarrow \Nmin$, $\Nmax=\Nmin+2\rightarrow \Nmin$ for the ``twist down".  To be a twist down, the linear function appearing as a coefficient must cancel with the numerator, which we now set about showing.  

Exploring $f_2$ the same way, and introducing the same convenient coefficient, we find 
\be\begin{split}
& (\Nmin-n_1+2)(n_3-\Nmin-1)f_2 \\
&= (\Nmin-n_1+2)(n_3-\Nmin-1)  \sum_{N=N_{\rm min}}^{\Nmin+2} b_N t^{(N+1)}P^{(N+1),-(n_1+n_3-N)}_{n_3-N-1}(1-2t) \\
&= b_{\Nmin}(\Nmin-n_1+2) t^{(\Nmin+1)}\Big((n_3-\Nmin-1) P^{(\Nmin+1),-(n_1+n_3-\Nmin)}_{n_3-\Nmin-1}(1-2t) \\
& \qquad\qquad -(n_1+n_3-\Nmin-1)\,t\,P^{(\Nmin+2),-(n_1+n_3-\Nmin-1)}_{n_3-\Nmin-2}(1-2t) \Big) \label{cloud2twistdownmid2} \\
&+ b_{(\Nmin+2)} (n_3-\Nmin-1)t^{\Nmin+2}\Big((\Nmin-n_1+2)\,t\,P^{(\Nmin+3),-(n_1+n_3-\Nmin-2)}_{n_3-\Nmin-3}(1-2t) \\
& \qquad\qquad -(\Nmin+2)P^{(\Nmin+2),-(n_1+n_3-\Nmin-1)}_{n_3-\Nmin-2}(1-2t)\Big)\ .
\end{split}\ee
To start, we take \eqref{xid6} and subtract off $(\gamma+\beta+1)$ times \eqref{xid3} to arrive at
\be
(\gamma+\alpha+1)(1-t)P_{\gamma}^{\alpha,\beta+1}(1-2t)=\beta t P_{\gamma}^{\alpha+1,\beta}(1-2t)+(\gamma+1)\Big(t P_{\gamma}^{\alpha+1,\beta}(1-2t)+P_{\gamma+1}^{\alpha,\beta}(1-2t)\Big) \ .
\ee
The term in parentheses above can be rewritten using \eqref{tPid1} to give
\begin{align}
(\gamma+\alpha+1)(1-t)P_{\gamma}^{\alpha,\beta+1}(1-2t)=\beta\, t\, P_{\gamma}^{\alpha+1,\beta}(1-2t)+(\gamma+1) P_{\gamma+1}^{\alpha,\beta-1}(1-2t)\ .
\end{align}
Substituting $\gamma=n_3-\Nmin-2$, $\alpha=\Nmin+1$, and $\beta=-(n_1+n_3-\Nmin-1)$ above we have
\be\begin{split}
n_3(1-t)P_{\gamma}^{\alpha,\beta+1}(1-2t)=&-(n_1+n_3-\Nmin-1)\, t\, P_{n_3-\Nmin-2}^{(\Nmin+2),-(n_1+n_3-\Nmin-1)}(1-2t) \\[2pt]
& +(n_3-\Nmin-1) P_{n_3-\Nmin-1}^{(\Nmin+1),-(n_1+n_3-\Nmin)}(1-2t)
\end{split}\ee
allowing us to identify the term proportional to $b_{\Nmin}$ in \eqref{cloud2twistdownmid2}.  We next consider the identity \eqref{tPid1} with indices shifted 
\be
t P_{\gamma}^{\alpha+1,\beta}(1-2t)=P_{\gamma+1}^{\alpha,\beta-1}(1-2t)-P_{\gamma+1}^{\alpha,\beta}(1-2t)\ .
\ee
We take identity \eqref{xid7} and subtract off $(\gamma+1)$ times the above equation to find
\be
(\gamma+\alpha+\beta+1) t P_{\gamma}^{\alpha+1,\beta}(1-2t) =(\gamma+\alpha+1)P_{\gamma}^{\alpha,\beta}(1-2t)-(\gamma+1)P_{\gamma+1}^{\alpha,\beta-1}(1-2t)\ .
\ee
We now rewrite the lowest degree Jacobi polynomial on the right hand side using \eqref{xid2}, finding
\be
(\gamma+\alpha+\beta+1) t P_{\gamma}^{\alpha+1,\beta}(1-2t) =-(\gamma+\alpha+1)P_{\gamma+1}^{\alpha-1,\beta}(1-2t)+\alpha P_{\gamma+1}^{\alpha,\beta-1}(1-2t)\ .
\ee
Identifying $\gamma=n_3-\Nmin-3$, $\alpha=\Nmin+2$, and $\beta=-(n_1+n_3-\Nmin-2)$, we have
\begin{align}
&(\Nmin-n_1+2)\,t\,P_{n_3-\Nmin-3}^{(\Nmin+3),-(n_1+n_3-\Nmin-2)}(1-2t)-(\Nmin+2) P_{n_3-\Nmin-2}^{(\Nmin+2),-(n_1+n_3-\Nmin-1)}(1-2t) \nn\\[3pt]
& = -n_3P_{n_3-\Nmin-2}^{(\Nmin+1),-(n_1+n_3-\Nmin-2)}(1-2t)
\end{align}
allowing us to identify the coefficient of $b_{\Nmin+2}$ in \eqref{cloud2twistdownmid2}.  We find
\begin{align}
 (\Nmin-n_1+2)(n_3-\Nmin-1)f_2& = -n_3t^{\Nmin+1}P_{n_3-\Nmin-2}^{(\Nmin+1),-(n_1+n_3-\Nmin-2)}(1-2t)\label{cloud2twistdownmid2fin} \\[3pt]
& \times(b_{\Nmin}(\Nmin-n_1+2)(t-1)+b_{\Nmin+2}(n_3-\Nmin-1)t)\ . \nn
\end{align}
Thus, starting with the covering space map $z=\frac{f_2}{f_1}$ in \eqref{near0} for the case $\Delta N=2$, imposing \eqref{cloud2twistdown}, and using \eqref{cloud2twistdownmid2fin} and \eqref{cloud2twistdownmid1fin}, we obtain
\be
z(t)=\frac{f_2}{f_1}= \frac{t^{\Nmin+1} P_{n_3-\Nmin-2}^{(\Nmin+1),-(n_1+n_3-\Nmin-2)}(1-2t)}{P^{-(\Nmin+1),-(n_1+n_3-\Nmin-2)}_{n_1-1}(1-2t)} \label{twistDownCloudMap_app}
\ee
which is the appropriate covering space map with no cloud of twist-2 operators, and ramifications $r_0=\Nmin$, $r_1=(n_1+n_3-(\Nmin+2))=(n_1-1+n_3-1-\Nmin)$, and $r_\infty=n_1-n_3-1=(n_1-1)-(n_3-1)-1$. Hence, the ramification of these points have not been affected.  In this case, the 3-point function is simply that given in \cite{Lunin:2000yv} for three long twists.  

Note that the original 5-point function, i.e three long twists and two twist-2, has two cross ratios.  One might be concerned that the above OPE limit is only a linear relationship between the $b_N$, and so should decrease the space of maps only by 1.  However, it is important to note that the function that has been canceled, namely 
\be
b_{\Nmin}(\Nmin-n_1+2)(t-1)+b_{\Nmin+2}(n_3-\Nmin-1)\,t\label{LinCancCloud}
\ee
determines the two coincident zeros of $Q(t)$, i.e. where the ramified points come together on the covering surface when \eqref{cloud2twistdown} is enforced.  This point is given by
\be
t_{\downarrow}= \frac{b_{\Nmin}(\Nmin-n_1+2)}{b_{\Nmin}(\Nmin-n_1+2)+b_{(\Nmin+2)}(n_3-\Nmin-1)} \label{tmAppx}
\ee 
matching \eqref{tmMain} in the main text.  This is a marked point on the cover where the ramified points approach each other, and we expect an OPE expansion, as explained in the main text after \eqref{tmMain}.

\section{Ramification subadditivity}\label{appx.radd}

In this appendix we briefly discuss ramification subadditivity under group multiplication in the symmetric group (or permutation groups).  

We recall several facts about the symmetric group.  First, every group element can be written as a product of disjoint cycles.  This is simply argued by considering the permutation group acting on a set of distinct elements in ordered positions, finding how these are acted on by permutation.  We start with the objects in their respective positions (first object in first position, etc).  We track where each element gets mapped in the following way.  We consider an element, say the first, and see where it gets mapped.  This displaces the object in that position and we can ask to what position that object is mapped, and so on.  Eventually, one must find the object that maps back to the first position.  This gives one of the cycles.  One then considers one of the elements outside of the cycle(s) already discussed and repeats this procedure.  Eventually, all elements are addressed, and each cycle refers to distinct elements and positions, and so the cycles are disjoint.  This decomposition is unique up to the ordering of these commuting cycles.  Trivial cycles are those that have only one element, corresponding to unmapped elements, and are usually omitted from notation.  We refer to this form of the group elements as their canonical form.  One may still regard the decomposition as unique if one includes the trivial cycles, however, each trivial cycle is included only once.  In this way we may guarantee that each index appears precisely once in each group element's product of cycles.  We call this form the canonical form as well, realizing that the trivial cycles may be dropped if one wishes.    

Next, every single cycle may be written as a product of two-cycles which only include the indices of the cycle itself.  By direct construction
\be
(1,2,3,\cdots, n)=(1,2)(2,3)(3,4)\cdots (n-1,n)\ .
\ee
Therefore, each non-trivial cycle of the canonical form may be decomposed into a product of two cycles which is unique using the above prescription, up to cyclic reordering of original cycle.  Each block of two cycles, corresponding to each non-trivial cycle of the canonical form, refers to a set of indices that is disjoint from every other block of two cycles.  We call this the decomposed form of the group element.  The total ramification of the group element is given by the sums of the ramifications of each cycle in the group element, and the ramification of a single cycle is the number of indices in the cycle minus 1.  This total ramification is the same as the total number of two cycles it takes to build the group element in the above way, and represents a minimal number of such two cycles: the two concepts coincide.  

Next, we consider the action of a group element $g_1$ on a group element $g_2$.  We consider decomposing $g_1$ into two cycles.  Of these two cycles, one may be regarded as being in the right-most position in $g_1$, and without loss of generality, we consider this cycle to be $(1,2)$.  We consider its action on $g_2$, and consider $g_2$ in canonical form, but explicitly writing the trivial cycles out, once each.  In this way, all indices appear in $g_2$ exactly once.  

There are different possibilities. First, it may be the case that the indices of $(1,2)$ appear distinct cycles in $g_2$.  In this case, we write\vs{-5}
\be
g_2=(1,3,4,\cdots,n)(2,n+1,\cdots,n+m) \ti{g}_2\ ,
\ee
where $\ti{g}_2$ is a product of disjoint cycles.  One may pull both of these two cycles to the left in $g_2$ because they commute with each other, and they each commute with all cycles in $\ti{g}_2$.    In this case,
\be
(1,2)g_2=(1,2)(1,3,4,\cdots,n)(2,n+1,\cdots,n+m)\ti{g}_2=(1,3,4,\cdots,n,2,n+1,\cdots,n+m) \ti{g}_2\ .
\ee
The right hand side is already in canonical form: the indices appearing in $(1,3,4,\cdots,n,2,n+1,\cdots,n+m)$ do not appear in $\ti{g}_2$.  This product even works in the case that either or both of the original cycles brought to the left of $g_2$ is trivial, simply by deleting the indices $3,\cdots,n$ from both sides, or deleting the indices $n+1,\cdots, n+m$ from both sides, or doing both simultaneously.  Before the product the total ramification was $1+r_{g_2}=1+n-2+m+r_{\ti{g}_2}=n+m-1+r_{\ti{g}_2}$.  After the product, the ramification is $n+m-1+r_{\ti{g_2}}$, and so the ramification is maintained before and after the product.  We refer to this operation as a ``join'' which joins two previously disjoint cycles, and is ``ramification preserving''.  Some of the cycles that have been joined may have been trivial cycles.  

The other possibility is that the indices $1$ and $2$ may both appear in the same cycle of $g_2$.  This may only happen when the cycle of $g_2$ in question is non-trivial.  In this case, we write 
\be
g_2=(1,3,4,\cdots,n,2,n+1,n+2,\cdots,n+m)\ti{g}_2
\ee
by bringing the cycle in $g_2$ that has the indices $1$ and $2$ to the left-most position in $g_2$.  Neither 1 nor 2 appear in the remaining cycles of $\ti{g}_2$.  Multiplying out we find
\begin{align}
(1,2)g_2& =(1,2)(1,3,4,\cdots,n,2,n+1,n+2,\cdots,n+m)\ti{g}_2 \nn \\
& =(1,3,4,\cdots,n)(2,n+1,n+2,\cdots,n+m)\ti{g}_2 \ .
\end{align}
The above again applies even when deleting the indices $3,\cdots, n$ from both sides, or deleting the indices $n+1, \cdots, n+m$ from both sides, or doing both simultaneously, in which case trivial cycles appear on the right hand side and may be omitted.  The product on the right hand side above is already in canonical form, and so the ramification is easy to read.  The total sum of ramifications of the individual group elements is $1+r_{g_2}=1+n+m-1+r_{\ti{g}_2}=n+m+r_{\ti{g}_2}$, and the ramification of the product is $n-2+m+r_{\ti{g}_2}$, and so the ramification has decreased by 2.  This counting is still valid even in the special cases where the indices are deleted, explained above.  This operation we regard as a ``split'', and is ``ramification decreasing''.

Iterating this with all of the two-cycles in the decomposed form of $g_1$ gives that the ramifications obey 
\be
r_{g_1}+r_{g_2}\geq r_{(g_1 g_2)}\ .
\ee
Iterating this again with a series of group elements we see that
\be
r_{g_1}+r_{g_2}+r_{g_3}+\cdots+r_{g_n}\geq r_{(g_1g_2\cdots g_n)}\ .
\ee
Thus, the group product is at best ramification preserving, but often reduces total ramification.

As an important consequence, if we have a group product $g_1 g_2 \cdots g_n=e$, where $e$ is the identity element, then the initial product $g_1 \cdots g_{n-1}=g_n^{-1}$.  The ramification of $g_n$ and $g_{n}^{-1}$ are the same (they are made out of the same size cycles).  Therefore, one must have that
\be
r_{g_1}+r_{g_2}+\cdots+r_{g_{n-1}}\geq r_{g_{n}}\ .
\ee
One may reach the same conclusion by moving any of the $r_{g_i}$ to the right, finding that the group product must obey
\be
\sum_{i\neq j} r_{g_i} \geq r_{j}
\ee
for all terms in the product whenever considering a case where $g_1 g_2 \cdots g_n= e$.  This must, in fact, be the case for individual cycles in each of the $g_i$ as well, given that we can decompose each group element into cycles.

\section{Finding the Wronskian for finite \texorpdfstring{$\Delta N\geq 3$}{TEXT}}\label{WronskGen}

The form of the Wronskian may be used to fix the coefficients $A_i$ in \eqref{Qform}, given the expansion of the Jacobi polynomials by looking at the first $\Delta N+1$ highest powers of $t$, which we now show. Expanding the right hand side of \eqref{wronskform} we find
\be\begin{split}
W&=A_0 t^{n_1+n_3-1}+(-(n_1+n_3-\Nmax-1)A_0+A_1)t^{n_1+n_3-2}  \\[3pt]
& +\bigg(\frac{(n_1+n_3-\Nmax-2)_{(2)}}{2!}A_0-(n_1+n_3-\Nmax-1)A_1+A_2\bigg)t^{n_1+n_3-3}+\ldots\ .
\end{split}\ee
Above we have only shown the first three terms although one can expand to any order.
We can see that the highest powers of $t$ in the polynomials $f_1$ and $f_2$ fix $A_0$.  With this in hand, the highest and second highest powers of $t$ in $f_1$ and $f_2$ fix $(-(n_1+n_3-\Nmax-1)A_0+A_1)$, which given the last step, gives $A_1$.  Repeating this process gives the $A_i$ in terms of the coefficients of the $\Delta N+1$ largest powers of $t$ appearing in $f_1$ and $f_2$. We now turn to finding these coefficients.

We expand
\begin{align}
P_{n_1}^{-(N+1),-(n_1+n_3-N)}(1-2t)&= \bigg(\frac{(n_3-n_1+1)_{(n_1)}}{n_1!}t^{n_1} +\frac{(n_3-n_1+2)_{(n_1-1)}(n_1-N-1) n_1 }{n_1!}t^{n_1-1} \nn  \\[3pt]
& +\frac{(n_3-n_1+3)_{(n_1-2)}(n_1-N-2)_{(2)}(n_1-1)_{(2)}}{n_1!2!} t^{n_1-2}+\cdots \bigg) 
\end{align}
\vs{-3}and
\begin{align}
& t^{N+1} P^{(N+1),-(n_1+n_3-N)}_{n_3-N-1}(1-2t)=(-1)^{n_3-N-1}\bigg(\frac{(N-n_1+1)_{(n_3-N-1)}}{(n_3-N-1)!}t^{n_3} \\[3pt]
&\qquad\qquad\qquad  -\frac{(N-n_1+1)_{(n_3-N-2)}n_3}{(n_3-N-2)!}t^{n_3-1}  +\frac{(N-n_1+1)_{(n_3-N-3)}(n_3-1)_{(2)}}{(n_3-N-3)!2!}t^{n_3-2}+\cdots \bigg) \nn
\end{align}
keeping track of the denominator term, interpreting it as being infinite when $(n_3-N-1)$ is a negative integer, i.e. removing the Jacobi polynomial following the rule \eqref{rule}.  Plugging these expressions into $f_1$ and $f_2$ we arrive at the expansions
\be
f_1=d_{1,0} t^{n_1}+d_{1,1}t^{n_1-1}+d_{1,2}t^{n_1-2}+\ldots \ ,\qquad\qquad f_2=d_{2,0} t^{n_3}+d_{2,1}t^{n_3-1}+d_{2,2}t^{n_3-2}+\ldots\ ,
\ee
with
\begin{align}
& d_{1,0}=\!\!\sum_{N=\Nmin}^{\Nmax} \frac{(n_3-n_1+1)_{(n_1)}}{n_1!} b_N\ , \qquad
d_{1,1}=\!\!\sum_{N=\Nmin}^{\Nmax} \frac{(n_3-n_1+2)_{(n_1-1)}(n_1-N-1) n_1 }{n_1!}b_N\ , \nn \\[3pt]
& d_{1,2}=\!\!\sum_{N=\Nmin}^{\Nmax} \frac{(n_3-n_1+3)_{(n_1-2)}(n_1-N-2)_{(2)}(n_1-1)_{(2)}}{n_1!2!} b_N\ ,\nn
\end{align} 
and
\begin{align}
& d_{2,0}=\!\!\sum_{N=\Nmin}^{\Nmax} (-1)^{n_3-N-1} \frac{(N-n_1+1)_{(n_3-N-1)}}{(n_3-N-1)!}b_N\ ,\nn\\
& d_{2,1}=-\!\!\sum_{N=\Nmin}^{\Nmax} (-1)^{n_3-N-1} \frac{(N-n_1+1)_{(n_3-N-2)}n_3}{(n_3-N-2)!} b_N\ ,\\
& d_{2,2}=\!\!\sum_{N=\Nmin}^{\Nmax} (-1)^{n_3-N-1} \frac{(N-n_1+1)_{(n_3-N-3)}(n_3-1)_{(2)}}{(n_3-N-3)!2!} b_N \ .  \nn
\end{align} 
Putting these expansions into the Wronskian, we obtain
\begin{align}
&A_0=(n_3-n_1)\,d_{1,0}\,d_{2,0}\ ,\nn \\[3pt]
&A_1=(n_1+n_3-\Nmax-1)\,A_0+(n_3-n_1-1)\,d_{1,0}\,d_{2,1}+(n_3-n_1+1)\,d_{1,1}\,d_{2,0}\ ,\nn \\[3pt]
&A_2=-\frac{(n_1+n_3-\Nmax-2)_2}{2!}\,A_0+(n_1+n_3-\Nmax-1)\,A_1 \nn \\
&  \qquad\, +(n_3-n_1-2)\,d_{1,0}\,d_{2,2}+(n_3-n_1)\,d_{1,1}\,d_{2,1}+(n_3-n_1+2)\,d_{1,2}\,d_{2,0}\ .  
\end{align}
The above $A_0$ agrees with the generic expression \eqref{A0gen}.  Extending these to higher order terms is straightforward and algorithmic.  One may also use similar procedures by expanding near $t=0$ or $t=1$.

\small\baselineskip=.87\baselineskip
\bibliographystyle{utphys}
\bibliography{coveringmaps}

\end{document}